\documentclass[journal=jcim,manuscript=article]{achemso}

\usepackage[version=3]{mhchem} % Formula subscripts using \ce{}
\usepackage{hyperref}
\usepackage{booktabs}
\usepackage{multirow}
\usepackage{graphicx}
\graphicspath{ {Figures/} }
\usepackage{caption}
\usepackage{subcaption}
\usepackage{float}
\usepackage{tabularx}

\author{María Virginia Sabando}
\affiliation{Institute for Computer Science and Engineering (UNS-CONICET), Bahía Blanca, Argentina}
\alsoaffiliation{Department of Computer Science and Engineering, Universidad Nacional del Sur, Bahía Blanca, Argentina}
\email{virginia.sabando@cs.uns.edu.ar}
\author{Ignacio Ponzoni}
\affiliation{Institute for Computer Science and Engineering (UNS-CONICET), Bahía Blanca, Argentina}
\alsoaffiliation{Department of Computer Science and Engineering, Universidad Nacional del Sur, Bahía Blanca, Argentina}
\author{Evangelos E. Milios}
\affiliation{Faculty of Computer Science, Dalhousie University, Halifax, NS, Canada}
\author{Axel J. Soto}
\affiliation{Institute for Computer Science and Engineering (UNS-CONICET), Bahía Blanca, Argentina}
\alsoaffiliation{Department of Computer Science and Engineering, Universidad Nacional del Sur, Bahía Blanca, Argentina}

\title{Using Molecular Embeddings in QSAR Modeling: Does it Make a Difference?}

\abbreviations{MVS,IP,EEM,AJS}

\begin{document}


\begin{abstract}

%%%%%%%%%%%%%%%%%%%%%%%%%%%%%%%%%%%%%%%%%%%%%%%%%%%%%%%%%%%
%%      ABSTRACT 
%%%%%%%%%%%%%%%%%%%%%%%%%%%%%%%%%%%%%%%%%%%%%%%%%%%%%%%%%%%
With the consolidation of deep learning in drug discovery, several novel algorithms for learning molecular representations have been proposed. Despite the interest of the community in developing new methods for learning molecular embeddings and their theoretical benefits, comparing molecular embeddings with each other and with traditional representations is not straightforward, which in turn hinders the process of choosing a suitable representation for QSAR modeling. A reason behind this issue is the difficulty of conducting a fair and thorough comparison of the different existing embedding approaches, which requires numerous experiments on various datasets and training scenarios. To close this gap, we reviewed the literature on methods for molecular embeddings and reproduced three unsupervised and two supervised molecular embedding techniques recently proposed in the literature. We compared these five methods concerning their performance in QSAR scenarios using different classification and regression datasets. We also compared these representations to traditional molecular representations, namely molecular descriptors and fingerprints. As opposed to the expected outcome, our experimental setup consisting of over $25,000$ trained models and statistical tests revealed that the predictive performance using molecular embeddings did not significantly surpass that of traditional representations. While supervised embeddings yielded competitive results compared to those using traditional molecular representations, unsupervised embeddings tended to perform worse than traditional representations. Our results highlight the need for conducting a careful comparison and analysis of the different embedding techniques prior to using them in drug design tasks, and motivate a discussion about the potential of molecular embeddings in computer-aided drug design.
\end{abstract}

\section{Introduction}\label{sec:introduction}

% P1: QSAR modeling uses traditional reps
Quantitative Structure-Activity Relationship (QSAR) models constitute the cornerstone of modern in silico drug discovery \cite{wu2020we, wu2021hyperbolic, cherkasov2014qsar}. QSAR models are regression or classification models that predict the relationship between molecular features of compounds and a target property. They are extensively used in modern drug discovery to accelerate drug candidate identification while reducing costs associated with molecular synthesis and wet lab experiments. 
QSAR models are usually trained using traditional molecular representations, such as molecular descriptors and fingerprints, which are computed using widely known algorithms \cite{todeschini2009molecular}. Although these molecular representations are well-established and often yield good classification results, %they are limited in the molecular information they encode. For instance, molecular descriptors such as molecular weight or \textit{cLogP} (partition coefficient octanol/water) value might have a high correlation to specific properties, such as water solubility or boiling point, but might not be appropriate for those tasks where capturing aspects of the molecular structure is crucial, such as molecular docking. Because of this, 
there is an increasing interest in exploring new, enriched representations that take into account multiple aspects of the molecule \cite{chuang2020learning}.

% P2:  Recent years, deep learning, mol embs, arise, advantages over traditional mol reps
Deep learning techniques applied to Natural Language Processing (NLP), originally developed for text, have experienced a prolific development during the last decade and are being increasingly applied to other domains \cite{elton2019deep, chen2018rise, bouhedjar2020natural}. In drug design, significant research efforts have been put on designing novel techniques for learning rich molecular representations during the last years \cite{chuang2020learning, elton2019deep, david2020molecular}. These representations, which we refer to as \textit{molecular embeddings}, can be trained in different ways to capture diverse information about physical-chemical properties, molecular structure and bioactivity. In addition, unsupervised representation learning methods inspired by NLP approaches can be adapted to use huge amounts of molecular information in SMILES format \cite{weininger1988smiles}, thus benefiting from the inherent diversity in large ensembles of unlabeled compounds.

\begin{figure*}
\centering
\includegraphics[width=\textwidth]{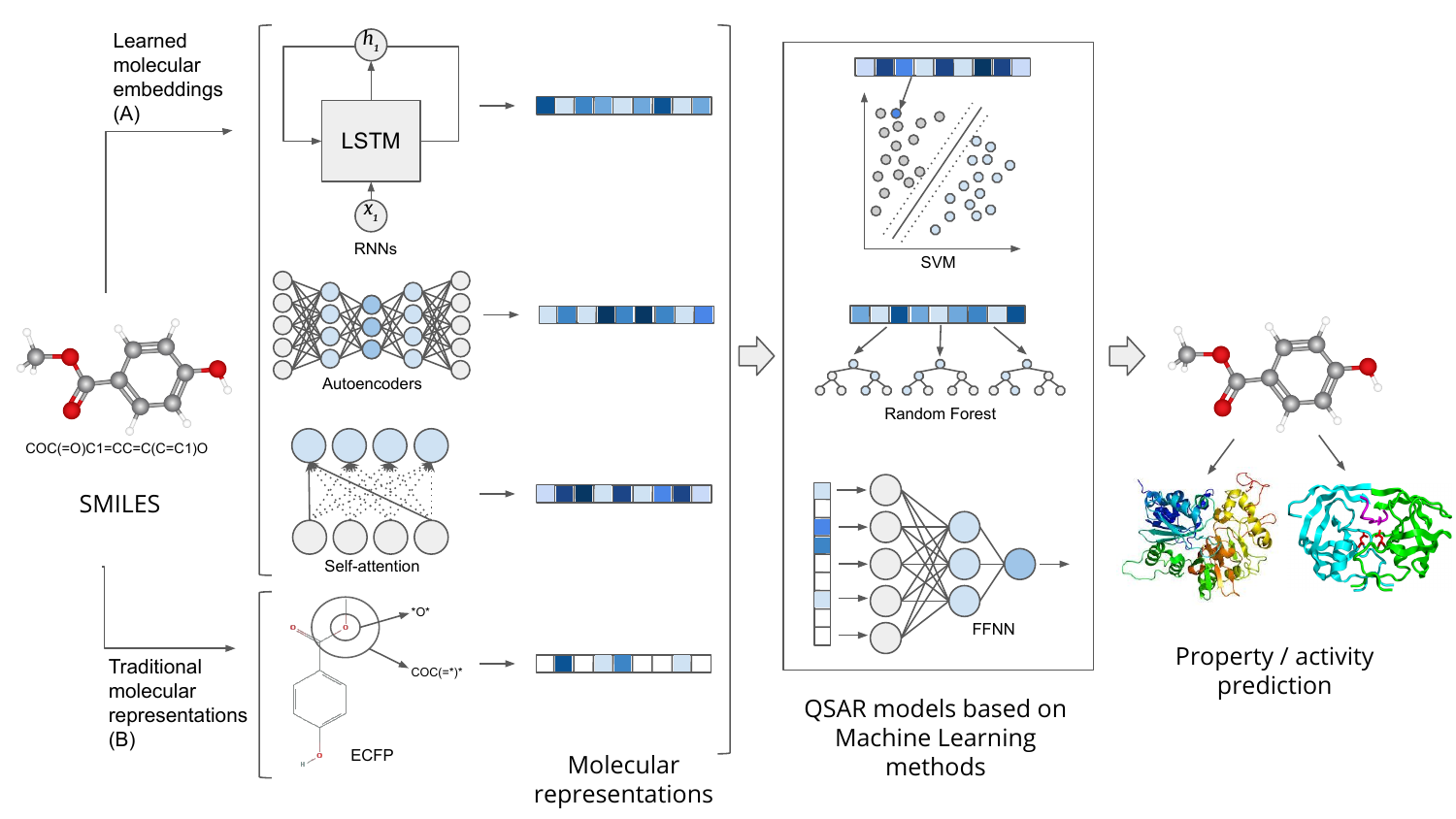} 
\caption{QSAR models can either be trained using learned molecular embeddings derived from deep learning and NLP techniques (A), or employing traditional molecular representations obtained from a \textit{feature engineering} process (B). Molecular embeddings are dense vectors of real numbers, whereas traditional molecular representations can be either real number vectors or bit vectors varying in sparsity. QSAR models are classification or regression models that are trained with these representations.}
\label{figure:qsar_molreps}
\end{figure*}

% P3: variety: intricacy, techniques employed. Unsupervised vs. supervised embeddings. Explanation of both categories. 
Algorithms for learning molecular embeddings employ a wide variety of state-of-the-art deep learning techniques \cite{elton2019deep}, ranging from neural-based autoencoders to graph neural networks \cite{bouhedjar2020natural, wu2020comprehensive} and self-attention \cite{vaswani2017attention}. Molecular embedding algorithms vary in complexity and intricacy, allowing to tailor the learned representations to specific tasks. For instance, many recent self-attention methods for molecular embeddings are designed to identify the molecular substructures that have a significant impact on their bioactivity profile \cite{oskooei2018paccmann, zheng2019identifying}. Nowadays, several QSAR studies employ molecular embeddings instead of traditional representations as training data \cite{chuang2020learning, wu2018moleculenet,  jiang2021could}. As depicted in Figure \ref{figure:qsar_molreps}, molecular embeddings are usually fixed-size dense vectors of real numbers, whereas traditional molecular representations can either consist of real number vectors or bit vectors varying in sparsity. These representations are fed to classification or regression models for QSAR modeling.

% FIGURA 1 VA AQUI
% \begin{figure*}
%   % \centering{{\color{black!20}\rule{100pt}{30pt}}}
%      \centering \rulebox{Figure \ref{figure:qsar_molreps} goes here}
% \caption{QSAR models can either be trained using learned molecular embeddings derived from deep learning and NLP techniques (A), or employing traditional molecular representations obtained from a \textit{feature engineering} process (B). Molecular embeddings are dense vectors of real numbers, whereas traditional molecular representations can be either real number vectors or bit vectors varying in sparsity. QSAR models are classification or regression models that are trained with these representations.}
% \label{figure:qsar_molreps}
% \end{figure*}

% CODIGO LATEX FIGURA 1 (DESCOMENTAR)
% FIN CODIGO LATEX FIGURA 1 (DESCOMENTAR)

% P4: Unsupervised techniques potential over supervised techniques. Advantages and disadvantages. 
At learning time, molecular embeddings can either be \textit{supervised}, meaning that they take into account labeled/external information about the bioactivity profile of the molecules during training, or \textit{unsupervised}, which means that the embeddings are built with no human-labeled information. While supervised embeddings encode bioactivity or physical-chemical information, which could favor them in predictive tasks, they are consequently less flexible than unsupervised embeddings, since new embeddings need to be learned for different biological targets. Moreover, in order to learn supervised embeddings it is necessary to have labeled datasets, which are often small and scarce, thus potentially affecting the embedding quality negatively. 

% P5: Presentation of our proposal, research questions
\sloppy Despite the proliferation of learned molecular representation methods, there is no empirical research on how to choose the most suitable representation method for QSAR analysis. Interestingly, some recent studies have been published where molecular embeddings appear to only match or slightly surpass traditional representations in QSAR modeling \cite{jiang2021could, jaeger2018mol2vec, gomez2018automatic, goh2017chemception, yang2019analyzing, erratumyang2019analyzing, chithrananda2020chemberta}. While establishing a fair comparison between molecular embeddings and traditional representations is not straightforward, we argue that such comparison is necessary and that it needs to be carried out through an extensive and careful experimental workflow. In order to establish a comparison on a fair ground, we propose a carefully designed evaluation where hyperparameters are optimized thoroughly, datasets of different characteristics are used, and several models are trained to account for their inherent stochasticity. In particular, this work aims to address the following research questions:
\begin{itemize}
    \item Q1: What are the main molecular embedding methods used for QSAR modeling in the literature? Do they outperform traditional molecular representations in a classification/regression task?
    \item Q2: Does incorporating information about the biological target into the molecular embedding (supervised embeddings) incur higher predictive performance than that obtained from unsupervised embeddings? 
    \item Q3: Do different preprocessing decisions, such as the canonical form of the SMILES formulas used as training data or the size of the final embeddings, significantly impact on the predictive performance of a QSAR model using molecular embeddings?
\end{itemize}

In this scenario, we conducted an extensive analysis of different molecular embedding techniques based on state-of-the-art deep learning methods focusing on their suitability for QSAR modeling. We reproduced five recently proposed embedding techniques: two \textit{supervised} and three \textit{unsupervised} techniques. All five techniques were trained using SMILES formulas as the raw molecular representation. To train the unsupervised methods, we employed a dataset of 40 million randomly selected and curated compounds retrieved from the ZINC database \cite{sterling2015zinc}. In addition, we tested two different ways of computing canonical SMILES formulas in an effort to leverage the way in which each unsupervised technique processes the SMILES formulas. 
We trained different classifiers using eight labeled datasets with varying class imbalance levels and sizes, corresponding to both classification and regression tasks. Finally, we contrasted the performance attained by molecular embeddings to the baseline results obtained using traditional molecular representations. Our experimental workflow consisted of over $25,000$ trained models, which comprised several stages of model selection and replications, followed by statistical analyses of the results.

\section{Related work}\label{sec:related_work}

% P1: Molecular representations: traditional mol reps: descriptors, ECFP, MACCs keys, brief overview, advantages and disadvantages. Examples of their usage in QSAR modeling. 

The trustworthiness of QSAR models and their predictive performance is linked to the choice of the molecular representation used for training \cite{chuang2020learning, sabando2021chemva}. For this reason, for many years the standard has been to manually engineer high-quality molecular representations. This process, known as \emph{feature engineering}, has led to widespread and commonly used molecular representations \cite{todeschini2009molecular, cereto2015molecular}. Traditional molecular representations vary mostly in the type of information they encode and their use depends on the specific task \cite{grisoni2018impact, schneider2010virtual}. 
Among the most commonly used traditional molecular representations, we can enumerate: \textit{molecular descriptors} \cite{todeschini2009molecular}, \textit{Extended Connectivity Fingerprints (ECFPs)} \cite{rogers2010extended} and \textit{Molecular ACCess System (MACCS)} keys \cite{durant2002reoptimization}. 

% P2: Learned mol reps: broad taxonomy: from feature engineering to feature learning. 
While a vast number of published QSAR models are based on these traditional representations \cite{seth2020qsar, yang2020qsar, gao20202d, sabando2019neural}, the process of obtaining such representations via feature engineering is costly and requires strong domain-specific expertise. Moreover, since each of them encodes different information, there is no single representation suitable for every task. %Traditional approaches for molecular representations often rely on a feature selection step, which introduces further design decisions before training a machine learning model \cite{idakwo2019review}. Furthermore, modern cheminformatics methods 
For this reason, there is an increasing trend towards the use of versatile molecular representations that capture diverse aspects of the chemical space \cite{chuang2020learning}, so many new methods for learning molecular embeddings have been developed recently, where most of them are based on deep learning techniques \cite{chuang2020learning, elton2019deep}. 

% P3: techniques based on smiles, why not other graph representations
Recent methods for learning molecular embeddings employ \textit{SMILES} formulas \cite{weininger1988smiles}, which is the most widely used linear representation for encoding molecular graph information. Since SMILES formulas directly encode the molecular graph into a sequence of ASCII characters, they can be used by deep learning techniques designed for sequential or graph-like data \cite{elton2019deep, jiang2021could}. Given their high availability---most molecular databases are stored in SMILES format---, many unsupervised methods for learning molecular representations have been developed and trained on large databases without the need for labeled information \cite{jaeger2018mol2vec, liu2018n, swann2018representing, ozturk2018novel, xu2017seq2seq}. While images or structured graph representations have also been employed for learning molecular embeddings in QSAR modeling \cite{goh2017chemception, kuzminykh20183d,shi2019molecular}, their usage is not as widespread and has limitations in terms of data availability. Thus, in this paper, we focused on SMILES-based embedding methods.

% P4 5 & 6: Techniques: autoencoders of many types, RNNs GNNs, Self-attention (is a way of GNN). Brief comment on generative models for de novo design (related to autoencoders). Focus on comparing self attention (GNN) to autoencoders. INTRODUCE OUR FIVE TECHNIQUES AND DESCRIBE THEM BRIEFLY.
The high availability of molecular data in SMILES format has motivated numerous approaches for learning molecular embeddings based on autoencoders, which enable to conduct unsupervised training procedures \cite{jaeger2018mol2vec,gomez2018automatic,xu2017seq2seq,ozturk2018chemical}. In particular, 
%\citet{ozturk2018novel} 
{\"O}zt{\"u}rk et al.~\cite{ozturk2018novel} introduced \textit{SMILESVec}, an unsupervised SMILES-based method that learns representations for small molecules using the popular \textit{word2vec} model by 
%\citet{mikolov2013efficient}
Mikolov et al.~\cite{mikolov2013efficient}. The authors performed a \textit{tokenization} step onto a dataset of SMILES formulas---i.e., converted each SMILES into a string of distinguishable tokens in an alphabet--- by computing overlapping SMILES substrings, which were then used to train their model. Finally, they computed an embedding for each molecule in the corpus by averaging the learned vectors for each of the tokens in its SMILES formula. 
%\citet{jaeger2018mol2vec}
Jaeger et al.~\cite{jaeger2018mol2vec} presented \textit{Mol2Vec}, which is also based on \textit{word2vec} \cite{mikolov2013efficient}. In this method, SMILES formulas undergo a preprocessing stage before being fed to the autoencoder, which consists of tokenizing the SMILES formulas using the Morgan algorithm for ECFPs \cite{rogers2010extended}. After an unsupervised training phase, the final representations are computed by summing the embedded vectors of all tokens in the molecule. %The preprocessing step carried out by this method involves exploring the topology of the molecule in a circular way, which resembles the way Graph Neural Networks (GNNs) learn from the topology of graph-like data \cite{NIPS2015_f9be311e}.

Since SMILES is a linear molecular notation of a sequential nature, methods for either generating new molecules or learning new molecular embeddings based on Recurrent Neural Networks (RNNs) are also found in the literature \cite{xu2017seq2seq,segler2018generating,popova2018deep}. Most generative models in the literature are based on RNN autoencoders trained on large sets of unlabeled data \cite{elton2019deep}, which constitutes an interesting approach for learning molecular representations. %given that they capture the graph information encoded throughout the whole SMILES string. 
In this direction, 
%\citet{xu2017seq2seq} 
Xu et al.~\cite{xu2017seq2seq} proposed \textit{Seq2Seq Fingerprint}, an unsupervised method based on a multi-layer RNN autoencoder built using Gated Recurrent Unit cells (GRUs) \cite{Cho2014}. \textit{Seq2Seq Fingerprint} learns directly from SMILES formulas leveraging information from long-term dependencies and sequential relationships between characters. The molecular embeddings are finally obtained by concatenating the hidden states of the autoencoder.

A recent trend in learned molecular representations is the use of self-attention \cite{vaswani2017attention}, which enables to capture rich information about relevant substructures of the molecule, while avoiding the lengthy training procedures of RNNs \cite{vaswani2017attention}. Self-attention strongly relates to the way GNNs operate on graph-like data \cite{joshi2020transformers} but are designed to learn from sequence data such as SMILES formulas. In this direction, 
% \citet{oskooei2018paccmann} 
Oskooei et al.~\cite{oskooei2018paccmann} developed \textit{PaccMann}, a novel approach for predicting anticancer compound sensitivity by means of multi-modal attention-based neural networks. The authors proposed a series of supervised encoders; the best performing one was a self-attentive encoder trained using a simple tokenization of SMILES strings and complemented with gene expression information. A shallow feed-forward neural network for property prediction was stacked to the embedding model so that the molecular embeddings were learned during the training phase of the classifier or regression model. 
%\citet{zheng2019identifying}
Zheng et al.~\cite{zheng2019identifying} developed \textit{SA-BiLSTM}, a supervised method based on a combination of a Bidirectional Long-Short Term Memory (LSTM) \cite{hochreiter1997long} RNN and a self-attention mechanism. This model is then plugged into a shallow feed-forward neural network for property prediction, thus the self-attentive embeddings are learned along with the prediction task. One interesting aspect of this method is that it does not learn from SMILES formulas directly but instead uses sequences of token embeddings from a pretrained \textit{Mol2Vec} model \cite{jaeger2018mol2vec}. 
%\citet{wang2019smiles} 
Wang et al.~\cite{wang2019smiles} developed \textit{SMILES-BERT}, a semi-supervised model consisting of a Transformer Layer \cite{vaswani2017attention} trained using a large unlabeled dataset through a masked SMILES recovery task, as in the popular NLP \textit{BERT} architecture \cite{devlin2018bert}. This model can be then fine-tuned by means of a labeled dataset. Although \textit{SMILES-BERT} seems a promising approach, important implementation details are missing to fully reproduce the method, especially in terms of the fine-tuning stage. 

\section{Materials and methods}\label{sec:materials_and_methods}

% P1: In this section: datasets (ZINC, 5 labeled DBs), 5 techniques, plus one canonicalization technique.

%In this section, we describe the datasets and their preprocessing stage applied in this paper. 
In order to enable reproducibility of this paper, in this section we provide a thorough explanation of our experimental setup and a workflow overview, including a description of the datasets and their preprocessing stage, the neural architectures of the five methods we reproduced and their training. Finally, we provide details on the classification and regression stages by which we evaluated the learned molecular embeddings. %All data and source code used in this paper are either cited to the original reference or made publicly available to facilitate the reproducibility of our work. In addition, all trained embedding models are also available so that they can be readily used. 
Further details about materials and source code can be found in Section ``Data availability statement".

% subsection 1: Datasets: brief description and citation of the datasets used in the experiments (ZINC and 5 others). Data retrieval, curation and preparation.
\subsection{\textbf{Datasets}}\label{subsec:datasets}

Following the steps needed to reproduce the unsupervised techniques, we collected and downloaded the SMILES formulas of approximately 200 million purchasable compounds from ZINC database \cite{sterling2015zinc}. We conducted a preprocessing stage that consisted of filtering out compounds that did not comply with Lipinski's Rule of 5 \cite{lipinski2004lead}. We only kept compounds having a molecular weight between $12$ and $600$, heavy atom count between $3$ and $50$ and \textit{cLogP} (partition coefficient octanol/water) value between $-5$ and $7$. We also filtered out compounds that presented non-drug-like atoms, such as heavy metals, and removed salts and solvents. Finally, we obtained the canonical SMILES formulas for each of the remaining compounds. This whole process was carried out using RDKit \cite{landrum2016rdkit}. After the preprocessing stage, we randomly selected a subset of 40 million compounds---twice as many compounds as the ones used in the reference papers \cite{jaeger2018mol2vec,ozturk2018novel,xu2017seq2seq}---, which were later used to train the unsupervised methods. 

We also selected eight different labeled datasets, five classification datasets and three regression datasets, which were used to train the supervised methods and to evaluate all embedding methods. All of the classification datasets pose binary classification tasks. We prioritized datasets that had initially been used in the reference papers \cite{oskooei2018paccmann,zheng2019identifying,jaeger2018mol2vec,ozturk2018novel,xu2017seq2seq} while we accounted for diversity of sizes and class imbalance scenarios. The selected datasets were preprocessed following the same procedure used for ZINC, described above. Further details about class imbalance and numbers of compounds are summarized in Table \ref{table:labeled_datasets}. Namely, these datasets are:

\begin{itemize}
\item \textit{SR-ARE}, a bioassay for small molecule agonists of the antioxidant response element (ARE) signaling pathway\footnote{\url{https://pubchem.ncbi.nlm.nih.gov/bioassay/743219}}. This assay of data is contained in the \emph{Tox21 challenge} dataset, consisting of qualitative toxicity measurements on twelve biological targets\footnote{\url{https://tripod.nih.gov/tox21/challenge/about.jsp}}.
\item \textit{SR-MMP}, a stress response assay for small molecule disruptors of the mitochondrial membrane potential (MMP)\footnote{\url{https://pubchem.ncbi.nlm.nih.gov/bioassay/720637}}. This dataset is also included in the \emph{Tox21 challenge} dataset.
\item \textit{SR-ATAD5}, a set of small molecules that induce genotoxicity in human embryonic kidney cells expressing luciferase-tagged ATAD5\footnote{\url{https://pubchem.ncbi.nlm.nih.gov/bioassay/720516}}. This dataset also belongs to the \emph{Tox21 challenge} dataset.
\item \textit{HIV}, a dataset introduced by the Drug Therapeutics Program (DTP) AIDS Antiviral Screen containing information about molecular ability to inhibit HIV replication. Screening results were categorized as confirmed inactive (CI), confirmed active (CA) and confirmed moderately active (CM). We merged the compounds categorized under the latter two labels, which yielded two classes: inactive (CI) and active (CA and CM)\footnote{\url{https://wiki.nci.nih.gov/display/NCIDTPdata/AIDS+Antiviral+Screen+Data}}.
\item \textit{PCBA-686978}, a PubChem bioassay containing information about molecular ability to inhibit the human tyrosyl-DNA phosphodiesterase 1 (TDP1)\footnote{\url{https://pubchem.ncbi.nlm.nih.gov/bioassay/686978}}.
\item \textit{ESOL}, a dataset consisting of water solubility data for 1128 compounds, used to train models that estimate solubility directly from chemical structures \cite{delaney2004esol}.
\item \textit{FreeSolv}---the \textit{Free Solvation Database}---, a dataset comprising experimental and calculated hydration free energy values for small molecules in water \cite{mobley2014freesolv}.
\item \textit{Lipophilicity}, a dataset curated from ChEMBL database \cite{bento2014chembl} including experimental results of octanol/water distribution coefficient (\textit{LogD 7.4}) for 4200 compounds \footnote{\url{https://www.ebi.ac.uk/chembl/document_report_card/CHEMBL3301361/}}.
\end{itemize}

% % TABLA 1 VA AQUI
% \begin{table*}
% \caption{Details of labeled datasets. The imbalance ratio for classification datasets is computed as the number of active compounds every 100 inactive compounds.}
% \label{table:labeled_datasets}
% \centering \rulebox{Table \ref{table:labeled_datasets} goes here}
% \end{table*}

% CODIGO LATEX TABLA 1
\begin{table*}
\begin{center}

\caption{Details of labeled datasets. The imbalance ratio for classification datasets is computed as the number of active compounds every 100 inactive compounds.}
\label{table:labeled_datasets}
\begin{tabular*}{\textwidth}{llcccc}
\toprule
Dataset  & Task   & \# Compounds & \# Active & \# Inactive & Imbalance ratio\\ \midrule
SR-ARE  & classification    & 5956         & 941       & 5015        & 18.76                                      \\
SR-MMP   & classification     & 5937         & 925       & 5012        & 3.68                                       \\
SR-ATAD5   & classification   & 7251         & 258       & 6993        & 18.46                                      \\
HIV      & classification     & 41127        & 1443      & 39684       & 27.50                                      \\
PCBA-686978  & classification  & 302175       & 62800     & 239375      & 3.81                                       \\ 
ESOL     & regression     & 1128       & -     & -      & -                                       \\ 
FreeSolv  & regression   & 642        & -     & -      & -                                      \\
Lipophilicity & regression & 4200        & -     & -      & -                                     \\ \bottomrule
\end{tabular*}
\end{center}
\end{table*}
% FIN CODIGO LATEX TABLA 1

\subsection{\textbf{Molecular embedding methods}}\label{subsec:molecular_embedding_methods}

After identifying the most prominent deep learning techniques for learning molecular representations based on SMILES formulas, we chose \textit{SMILESVec} \cite{ozturk2018novel}, \textit{Mol2Vec} \cite{jaeger2018mol2vec}, \textit{Seq2Seq Fingerprint} \cite{xu2017seq2seq}--- which we refer to as \textit{Seq2Seq} in this paper\mbox{---,} \textit{PaccMann's} self attentive encoder \cite{oskooei2018paccmann}---named \textit{PaccMann} throughout this paper--- and \textit{SA-BiLSTM} encoder \cite{zheng2019identifying} as our reference embedding methods. 
We decided not to experiment with \textit{SMILES-BERT}, due to the reproducibility issues pointed out in the previous section and because both \textit{SA-BiLSTM} and \textit{PaccMann} are also self-attention-based encoders.
We selected these methods to test different deep learning architectures, model sizes in terms of trainable parameters and considering both supervised and unsupervised training scenarios. %The goal of our experimental process was to thoroughly test each of these methods and their suitability for QSAR modeling while taking into consideration different aspects that might influence their performance. 
Moreover, we aimed at conducting a training phase consisting of multiple runs of each method in order to account for their intrinsic variance.
We conducted our benchmarking process by testing an extensive range of combinations of hyperparameters for each embedding method and classification/regression model and by leveraging different canonical forms of the SMILES formulas in the datasets. All of the unsupervised methods were trained on the 40 million compound dataset retrieved and curated from ZINC, whereas the supervised methods were trained using each of the eight labeled datasets described previously. We hereby describe the neural-based architectures of each embedding method and the main characteristics of the obtained molecular representations. We provide further details on the parameterization of each model in Appendices A and B.

\subsubsection{Unsupervised methods}\label{subsubsec:unsup_methods}
%\sloppy \textit{SMILESVec} \cite{ozturk2018novel} is an unsupervised method for learning molecular representations from SMILES formulas, initially intended to represent proteins based on the combination of their ligands' vectors. 
\sloppy The architecture of \textit{SMILESVec} \cite{ozturk2018novel} consists of a feed-forward neural-based autoencoder \footnote{\url{https://github.com/hkmztrk/SMILESVecProteinRepresentation/tree/master/source/word2vec}.}, as proposed by 
%\citet{mikolov2013efficient}
Mikolov et al.~\cite{mikolov2013efficient}. According to the original paper, we first tokenized the SMILES formulas into overlapping substrings of eight characters and compiled a vocabulary set with all the tokens in the training set. Considering the long-term dependencies often present in the SMILES sequences, such as branches or stereochemistry information, and that the eight-gram tokenization might limit the ability of the model to learn from them, we employed two different canonicalization processes on the ZINC training data, thus obtaining two different training sets for this method. On the one hand, we employed RDKit canonical SMILES \cite{landrum2016rdkit}, whereas, on the other hand, we tested \textit{DeepSMILES} canonical formulas \cite{dalke2018deepsmiles}, which are built by eliminating cycles in the molecular graph in order to reduce long-term dependencies in the sequences. \textit{DeepSMILES} has also been employed by the authors who proposed \textit{SMILESVec} on a follow-up study \cite{ozturk2018chemical}, which showed improved performance over the results obtained with traditional SMILES formulas. 

As proposed by the \textit{SMILEVec} authors \cite{ozturk2018novel}, a Skip-gram \textit{word2vec} model \cite{mikolov2013efficient} was used, in which the size of the embeddings is determined by the number of nodes in its hidden layer. We trained two different embedding sizes for each canonicalization technique mentioned above: 100 and 300. While the embedding size used in the original paper was 100, we decided to test 300-dimensional embeddings to compare \textit{SMILESVec} against higher-dimensional embeddings obtained by other methods tested in this paper. As performed by 
%\citet{ozturk2018novel} 
{\"O}zt{\"u}rk et al.~\cite{ozturk2018novel}, the unsupervised embedding models were trained for 20 epochs and the final molecular representations were obtained by computing the average of the learned vectors for each token in the SMILES formulas.

\textit{Mol2Vec} \cite{jaeger2018mol2vec} is also an unsupervised molecular embedding method based on \textit{word2vec} \cite{mikolov2013efficient}. The architecture of the model reported by the authors \footnote{\url{https://github.com/samoturk/mol2vec}.} consists of a Skip-gram \textit{word2vec} model, where the embedding size matches the size of the hidden layer of the autoencoder. Following the preprocessing steps indicated by %\citet{jaeger2018mol2vec}
Jaeger et al.~\cite{jaeger2018mol2vec}, we tokenized the RDKit canonical SMILES formulas in the ZINC database into sequences of chemical words obtained by employing an adaptation of the Morgan algorithm for computing ECFPs \cite{rogers2010extended}. We did not use \textit{DeepSMILES} canonical SMILES to train this method since the Morgan algorithm cannot reconstruct the molecular graph from them. Following the reference paper, we compiled a vocabulary set including all chemical words obtained from the ZINC training dataset and trained two models for 5 epochs each, with embedding sizes 100 and 300, respectively. Finally, we obtained the molecular embeddings by computing the sum of the learned vectors of all tokens in each molecule.

The third and last unsupervised method we reproduced is \textit{Seq2Seq} \cite{xu2017seq2seq}, an embedding model consisting of an RNN-based autoencoder\footnote{\url{https://github.com/XericZephyr/seq2seq-fingerprint}.}. The SMILES formulas in the ZINC training set were canonicalized to both RDKit and \textit{DeepSMILES} canonical SMILES. While RDKit canonical SMILES were the only ones used in the reference paper \cite{xu2017seq2seq}, we decided to additionally employ \textit{DeepSMILES} canonical formulas, aiming to test whether they contributed to the learning process of the model. The sequences in both canonical datasets were tokenized by splitting the SMILES strings into separate characters and padded to match the length of the longest sequence in the dataset, as performed in the reference paper. According to %\citet{xu2017seq2seq}
Xu et al.~\cite{xu2017seq2seq}, the architecture of the model consists of a multi-layer bidirectional GRU autoencoder, and the learned embeddings are obtained by concatenating the hidden states of the trained model. Thus, their size is determined by the number of hidden units and the number of autoencoder layers. We trained two different models for each of the canonical forms of SMILES strings, obtaining 100-dimensional and 384-dimensional embeddings. Size 384 was the default embedding size reported by the authors of the method, and we also tested size 100 to account for a fair comparison with \textit{SMILESVec} and \textit{Mol2Vec}. We provide a summary of the molecular embeddings we obtained after training all the mentioned variants of the unsupervised methods in Table \ref{table:summary-unsupervised}.

% TABLA 2 VA AQUI
% \begin{table*}
% \caption{Summary of all unsupervised embedding methods. The column \textit{Denomination} denotes the name by which we refer to each of the learned embeddings throughout this paper. The denominations marked with an asterisk (*) correspond to the embeddings we decided to experiment with in addition to the ones proposed in the reference papers.}
% \label{table:summary-unsupervised}
% \centering \rulebox{Table \ref{table:summary-unsupervised} goes here}
% \end{table*}

% CODIGO LATEX TABLA 2
\begin{table*}
\begin{center}
\caption{Summary of all unsupervised embedding methods. The column \textit{Denomination} denotes the name by which we refer to each of the learned embeddings throughout this paper. The denominations marked with an asterisk (*) correspond to the embeddings we decided to experiment with in addition to the ones proposed in the reference papers.}
\label{table:summary-unsupervised}
\begin{tabular*}{0.9\textwidth}{@{\extracolsep{\fill}}cccl@{\extracolsep{\fill}}}
\toprule
Method                     & Canonicalization            & Embedding size & Denomination         \\ \midrule
\multirow{4}{*}{SMILESVec} & \multirow{2}{*}{RDKit}      & 100            & SMILESVec\_100       \\
                          &                             & 300            & SMILESVec\_300 (*)     \\ \cline{2-4}  
                          & \multirow{2}{*}{DeepSMILES} & 100            & Deep\_SMILESVec\_100 \\
                          &                             & 300            & Deep\_SMILESVec\_300 (*) \\ \midrule
\multirow{2}{*}{Mol2Vec}   & \multirow{2}{*}{RDKit}      & 100            & Mol2Vec\_100         \\
                          &                             & 300            & Mol2Vec\_300         \\ \midrule
\multirow{4}{*}{Seq2Seq}   & \multirow{2}{*}{RDKit}      & 100            & Seq2Seq\_100 (*)       \\
                          &                             & 384            & Seq2Seq\_384         \\ \cline{2-4} 
                          & \multirow{2}{*}{DeepSMILES} & 100            & Deep\_Seq2seq\_100 (*)  \\
                          &                             & 384            & Deep\_Seq2seq\_384 (*)  \\ \bottomrule
\end{tabular*}
\end{center}
\end{table*}
% CODIGO LATEX TABLA 2

\subsubsection{Supervised methods}\label{subsubsec:sup_methods}

\textit{PaccMann} \cite{oskooei2018paccmann} is a supervised embedding method based on a set of multi-modal neural-based encoders that learn from both SMILES formulas and gene expression information, originally designed for predicting anticancer compound sensitivity. In particular, we reproduced the \textit{self-attention (SA)} encoder proposed by the authors, which reportedly yielded the best results. Unlike the reference paper, we did not employ gene expression information in the training phase and instead trained the self-attentive encoder using only RDKit canonical SMILES formulas. In order to attain such a model, we adapted the source code provided by %\citet{oskooei2018paccmann}
Oskooei et al.~\cite{oskooei2018paccmann}\footnote{\url{https://github.com/drugilsberg/paccmann}.}. According to the reference paper, the SMILES formulas were tokenized following an algorithm \cite{schwaller2018found} that consists of separating the SMILES formulas into characters and performing a basic filtering step. After such tokenization, each sequence was padded to match the longest sequence in the dataset.

The architecture of \textit{PaccMann} consists of an input layer by which the tokenized SMILES formulas are fed to the model. Next, a sinusoidal positional encoding function \cite{vaswani2017attention} is optionally applied to these inputs, followed by a single-headed self-attention layer, whose implementation details are in the reference paper \cite{oskooei2018paccmann}. The outcome of the self-attentive layer is fed to a shallow feed-forward neural network for property prediction.  
The learned molecular embeddings were extracted from the output of the self-attention layer, and their size is computed as the product between the number of hidden units $u$ in the self-attention layer and the length of the input sequence $n$. Consequently, the embedding size $n\times u$ is different for each of the eight labeled datasets on which this method was trained. A summary of the embedding dimensionality for each labeled dataset is provided in Table \ref{table:summary-supervised}. 

The second supervised method we reproduced in this paper is \textit{SA-BiLSTM} \cite{zheng2019identifying}. This method consists of a self-attentive model combined with a bidirectional RNN using LSTM nodes and followed by a shallow feed-forward neural network for property prediction.  
According to the reference paper, we trained this model using \textit{Mol2Vec} embeddings \cite{jaeger2018mol2vec} as input data, which were computed for the eight labeled datasets described previously and using a pretrained model provided by 
%\citet{jaeger2018mol2vec}
Jaeger et al.~\cite{jaeger2018mol2vec}. %The SMILES formulas were canonicalized using RDKit before computing the embeddings. 
We employed RDKit canonical SMILES to train this method. Unlike the other methods tested in this paper, the authors of \textit{SA-BiLSTM} have not provided the source code or an installable package. Therefore, we implemented the model from the equations and details provided in their paper. 

According to the authors, the architecture of \textit{SA-BiLSTM} consists of an input layer that receives the \textit{Mol2Vec} embeddings, followed by a single-layer bidirectional LSTM RNN. The output of such a layer is then passed to a multi-head self-attention layer, which yields the final self-attentive embeddings of the compounds. These embeddings are fed to a shallow feed-forward neural network for property prediction, consisting of a single fully-connected layer. %Further details about this model can be found in the reference paper \cite{zheng2019identifying}.  
The self-attentive embeddings obtained from this model consist of a set of $r$ vectors of size $2u$, where $r$ is the number of attention heads and $u$ is the number of hidden units in the LSTM layer (the factor of $2$ is because the RNN is bidirectional). According to the reference paper, these embeddings (the $r$ vectors) can be analyzed separately. However, since we wanted to use the learned information from all the attention heads for prediction, we concatenated them for the subsequent classification/regression step. Thus, the resulting embeddings consist of vectors of size $r\times2u$. Similarly to the case of \textit{PaccMann}, because of the fitting process the size of the embedding varies for each of the eight labeled datasets used to train \textit{SA-BiLSTM}. A summary of the \textit{SA-BiLSTM} embedding dimensionality for each labeled dataset is provided in Table \ref{table:summary-supervised}.

% TABLA 3 VA AQUI
% \begin{table}
% \caption{Summary of all supervised embedding methods and their embedding sizes.}
% \label{table:summary-supervised}
% \begin{tabular*}{\columnwidth}{@{\extracolsep{\fill}}c@{\extracolsep{\fill}}}
% \centering \rulebox{Table \ref{table:summary-supervised} goes here}
% \end{tabular*}
% \end{table}

% CODIGO LATEX TABLA 3
\begin{table}
\caption{Summary of all supervised embedding methods and their embedding sizes.}
\label{table:summary-supervised}
\begin{tabular*}{\columnwidth}{@{\extracolsep{\fill}}clc@{\extracolsep{\fill}}}
\toprule
Method/Denomination                               & Dataset     & Embedding size \\ \midrule
\multirow{8}{*}{PaccMann} & SR-ARE      & 12000          \\
                                     & SR-MMP      & 24000          \\
                                     & SR-ATAD5    & 24000          \\
                                     & HIV         & 24200          \\
                                     & PCBA-686978 & 6976           \\
                                     & ESOL      & 9700          \\
                                     & FreeSolv      & 5300          \\
                                     & Lipophilicity    & 10250          \\\midrule
\multirow{8}{*}{SA-BiLSTM}           & SR-ARE      & 2560           \\
                                     & SR-MMP      & 1920           \\
                                     & SR-ATAD5    & 1280           \\
                                     & HIV         & 2560           \\
                                     & PCBA-686978 & 1280           \\
                                     & ESOL      & 1280          \\
                                     & FreeSolv      & 1280          \\
                                     & Lipophilicity    & 1280          \\\bottomrule
\end{tabular*}
\end{table}
% CODIGO LATEX TABLA 3

\subsection{\textbf{Experimental design}}\label{subsec:experimental_design}

\begin{figure*}
\centering
\includegraphics[width=0.8\textwidth]{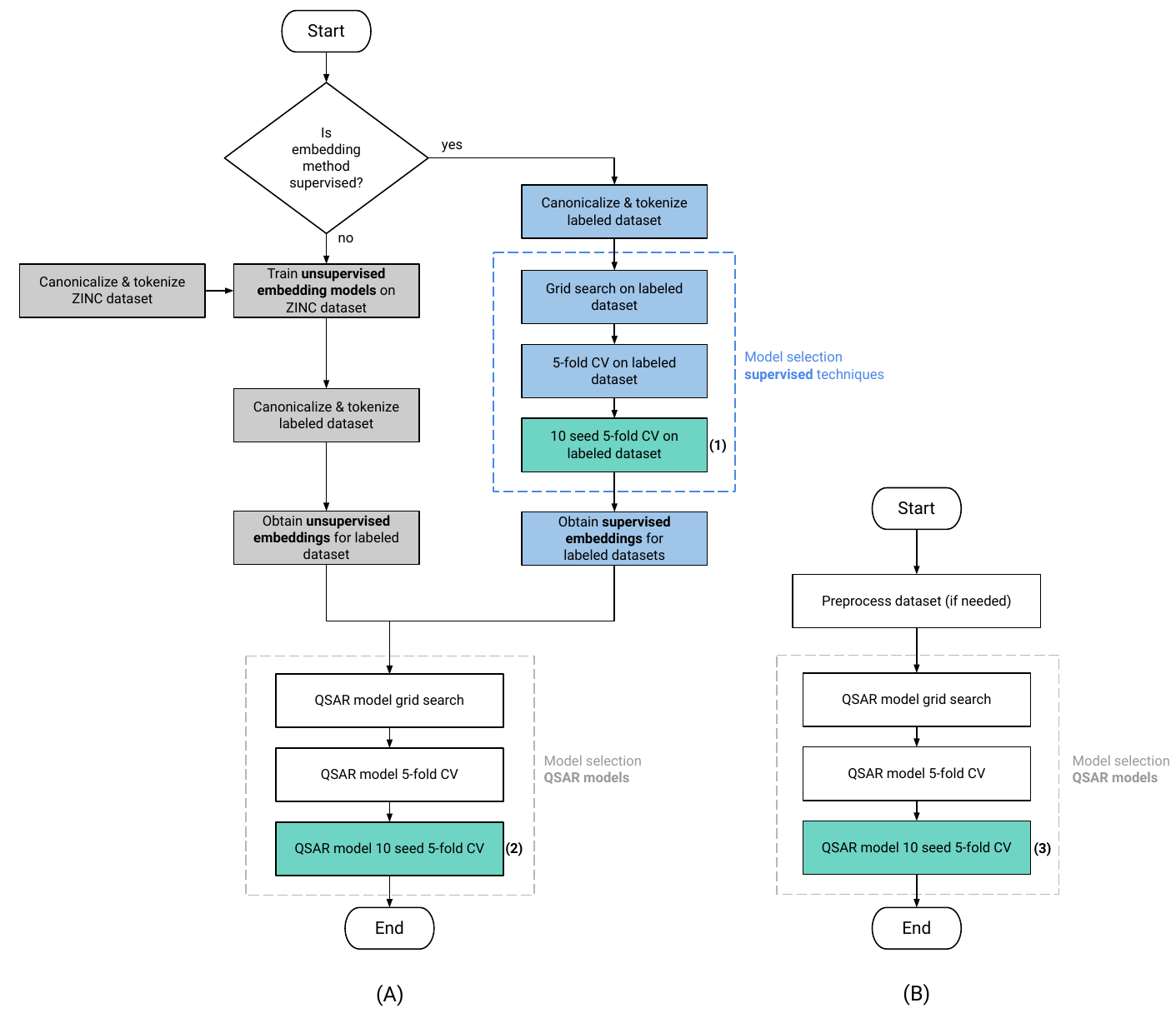} 
\caption{Overview of the experimental workflow: (A) corresponds to the workflow for embedding-based representations, whereas (B) depicts the workflow for traditional molecular representations. The diagram cells in grey correspond to the experiments on the unsupervised embedding techniques, whereas the diagram cells in blue correspond to the experiments on the supervised embedding techniques. The three final cells correspond to the prediction stage. The cells in green indicate the reported classification/regression results: (1) the \textit{fitting} results, i.e., the classification/regression results obtained from the supervised embedding methods as a result of their training process (2) the classification/regression results obtained from the QSAR models trained using either supervised or unsupervised embeddings. (3) the classification/regression results obtained from the QSAR models trained using traditional molecular representations.}
\label{figure:summary-workflow}
\end{figure*}

In this section, we give an overview of the experimental workflow we conducted to obtain and evaluate the molecular embeddings computed from each of the reviewed methods. %We provide a thorough explanation of the training stage of each method and their subsequent embedding extraction stage. Afterwards, we describe the evaluation process, which consists of different classifiers that were trained using the learned molecular embeddings and traditional molecular representations. Finally, we list the metrics used to evaluate all intermediate and final results. 
A summary of such workflow can be found in Figure \ref{figure:summary-workflow}.

% FIGURA 2 VA AQUI
% \begin{figure*}
%   % \centering{{\color{black!20}\rule{100pt}{30pt}}}
%     \centering \rulebox{Figure \ref{figure:summary-workflow} goes here}
% \caption{Overview of the experimental workflow: (A) corresponds to the workflow for embedding-based representations, whereas (B) depicts the workflow for traditional molecular representations. The diagram cells in grey correspond to the experiments on the unsupervised embedding techniques, whereas the diagram cells in blue correspond to the experiments on the supervised embedding techniques. The three final cells correspond to the prediction stage. The cells in green indicate the reported classification/regression results: (1) the \textit{fitting} results, i.e., the classification/regression results obtained from the supervised embedding methods as a result of their training process (2) the classification/regression results obtained from the QSAR models trained using either supervised or unsupervised embeddings. (3) the classification/regression results obtained from the QSAR models trained using traditional molecular representations.}
% \label{figure:summary-workflow}
% \end{figure*}

% CODIGO LATEX FIGURA 2 (DESCOMENTAR)
% FIN CODIGO LATEX FIGURA 2 (DESCOMENTAR)

\subsubsection{Training and embedding extraction}\label{subsubsec:training_embedding_extraction}

The first step of our experimental setup consisted of training a model for each of the reviewed embedding methods. The training phase of the unsupervised methods (\textit{SMILESVec, Mol2Vec} and \textit{Seq2Seq)} was simpler than that of the supervised methods since the training dataset is unlabeled, and thus no early stopping criteria to avoid overfitting is needed. We trained each method following the amount of time or epochs and hyperparameterization specified in each reference paper. We obtained ten unsupervised embedding models for each combination of reviewed unsupervised method, embedding size and SMILES canonicalization of the ZINC training dataset, as summarized in Table \ref{table:summary-unsupervised}. %Finally, we extracted the molecular embeddings for each of the five labeled datasets from the ten embedding models, whose SMILES formulas had been previously tokenized as required by each approach.
Finally, we tokenized the SMILES formulas in each labeled dataset as required by each embedding approach, and extracted the molecular embeddings from the ten embedding models.

Regarding the supervised methods (\textit{PaccMann} and \textit{SA-BiLSTM}), we conducted a broader range of experiments during their training stage compared to the experiments on the unsupervised methods. Our goal was to attain embedding models whose performance could not be attributed solely to variance in the data partitions, in the initialization of the weights of the model or to nuances in the chosen hyperparameters.
For this reason, we designed the following training workflow:

\begin{enumerate}
    \item We filtered and tokenized the compounds of the five labeled datasets according to the requirements of each of the supervised methods. 
    \item We performed a model selection stage consisting of two steps:
    \begin{enumerate}
        \item We conducted a hyperparameter \textit{grid search}: an exploratory search for the best performing hyperparameter combination for each method. The ranges of hyperparameter values included those tested in the reference papers \cite{oskooei2018paccmann,zheng2019identifying} plus other values that might improve the training results. We selected the combinations of hyperparameters that yielded top results on each labeled dataset. A complete list of all hyperparameters tested on \textit{PaccMann} and \textit{SA-BiLSTM} is provided in Appendix A.  
        
        \item Each of the hyperparameter combinations selected in the previous step was further tested on a stratified five-fold cross-validation process, in order to confirm that the results found during the grid search were not imputable only to the data partition used during that process. As a result, we selected a single combination of hyperparameters for each labeled dataset based on the average validation results of the five-fold cross-validation process. 
    \end{enumerate}
    
    \item We performed a \textit{replication stage}: we trained ten trials of five-fold cross-validation of the model selected during the model selection stage. Each trial used a different random seed for weights initialization. This step was carried out to rule out any artifacts in the results due to the random weight initialization of the embedding models in the previous steps. As a result of this step, we obtained the classification/regression results obtained from the inherent training process of the supervised embedding methods, which we refer to as \textit{fitting} results.
    
    \item Since a single embedding model was needed to extract the supervised embeddings, we trained an embedding model using a stratified data partition (80\% train -- 20\% internal validation for early stopping). Finally, we extracted the molecular representations for each labeled dataset from these embedding models.
\end{enumerate}

\begin{figure*}
\scriptsize
\centering
\begin{tabular}{ccc}
 \includegraphics[trim=0 10 0 75,clip,width=\textwidth]{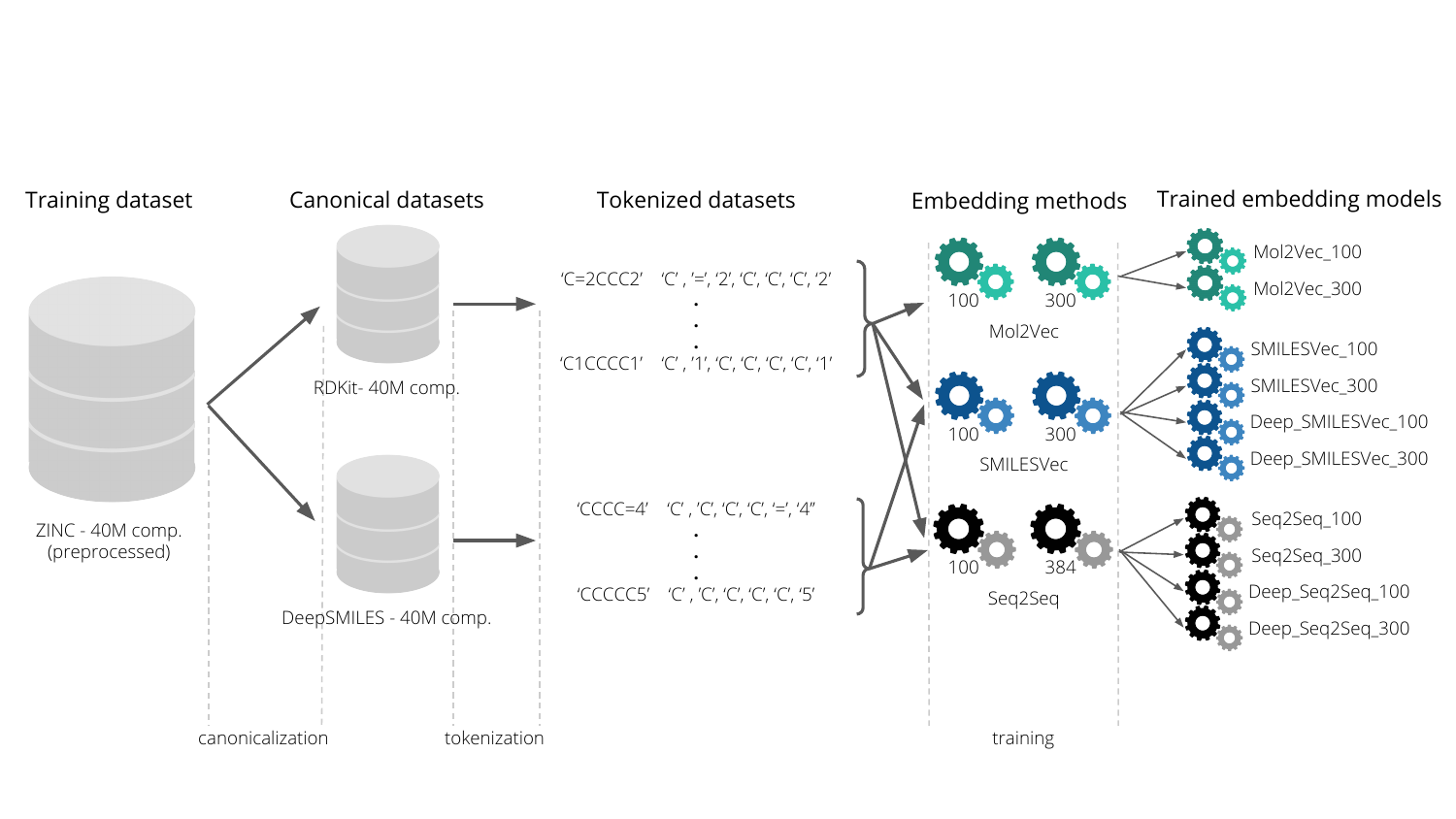} \\
(a) Unsupervised training using 40 million compounds from ZINC dataset \\[3pt]
\end{tabular}
\begin{tabular}{ccc}
 \includegraphics[trim=0 10 0 60,clip,width=\textwidth]{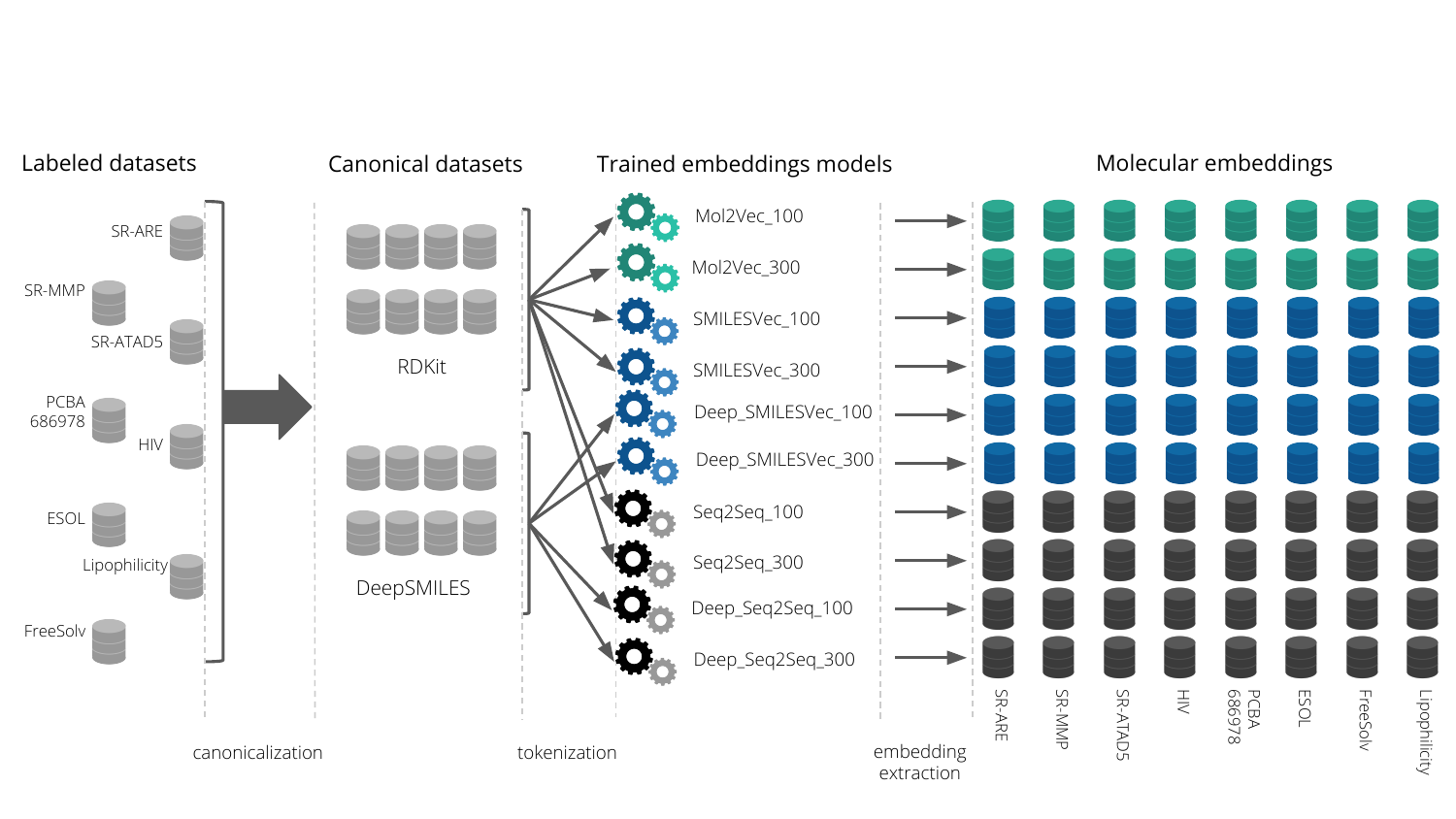} \\
 (b) Embedding extraction \\[3pt]
\end{tabular}
\caption{Training and embedding extraction from unsupervised embedding methods: (a) The ZINC dataset was preprocessed and canonicalized following two different canonicalization procedures. The compounds in the two canonical datasets were tokenized as required by each of the unsupervised techniques, and then each unsupervised embedding model was trained. (b) The eight labeled datasets were canonicalized and tokenized according to the requirements of each method. Afterward, molecular embeddings were obtained for each labeled dataset from all the unsupervised embedding models trained in the previous step.}
\label{figure:training_unsupervised}
\end{figure*}

\begin{figure*}
\scriptsize
\centering
\begin{tabular}{ccc}
\includegraphics[trim=35 10 0 35,clip, width=\textwidth]{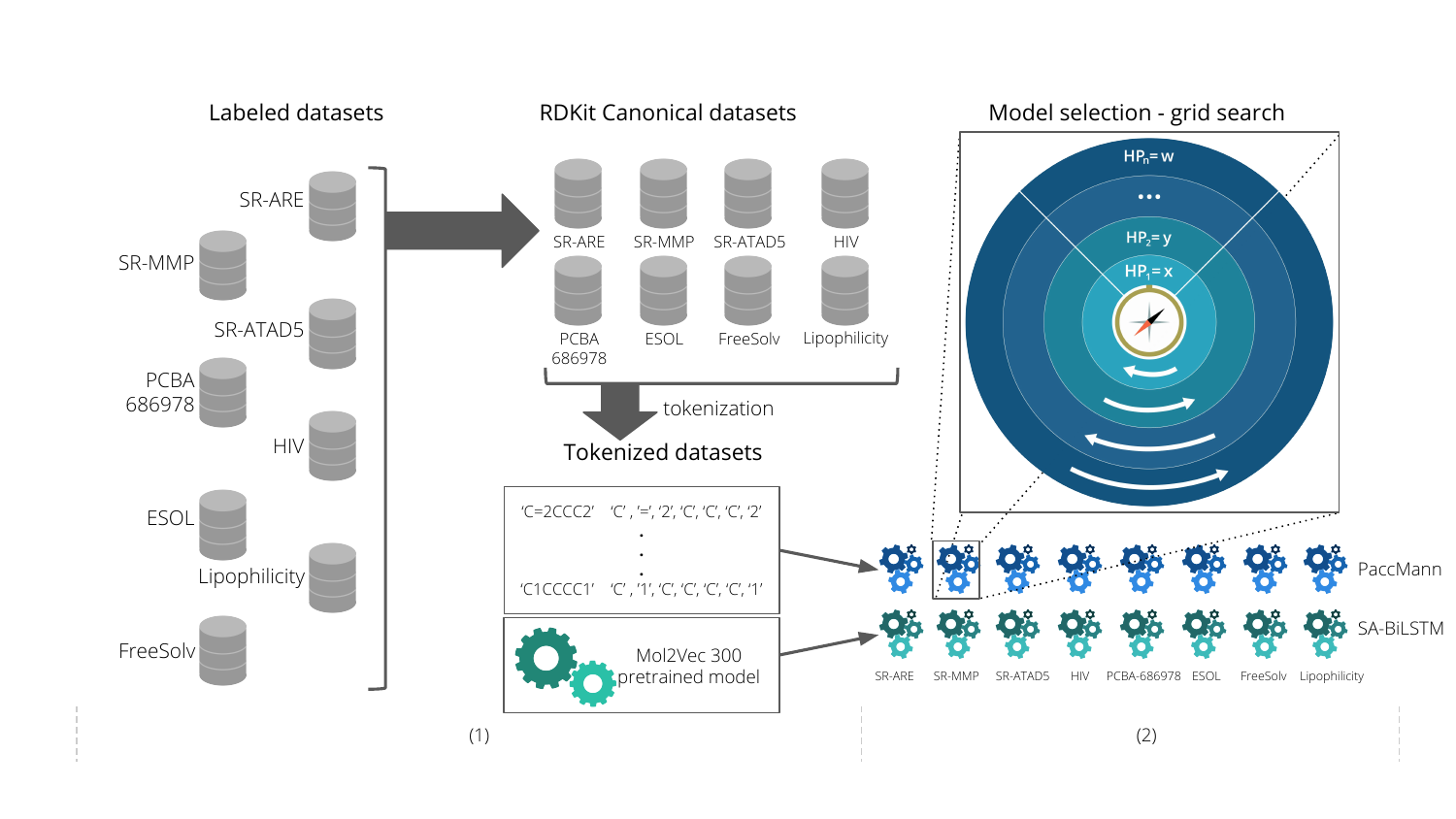} \\
\end{tabular}
\begin{tabular}{ccc}
 \includegraphics[trim=35 10 0 50,clip,width=0.98\textwidth]{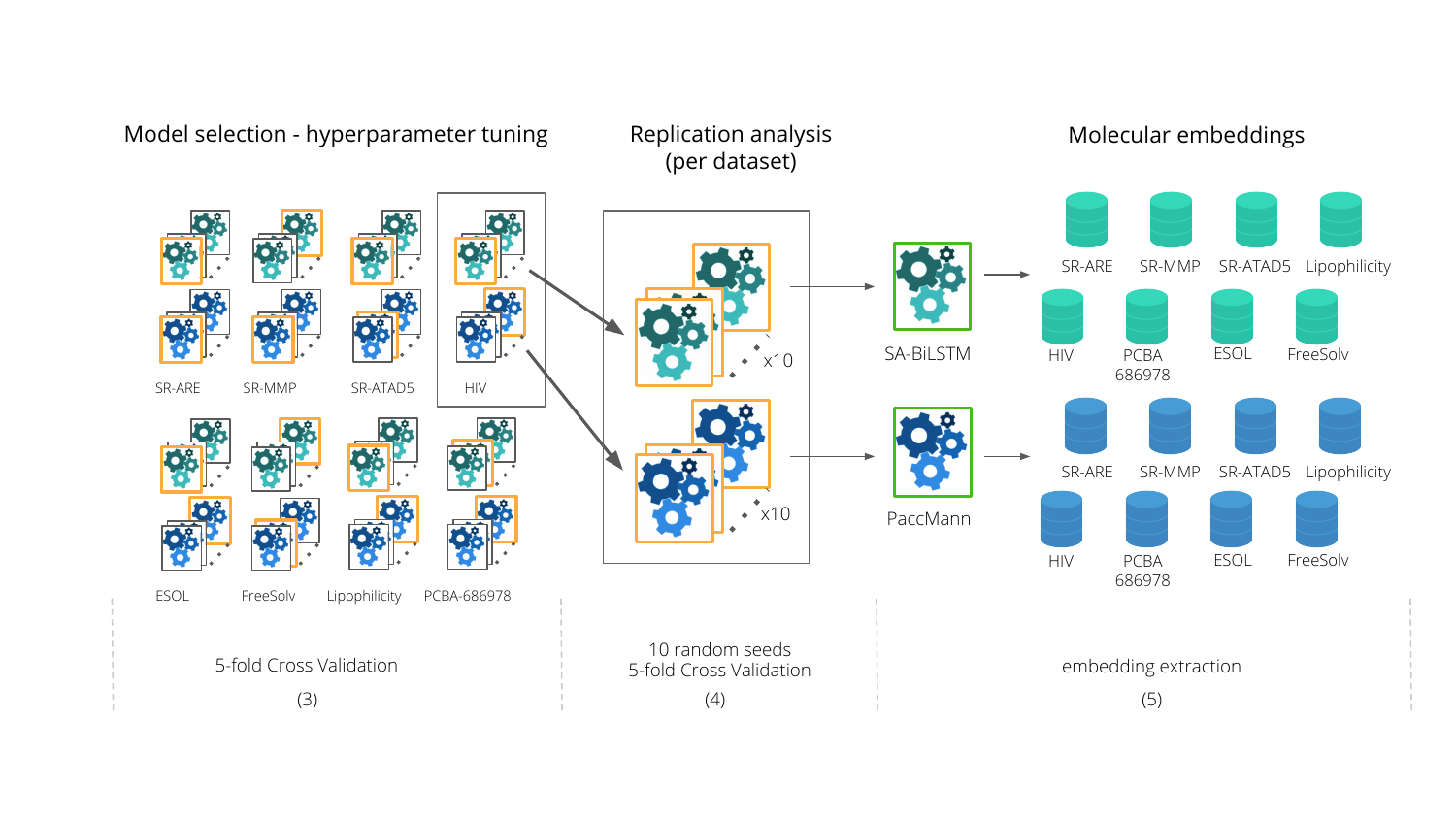} \\
\end{tabular}
\caption{Training and embedding extraction from supervised methods: (1) The compounds in the eight labeled datasets were preprocessed and tokenized. (2) The first step of model selection on each labeled dataset consisted of an exploratory grid search for the best performing hyperparameter combination for each method. As a result, a subset of hyperparameter combinations was chosen for each labeled dataset. (3) In the second step of the model selection stage, each hyperparameter combination in the subset was tested on a five-fold cross-validation process. As a result, only one combination of hyperparameters was selected for each labeled dataset (highlighted in yellow). (4) Ten trials of the selected combination using different random seeds on five-fold cross-validation were trained. (5) An embedding model was trained using a stratified data partition and the embeddings for the eight labeled datasets were extracted from such model.}
\label{figure:training_supervised}
\end{figure*}

In addition to Figure \ref{figure:summary-workflow}, which summarizes the whole experimental workflow of our paper, we provide a more detailed graphical summary of the training stage of both unsupervised and supervised methods in Figures \ref{figure:training_unsupervised} and \ref{figure:training_supervised}, respectively.  The source code needed to extract the embeddings from the trained embedding models is either provided by the original authors, as in the case of \textit{Seq2Seq}\footnote{\url{https://github.com/XericZephyr/seq2seq-fingerprint}.}, or by us. Please refer to Section ``Data availability statement" for further details.

% FIGURA 3 VA AQUI
% \begin{figure*}
%   % \centering{{\color{black!20}\rule{100pt}{30pt}}}
%     \centering \rulebox{Figure \ref{figure:training_unsupervised} goes here}
% \caption{Training and embedding extraction from unsupervised embedding methods: (a) The ZINC dataset was preprocessed and canonicalized following two different canonicalization procedures. The compounds in the two canonical datasets were tokenized as required by each of the unsupervised techniques, and then each unsupervised embedding model was trained. (b) The eight labeled datasets were canonicalized and tokenized according to the requirements of each method. Afterward, molecular embeddings were obtained for each labeled dataset from all the unsupervised embedding models trained in the previous step.}
% \label{figure:training_unsupervised}
% \end{figure*}

% CODIGO LATEX FIGURA 3 (DESCOMENTAR)
% FIN CODIGO LATEX FIGURA 3 (DESCOMENTAR)

% FIGURA 4 VA AQUI
% \begin{figure*}
%   % \centering{{\color{black!20}\rule{100pt}{30pt}}}
%     \centering \rulebox{Figure \ref{figure:training_supervised} goes here}
% \caption{Training and embedding extraction from supervised methods: (1) The compounds in the eight labeled datasets were preprocessed and tokenized. (2) The first step of model selection on each labeled dataset consisted of an exploratory grid search for the best performing hyperparameter combination for each method. As a result, a subset of hyperparameter combinations was chosen for each labeled dataset. (3) In the second step of the model selection stage, each hyperparameter combination in the subset was tested on a five-fold cross-validation process. As a result, only one combination of hyperparameters was selected for each labeled dataset (highlighted in yellow). (4) Ten trials of the selected combination using different random seeds on five-fold cross-validation were trained. (5) An embedding model was trained using a stratified data partition and the embeddings for the eight labeled datasets were extracted from such model.}
% \label{figure:training_supervised}
% \end{figure*}

% CODIGO LATEX FIGURA 4
% FIN CODIGO LATEX FIGURA 4

\subsubsection{Evaluation of the molecular embeddings}\label{evaluation_mol_embeddings}

In order to evaluate the different learned molecular embeddings in QSAR modeling, we tested them on five classification and three regression tasks, defined by the eight labeled datasets described in Table \ref{table:labeled_datasets}. We obtained a total of ten different unsupervised molecular embeddings and two different supervised molecular embeddings per labeled dataset, as shown in Figures \ref{figure:training_unsupervised} and \ref{figure:training_supervised}.
In addition, we computed three traditional molecular representations for each labeled dataset: a 1024-bit ECFP4 fingerprint using RDKit \cite{landrum2016rdkit}, a 166-bit MACCS keys fingerprint also computed with RDKit, and a vector of molecular descriptors computed using Mordred \cite{moriwaki2018mordred}. We computed 0D, 1D and 2D descriptors for all datasets and discarded those having more than $5\%$ of \textit{NaN} entries. The final number of molecular descriptors is as follows: $1018$ for SR-ARE, $1016$ for SR-MMP and SR-ATAD5, $1150$ for HIV, $1428$ for PCBA-686978, $1226$ for ESOL, $1176$ for FreeSolv and $1429$ for Lipophilicity. All the traditional representations were computed from RDKit canonical SMILES formulas.

In summary, we computed a total of $15$ ($10+2+3$) representations for each dataset, which we then used to train our QSAR models. For each of the classification tasks we trained four different classifiers, whereas in the case of the regression tasks we trained three different regression models. In order to account for a fair comparison among all representations, we repeated the following steps for each representation:
\begin{enumerate}
    \item For the classification tasks, we built and trained a Naïve Bayes classifier (NB), a Support Vector Machine (SVM) with an RBF kernel \cite{scholkopf2004kernel}, a Random Forest classifier (RF) and a shallow feed-forward neural network (FFNN). We used the Complement Naïve Bayes, SVM and RF implementations provided by Scikit-learn \cite{scikit-learn}, whereas the FFNNs were built and trained using Keras \cite{chollet2015keras} and Tensorflow \cite{tensorflow2015-whitepaper}. For the smallest datasets (\textit{SR-ARE, SR-MMP} and \textit{SR-ATAD5}), we trained single SVMs, whereas for the two largest labeled datasets (\textit{HIV} and \textit{PCBA-686978}), %for which training a standard non-linear kernel SVM on all compounds would have taken an inadmissible amount of time, 
    we employed the Scikit-learn implementation of a bagging classifier of ten SVMs, each trained on a stratified sample consisting of one-tenth of the dataset. For each classification method, we conducted a hyperparameter grid search using a stratified data partition. Full details about the hyperparameters tested on each classifier are provided in Appendix B.
    
    \item In the case of the regression tasks, we trained a Ridge regression (Ridge), a Gradient Boosting Regressor (GBR) and a shallow feed-forward neural network (FFNN). We used the Ridge and GRB implementations provided by Scikit-learn \cite{scikit-learn}, and the FFNNs were built and trained using Keras \cite{chollet2015keras} and Tensorflow \cite{tensorflow2015-whitepaper}. We conducted a hyperparameter grid search on a stratified data partition for each regression model. All details about the hyperparameters tested on each model are provided in Appendix B.
    
    \item We selected the best parameterizations of each method and ran a five-fold cross-validation training phase. %Similar to the training stage of the supervised embedding methods, the goal of this phase was to confirm that the results found during the grid search were not bound to the specific data partition used in the previous step.
    
    \item We selected the single best parameterization based on the results obtained during the five-fold cross-validation stage. Afterward, we ran ten trials of five-fold cross-validation using the same folds as before and using a different random seed for each run. This step was carried out only for the stochastic methods (i.e., RF, GBR, FFNN and ensembles of SVMs). %In this way, we ensured that the obtained classification results were not just due to a lucky random initialization of the weights of the classification models. 

\end{enumerate}

During the training process we used the same folds for each labeled dataset in every stage of the experiments. In all cases, we measured the average results of the five validation folds and their $95\%$ confidence intervals. The main reason behind this evaluation strategy is to account for a fair comparison of the results, minimizing the potential biases introduced by the variance of a particular random data partition, which can severely influence the results especially in small datasets, as it has been reported in previous studies \cite{jiang2021could, baumann2014reliable}.

\section{Results and discussion}\label{sec:results_discussion}

This section discusses the results obtained for the five classification and three regression tasks defined by each of the labeled datasets described in Table \ref{table:labeled_datasets}. We compare the results using all the molecular embedding techniques on the four classification and three regression methods described in the previous section. We also report the results obtained by the supervised methods during the embedding learning stage, i.e., the \textit{fitting} results, and the results obtained from traditional molecular representations (i.e.~\textit{molecular descriptors, ECFPs and MACCs keys}).

Throughout all stages of our experimental workflow, we measured the performance of our models using different metrics. The results on the regression tasks were evaluated by means of four metrics: \textit{Root Mean Squared Error ($RMSE$), Mean Absolute Error ($MAE$) and Coefficient of Determination ($R^2$)}. In turn, the classification results were evaluated by eight metrics: \textit{Sensitivity (Sn), Specificity (Sp), Precision, Accuracy (Acc), Balanced Accuracy (BA)}, $F_1$ score, $H_1$ score and \textit{Area Under the ROC Curve} $(AUC)$. $BA$ is the arithmetic mean of $Sn$ and $Sp$ and $F_1$ score is the harmonic mean of $Sn$ and $Precision$. To avoid overriding the notation for $F_1$ score, we hereby redefine the harmonic mean of $Sn$ and $Sp$ as $H_1$ score, which is also called $F_1$ or $F$ score elsewhere in the literature \cite{sokolova2006beyond}. We prioritized $RMSE$ for hyperparameter selections in regression settings, and $F_1$ score and $H_1$ score in classification tasks. $F_1$ and $H_1$ scores are suitable in contexts of highly imbalanced datasets (such as in the case of HIV and SR-ATAD5) in contrast to metrics like $Acc$ \cite{sokolova2006beyond, chawla2002smote}. All the results obtained for each step of our experimental workflow in all eight labeled datasets can be found in the Supplementary Material.

In order to assess the statistical significance of the results, we conducted a series of statistical tests through which we compared the $F_1$ score and $RMSE$ results obtained by the different molecular representations and QSAR models on each dataset. We first conducted a two-way ANOVA where the molecular representation and the choice of the prediction model were considered the two independent variables. Upon eventually finding that the results of different molecular representations were significantly different, we conducted a \textit{post-hoc} pairwise Tukey test \cite{tukey1977exploratory} with a global confidence level of $95\%$. %We present the results of the Tukey test on FFNN in Figure \ref{fig:tukey}. 
Full tables with the results of these statistical tests can be found in the Supplementary Material.

We accounted for the potential negative effects of class imbalance in classification tasks by three means. First, by training and evaluating the QSAR models using stratified partitions, in which the global proportions of active and inactive compounds (as described in Table \ref{table:labeled_datasets}) is present in each partition or fold. Second, by using weighed cost functions while training the supervised embedding models (\textit{SA-BiLSTM} and \textit{PaccMann}) and FFNNs, in a way such that the loss of the network during training is adjusted with a class weight proportional to the class imbalance. Third, by carefully selecting the metrics for model selection and evaluation of the results, as discussed above.

Our first research question (Q1) aimed at determining whether learned molecular representations could outperform traditional molecular representations in QSAR modeling. To answer this question, we compared the results obtained using the traditional molecular representations---\textit{molecular descriptors}, \textit{ECFPs} and \textit{MACCS keys}---to the results obtained using both supervised and unsupervised molecular embeddings. After conducting the two-way ANOVA test, we proceeded to compare the fifteen representations under study, one QSAR model at a time, using a pairwise Tukey test. 

\begin{figure*}
\scriptsize
\begin{tabular}{cc}
  \includegraphics[trim=0 13 180 0,clip,width=0.42\textwidth]{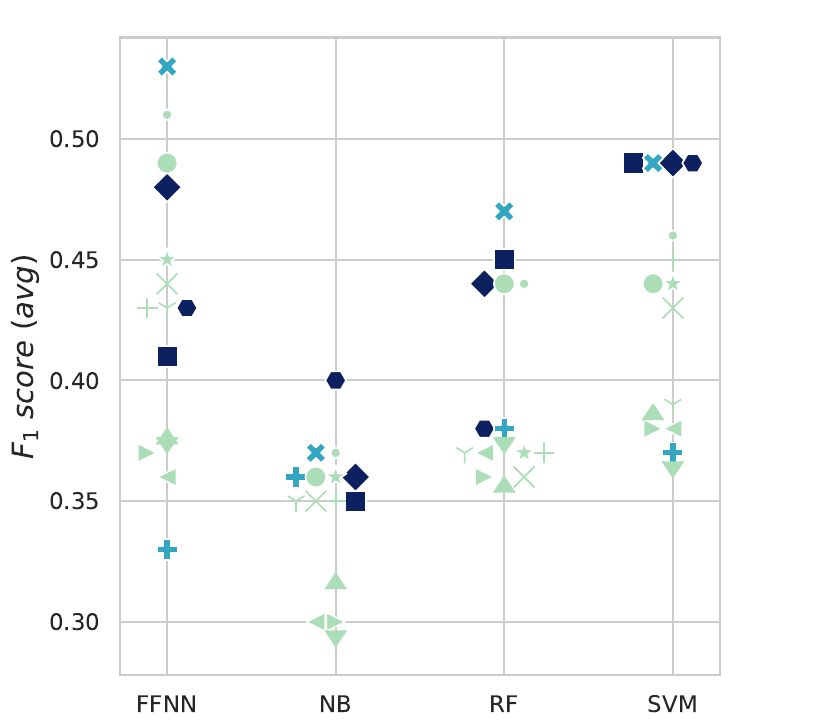} &   \includegraphics[trim=0 13 180 0,clip,width=0.42\textwidth]{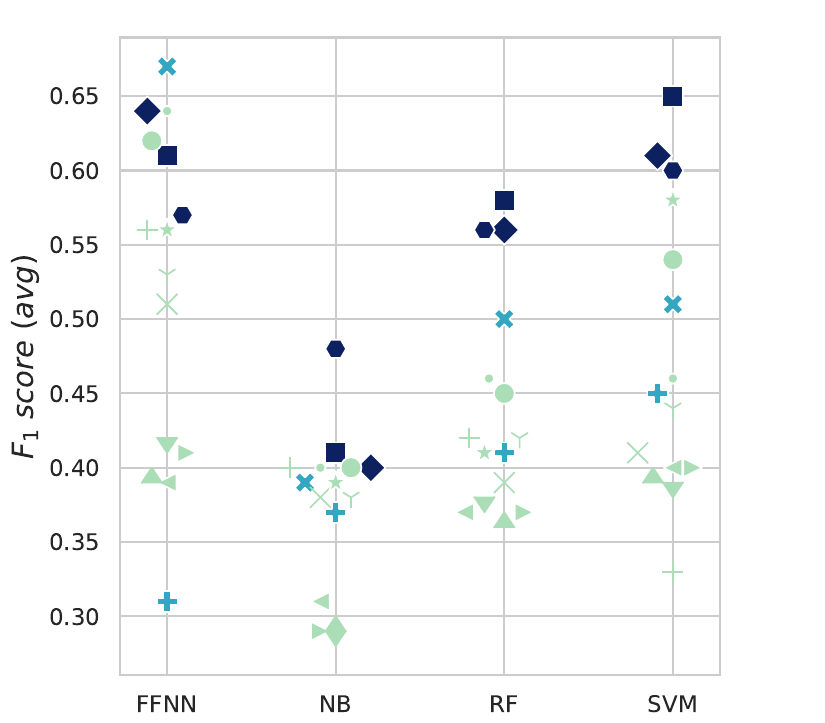} \\
(a) SR-ARE & (b) SR-MMP \\[3pt]
 \includegraphics[trim=0 13 180 0,clip,width=0.42\textwidth]{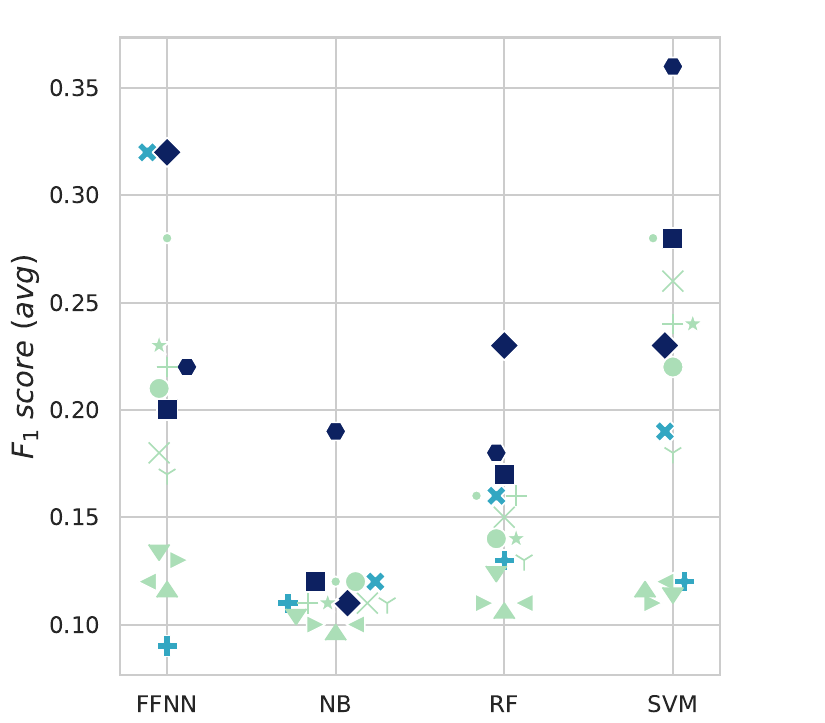} &   \includegraphics[trim=0 13 180 0,clip,width=0.42\textwidth]{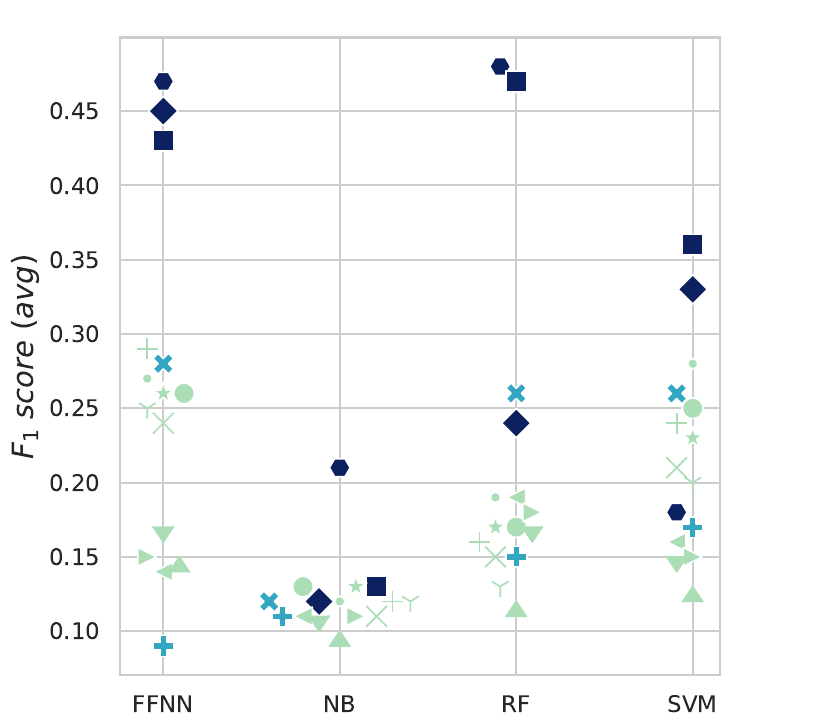} \\
(c) SR-ATAD5 & (d) HIV \\[3pt]
\includegraphics[trim=0 12 180 0,clip,width=0.42\textwidth]{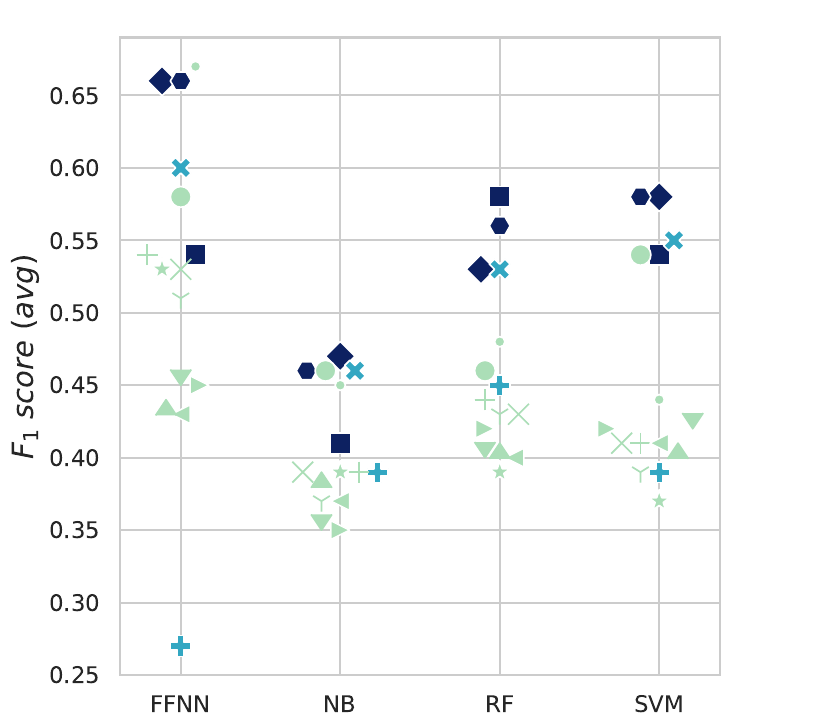} &
\includegraphics[trim=350 12 0 0,clip,width=0.22\textwidth]{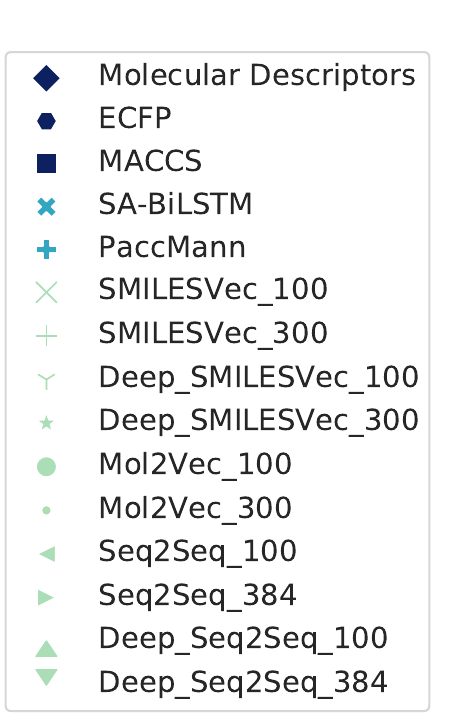}\\
(e) PCBA-686978
\end{tabular}
\caption{$F_1$ scores for the five classification datasets. Dark blue denotes traditional representations, light blue shows supervised embeddings, and green denotes unsupervised embeddings. Random horizontal jitter is applied to the markers to avoid overlap.
Traditional representations attained the best performances in most cases, matched or followed by \textit{SA-BiLSTM} and \textit{Mol2Vec} embeddings.}
\label{fig:q1_scatter_classification}
\end{figure*}

\begin{figure*}
\scriptsize
\centering
\begin{tabular}{ccc}
\includegraphics[trim=0 0 15 0,clip,width=0.3\textwidth]{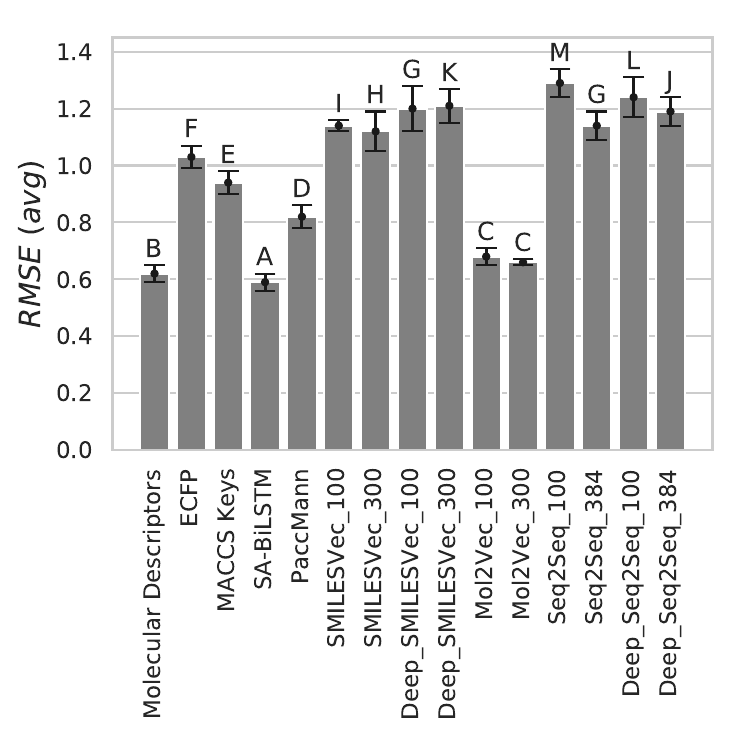} &
\includegraphics[trim=0 0 15 0,clip,width=0.3\textwidth]{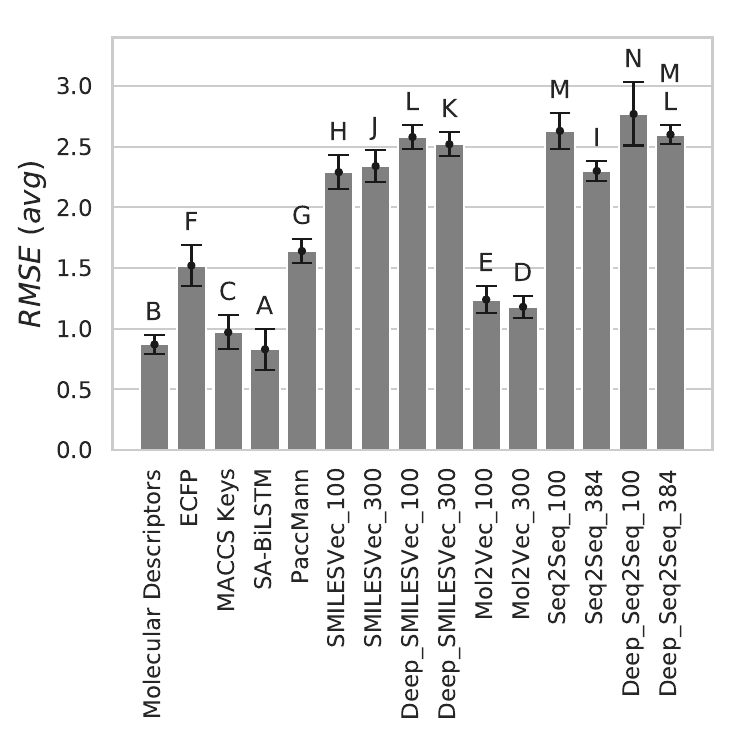} &
\includegraphics[trim=0 0 15 0,clip,width=0.3\textwidth]{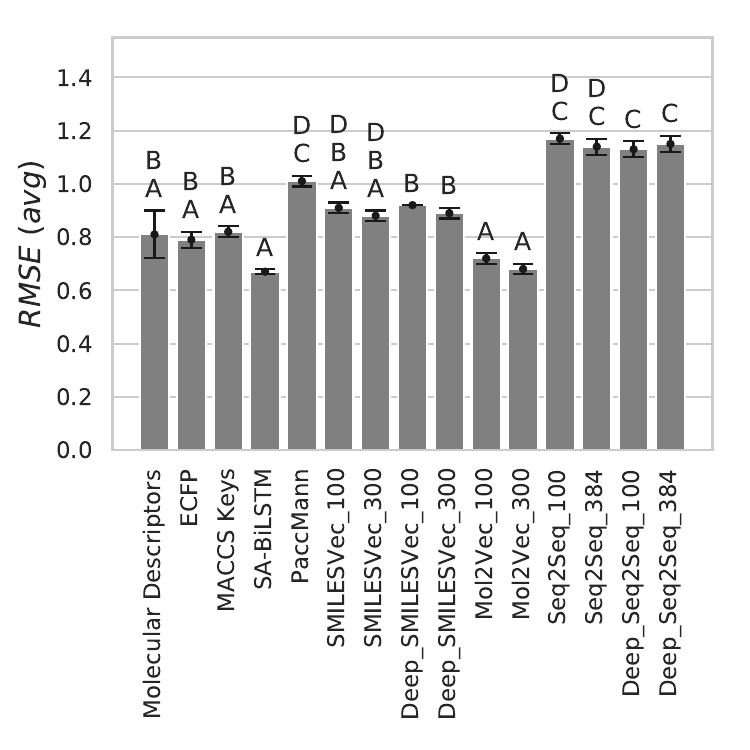}
\\
(a) ESOL  & (b) FreeSolv & (c) Lipophilicity\\[3pt]
\end{tabular}
\begin{tabular}{ccc}
\includegraphics[trim=0 0 15 0,clip,width=0.3\textwidth]{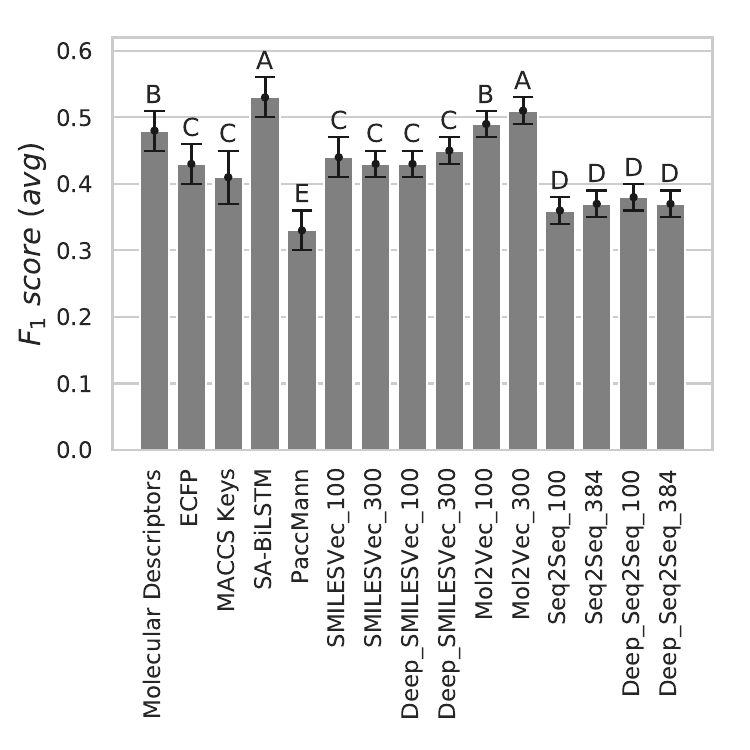} &
\includegraphics[trim=0 0 15 0,clip,width=0.3\textwidth]{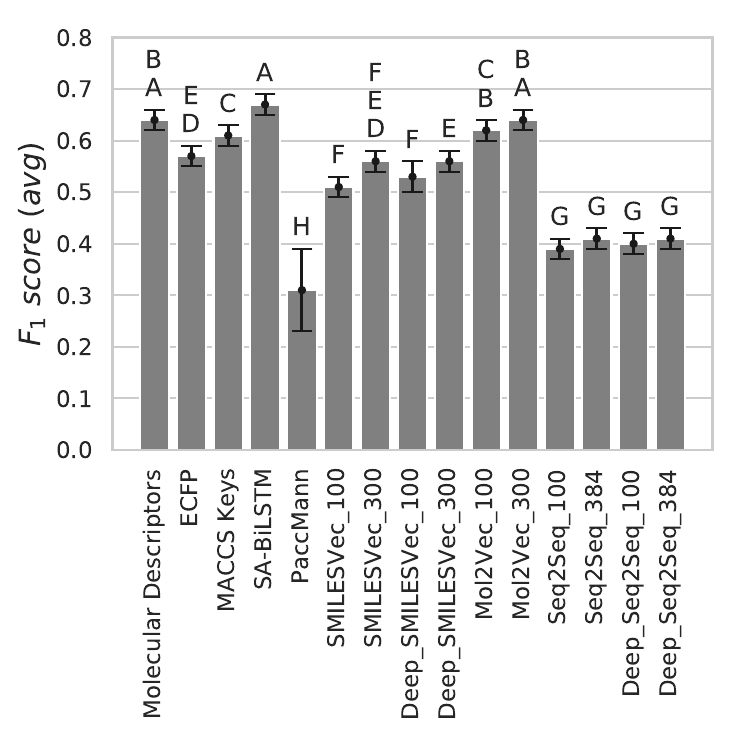} &
\includegraphics[trim=0 0 15 0,clip,width=0.3\textwidth]{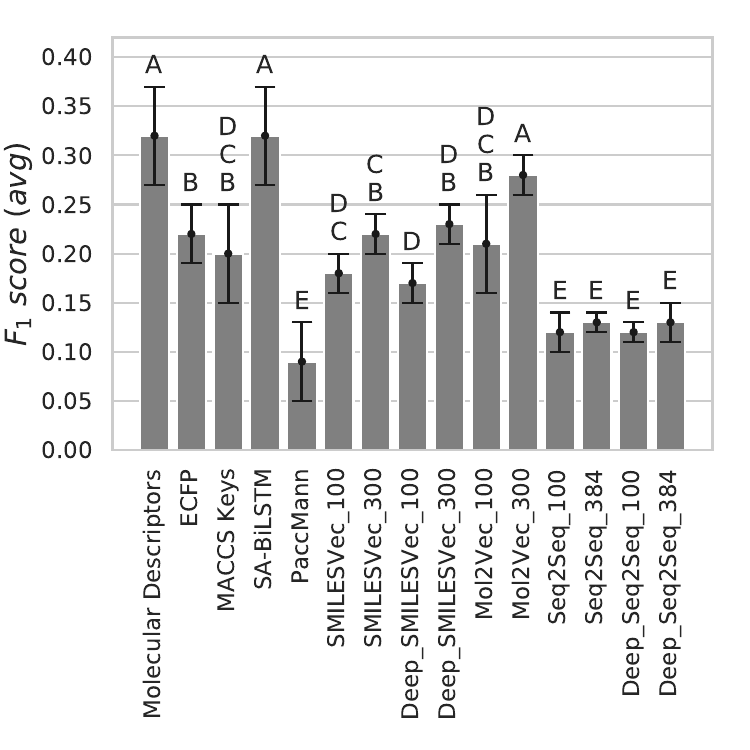} 
\\
 (d) SR-ARE & (e) SR-MMP & (f) SR-ATAD5 \\[3pt]
\end{tabular}
\begin{tabular}{cc}
\includegraphics[trim=0 0 15 0,clip,width=0.3\textwidth]{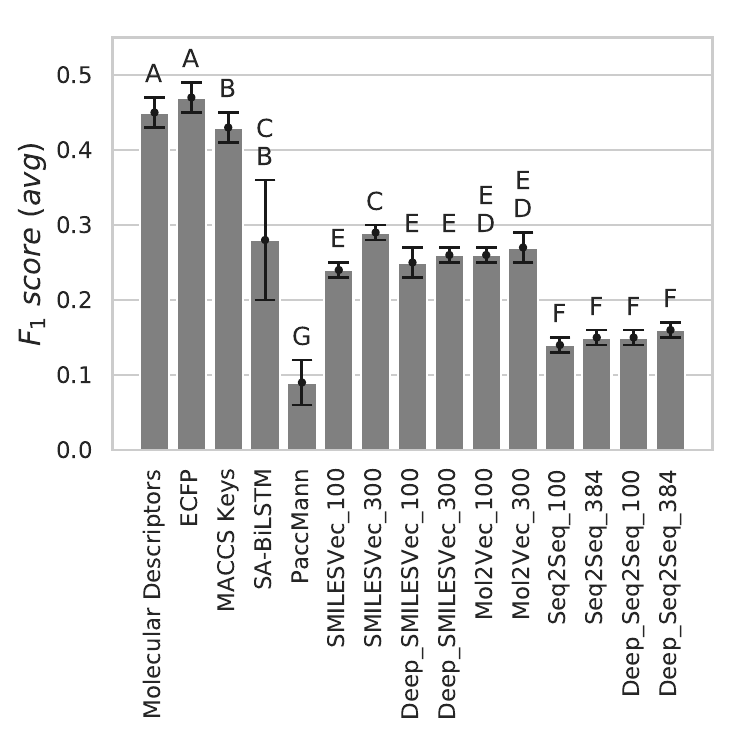} & 
\includegraphics[trim=0 0 15 0,clip,width=0.3\textwidth]{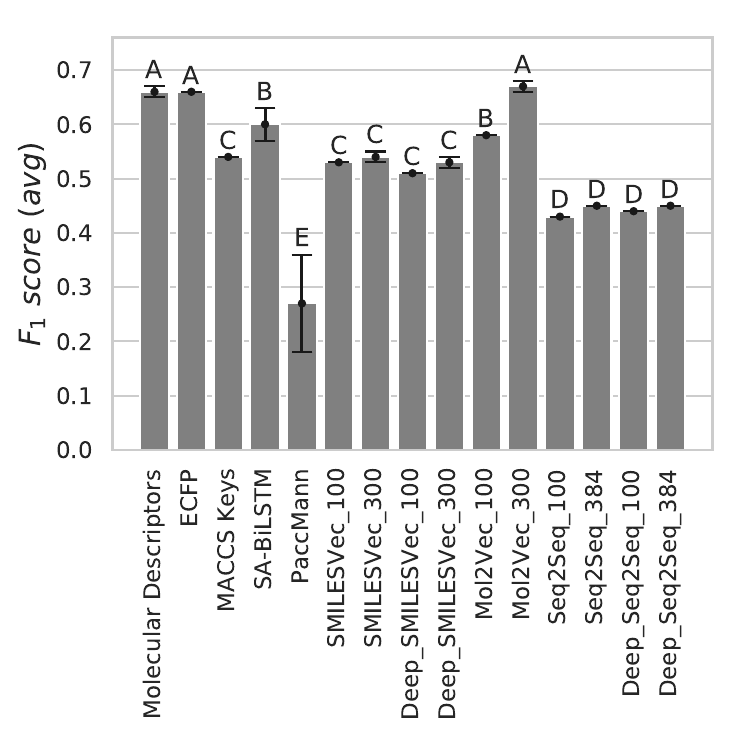}
\\
(g) HIV  & (h) PCBA-686978  \\[3pt]
\end{tabular}
\caption{$RMSE$ and $F_1$ scores  obtained by all molecular representations in FFNN for regression (a, b, c) and classification datasets (d, e, f, g, h), respectively. Means with the same letter are not significantly different, according to the pairwise Tukey test. In the classification datasets, the results obtained using \textit{Seq2Seq} embeddings are not statistically different in any of the datasets. \textit{Mol2Vec} tends to show different results depending on their embedding size and are not significantly different to \textit{SA-BiLSTM} embeddings or traditional representations. In datasets \textit{ESOL} and \textit{FreeSolv}, the results yielded by most representations were significantly different, whereas in dataset \textit{Lipophilicity} there were no statistically significant differences among traditional representations, supervised embeddings, and unsupervised embeddings \textit{Mol2Vec} and \textit{SMILESVec}.}
\label{fig:tukey}
\end{figure*}

For the classification tasks, and as it can be seen in Figure \ref{fig:q1_scatter_classification}, traditional representations were among the top-performing representations, followed by \textit{SA-BiLSTM} and \textit{Mol2Vec\_300}. Traditional molecular representations yielded the best results for all datasets in NB, SVM and RF classifiers, which were significantly better than most learned embeddings. In the case of the imbalanced datasets---\textit{SR-ATAD5} and \textit{HIV}---, \textit{ECFP} was significantly better than other representations using NB, and among the top-performing representation for the other classifiers, as shown in Figures \ref{fig:q1_scatter_classification} (c) and (d). \textit{MACCS keys} and \textit{molecular descriptors} also yielded significantly better results than learned embeddings in most datasets and classifiers. In the case of \textit{PCBA-686978} (Figure \ref{fig:q1_scatter_classification} (e)), the best results were obtained by traditional representations in all classifiers except for FFNN. There were no significant differences among the results obtained using the three traditional representations. The best results for FFNN were generally obtained using \textit{SA-BiLSTM}: this was observed for datasets \textit{SR-ARE}, \textit{SR-MMP} and \textit{SR-ATAD5}. For dataset \textit{HIV}, the best FFNN results were obtained by \textit{ECFP}, showing a significant difference to all learned embeddings, whereas \textit{Mol2Vec\_300} yielded the best results on FFNN for dataset \textit{PCBA-686978}. In all cases, these results were significantly better than those obtained using other learned embeddings, as shown in Figure \ref{fig:tukey}. In general, the results using unsupervised embeddings did not match the results using traditional representations, except for \textit{Mol2Vec} embeddings in FFNN classifiers. The statistical tests showed in all cases that the differences among these results were significant. 

\begin{figure*}
\scriptsize
\begin{tabular}{cc}
  \includegraphics[trim=0 13 180 0,clip,width=0.42\textwidth]{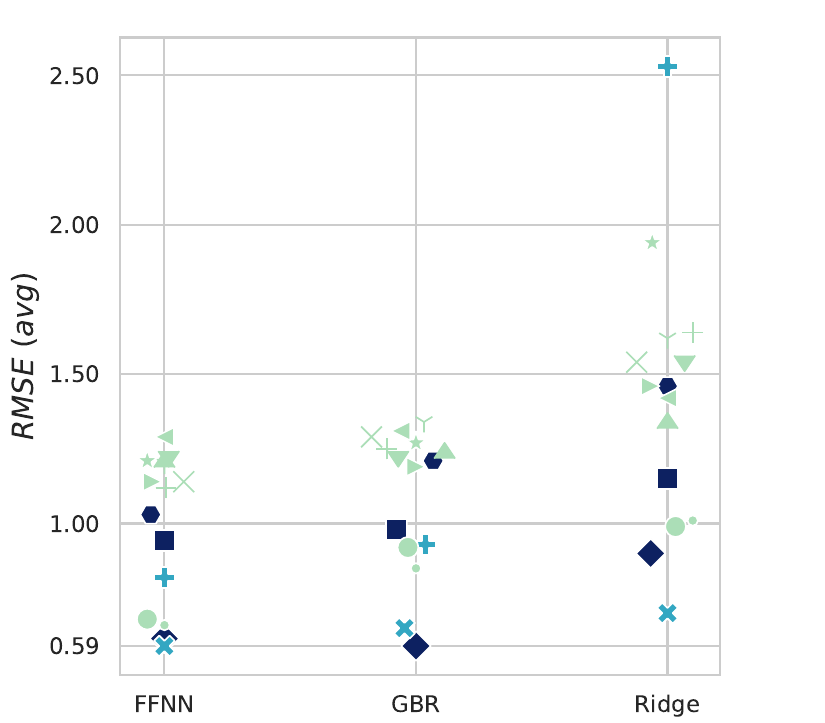} &   \includegraphics[trim=0 13 180 0,clip,width=0.42\textwidth]{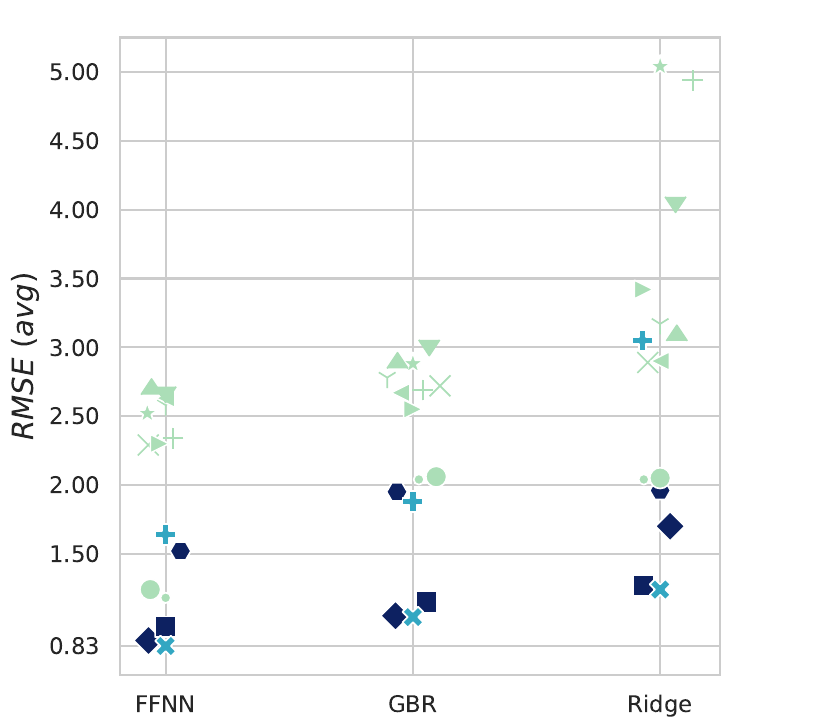}\\
(a) ESOL & (b) FreeSolv \\[3pt]
\includegraphics[trim=0 12 180 0,clip,width=0.42\textwidth]{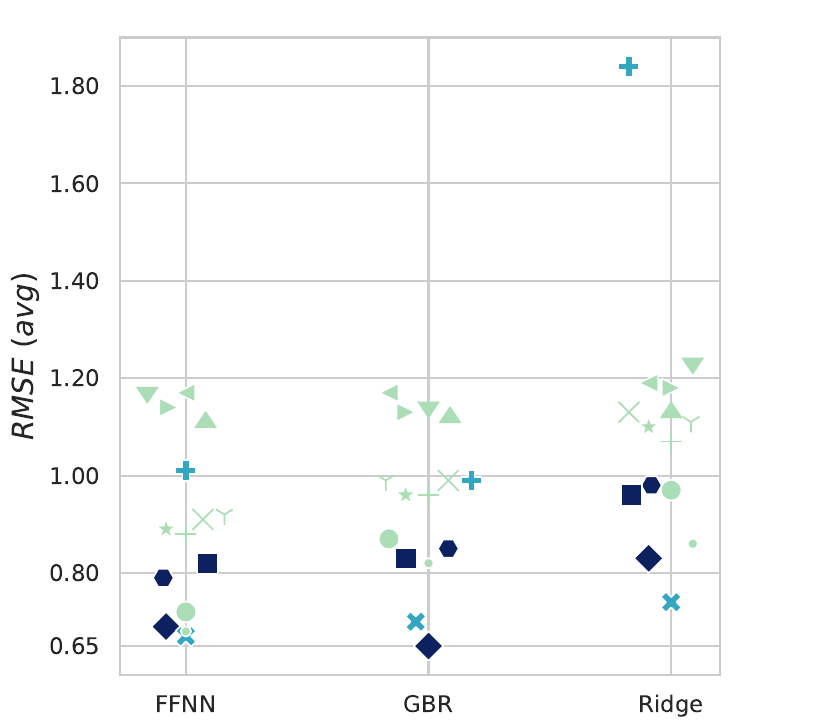} 
& \includegraphics[trim=350 12 0 0,clip,width=0.22\textwidth]{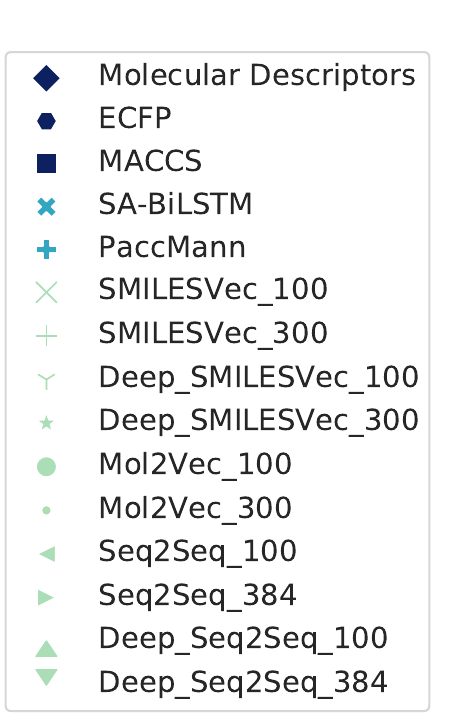} 
\\
(c) Lipophilicity
\end{tabular}
\caption{$RMSE$ values for the three regression datasets. Dark blue denotes traditional representations, light blue shows supervised embeddings, and green denotes unsupervised embeddings. Random horizontal jitter is applied to the markers to avoid overlap. Traditional representations generally yielded the top results, matched or followed by \textit{SA-BiLSTM} and \textit{Mol2Vec} embeddings. In particular, \textit{molecular descriptors} and \textit{SA-BiLSTM} yielded the best results on FFNN and GBR in all three datasets.}
\label{fig:q1_scatter_regression}
\end{figure*}

For regression tasks, as shown in Figure \ref{fig:q1_scatter_regression}, a similar scenario is observed: traditional molecular representations were among the top performing representations in all datasets and for all regression methods. In particular, \textit{molecular descriptors} and \textit{SA-BiLSTM} generally attained the best results on GBR and FFNN in all three datasets. In particular for GBR, the differences between the results of these two representations were not significant. \textit{SA-BiLSTM}, which was the top performing representation in FFNN, yielded significantly better results than those obtained by the unsupervised representations. A similar phenomenon was observed for traditional representations, except for \textit{Mol2Vec\_100} and \textit{Mol2vec\_300} which generally yielded results on a par with the traditional representations. In dataset \textit{Lipophilicity}, although the results for FFNN were different among the representations, these differences were no statistically significant among traditional representations, supervised embeddings, and unsupervised embeddings \textit{Mol2Vec} and \textit{SMILESVec}, as it can be seen in Figure \ref{fig:tukey} (c).

% FIGURA 5 VA AQUI
% \begin{figure*}
%   % \centering{{\color{black!20}\rule{100pt}{30pt}}}
%     \centering \rulebox{Figure \ref{fig:q1_scatter_classification} goes here}
% \caption{$F_1$ scores for the five classification datasets. Dark blue denotes traditional representations, light blue shows supervised embeddings, and green denotes unsupervised embeddings. Random horizontal jitter is applied to the markers to avoid overlap.
% Traditional representations attained the best performances in most cases, matched or followed by \textit{SA-BiLSTM} and \textit{Mol2Vec} embeddings.}
% \label{fig:q1_scatter_classification}
% \end{figure*}

% CODIGO LATEX FIGURA 5 (DESCOMENTAR)
% FIN CODIGO LATEX FIGURA 5 (DESCOMENTAR)

% FIGURA 6 VA AQUI
% \begin{figure*}
%   % \centering{{\color{black!20}\rule{100pt}{30pt}}}
%     \centering \rulebox{Figure \ref{fig:q1_scatter_regression} goes here}
% \caption{$RMSE$ values for the three regression datasets. Dark blue denotes traditional representations, light blue shows supervised embeddings, and green denotes unsupervised embeddings. Random horizontal jitter is applied to the markers to avoid overlap. Traditional representations generally yielded the top results, matched or followed by \textit{SA-BiLSTM} and \textit{Mol2Vec} embeddings. In particular, \textit{molecular descriptors} and \textit{SA-BiLSTM} yielded the best results on FFNN and GBR in all three datasets.}
% \label{fig:q1_scatter_regression}
% \end{figure*}

% FIGURA 6

% FIGURA 7 VA AQUI
% \begin{figure*}[b]
%   % \centering{{\color{black!20}\rule{100pt}{30pt}}}
%     \centering \rulebox{Figure \ref{fig:tukey} goes here}
% \caption{$RMSE$ and $F_1$ scores  obtained by all molecular representations in FFNN for regression (a, b, c) and classification datasets (d, e, f, g, h), respectively. Means with the same letter are not significantly different, according to the pairwise Tukey test. In the classification datasets, the results obtained using \textit{Seq2Seq} embeddings are not statistically different in any of the datasets. \textit{Mol2Vec} tends to show different results depending on their embedding size and are not significantly different to \textit{SA-BiLSTM} embeddings or traditional representations. In datasets \textit{ESOL} and \textit{FreeSolv}, the results yielded by most representations were significantly different, whereas in dataset \textit{Lipophilicity} there were no statistically significant differences among traditional representations, supervised embeddings, and unsupervised embeddings \textit{Mol2Vec} and \textit{SMILESVec}.}
% \label{fig:tukey}
% \end{figure*}

% CODIGO LATEX FIGURA 7 (DESCOMENTAR)
% FIN CODIGO LATEX FIGURA 7 (DESCOMENTAR)

Our second research question (Q2) intended to determine whether supervised molecular embeddings could surpass unsupervised molecular embeddings in classification and regression tasks. We conducted a two-way ANOVA test among all learned molecular embeddings. Afterward, we performed a pairwise Tukey test to compare the results using the supervised embeddings against the unsupervised embeddings on each classifier.

As shown in Figures \ref{fig:q1_scatter_classification} and \ref{fig:q1_scatter_regression}, the results obtained using the supervised representation \textit{SA-BiLSTM} were generally significantly better than the results obtained employing unsupervised embeddings. In classification tasks, \textit{SA-BiLSTM} was the top performing learned embedding in all balanced datasets (\textit{SR-ARE}, \textit{SR-MMP} and \textit{PCBA-686978}). This observation, however, did not hold in the case of the NB-based classifiers, for which no significant differences among classification results were observed in any of the datasets. In the case of datasets \textit{SR-ATAD5}, \textit{HIV} and \textit{PCBA-686978}, \textit{SA-BiLSTM} attained similar results to those obtained by \textit{Mol2Vec\_100} and \textit{Mol2Vec\_300} in all classifiers except RF. This might be explained by the fact that \textit{SA-BiLSTM} is trained using a form of \textit{Mol2Vec} embeddings. In the case of regression tasks, \textit{SA-BiLSTM} was also the top performing embedding in all datasets, showing significantly better results than those obtained by unsupervised representations. In dataset \textit{Lipophilicity}, no significant differences were observed between the results of \textit{PaccMann} and \textit{SMILESVec} embeddings in GBR, and between the results of \textit{Mol2Vec} and \textit{SA-BiLSTM} embeddings in FFNN.

In the case of \textit{PaccMann}, the results on the classification datasets did not show significant differences to the results obtained employing any of the unsupervised embeddings in the majority of the cases, except for FFNN-based classifiers, where the results are significantly worse than all other representations. This can be observed in Figures \ref{fig:q1_scatter_classification} and \ref{fig:tukey}. A different scenario is observed in the regression datasets, where \textit{PaccMann} embeddings exhibited a significantly better performance than all unsupervised embeddings, except for \textit{SMILESVec\_100} and \textit{Deep\_SMILESVec\_100} in \textit{Lipophilicity}, and \textit{Mol2Vec} in \textit{ESOL} and \textit{FreeSolv}, whose results in GBR were not significantly different. In Ridge regression, \textit{PaccMann} results were significantly worse than those of the remaining representations.

\begin{figure*}
\scriptsize
\centering
\begin{tabular}{cccc}
\includegraphics[trim=0 13 150 0,clip,width=0.30\textwidth]{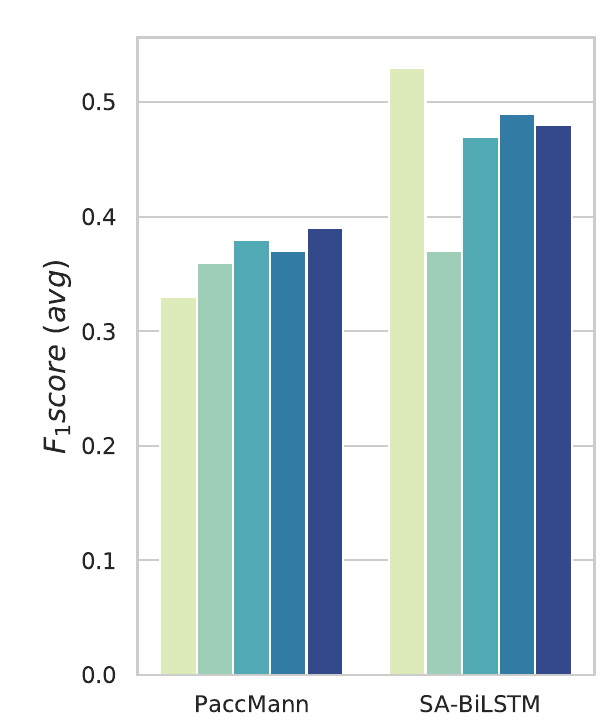} &
\includegraphics[trim=0 13 150 0,clip,width=0.30\textwidth]{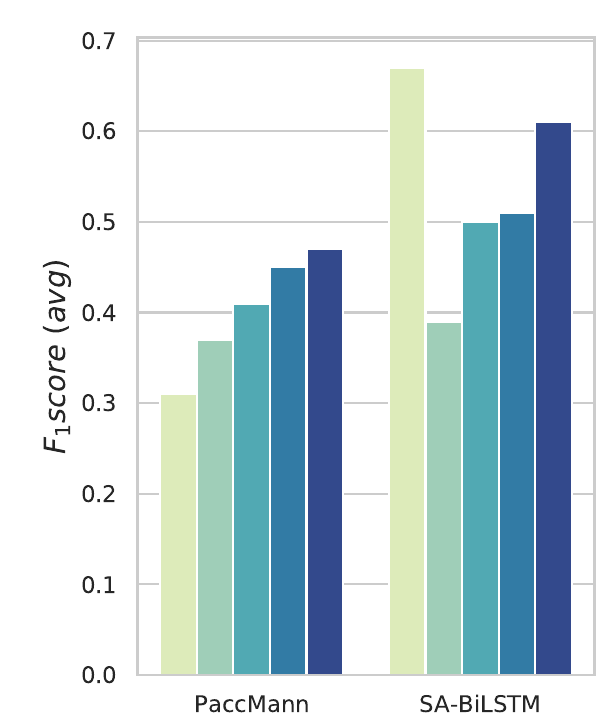} &
\includegraphics[trim=0 13 150 0,clip,width=0.30\textwidth]{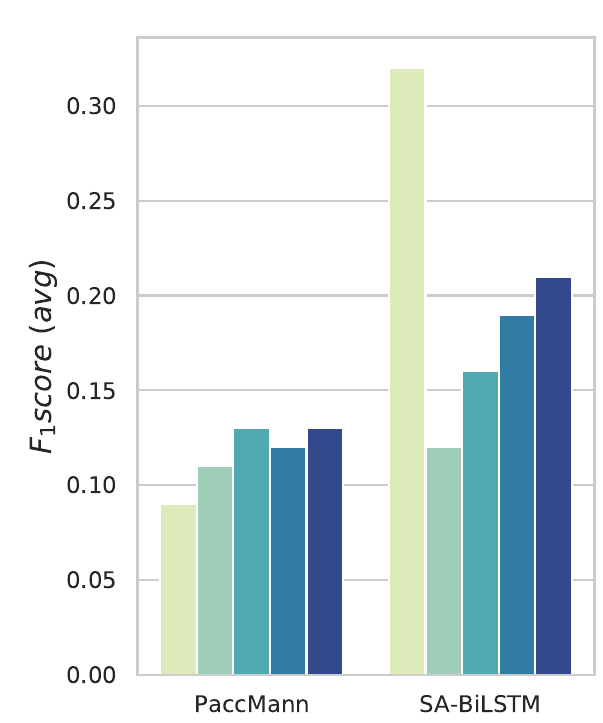} \\
(a) SR-ARE  & (b) SR-MMP & (c) SR-ATAD5  \\[3pt]
\end{tabular}
\begin{tabular}{cc}
\includegraphics[trim=0 13 150 0,clip,width=0.30\textwidth]{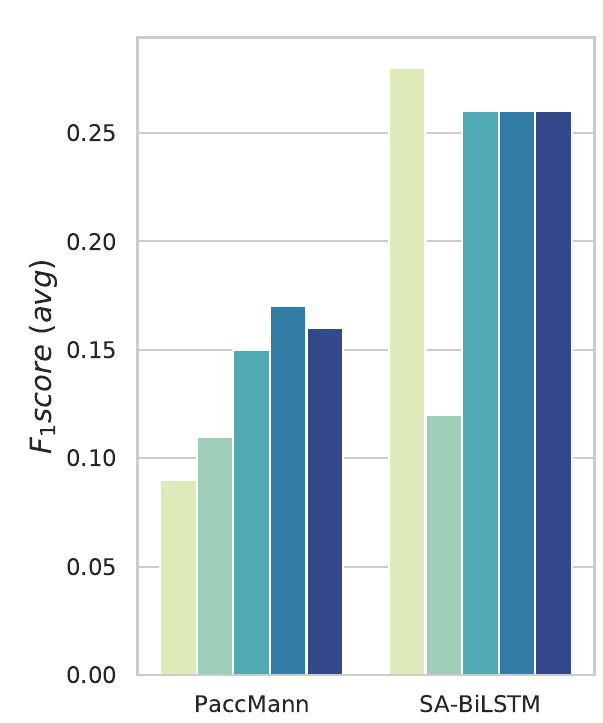} &
\includegraphics[trim=0 13 0 0,clip,width=0.46\textwidth]{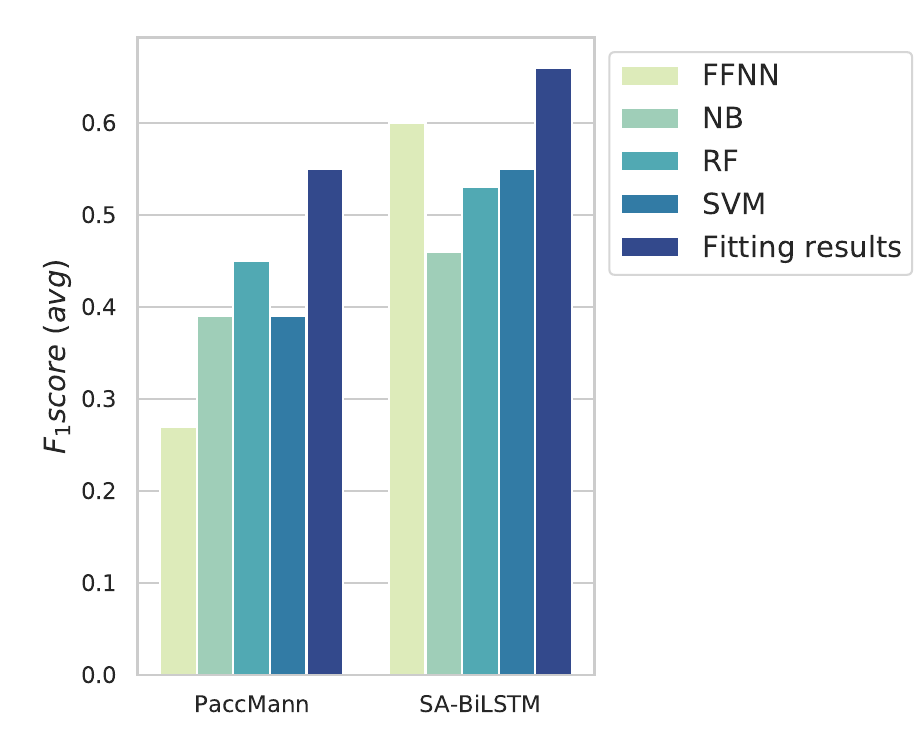} \\
(d) HIV  & (e) PCBA-686978  \\[3pt]
\end{tabular}
\caption{$F_1$ scores for the five classification datasets using supervised embeddings. \textit{SA-BiLSTM} obtained top results among the supervised embeddings, especially on FFNN. While \textit{PaccMann} results were significantly worse than those obtained by \textit{SA-BiLSTM}, its \textit{fitting} results---i.e., the classification results obtained from training the embedding model--- were usually better than those obtained on other classifiers.}
\label{fig:q2_bars_classification}
\end{figure*}

\begin{figure*}
\scriptsize
\centering
\begin{tabular}{ccc}
\includegraphics[trim=0 13 150 0,clip,width=0.28\textwidth]{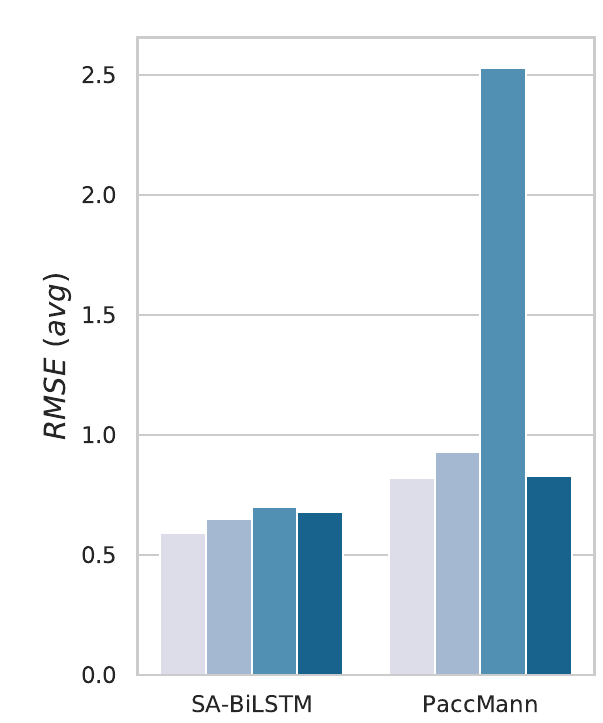} &
\includegraphics[trim=0 13 150 0,clip,width=0.28\textwidth]{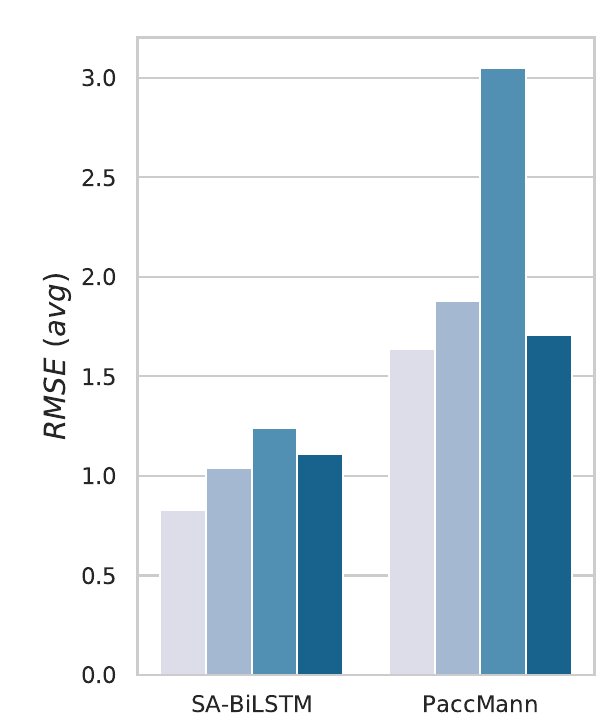} &
\includegraphics[trim=0 13 0 0,clip,width=0.425\textwidth]{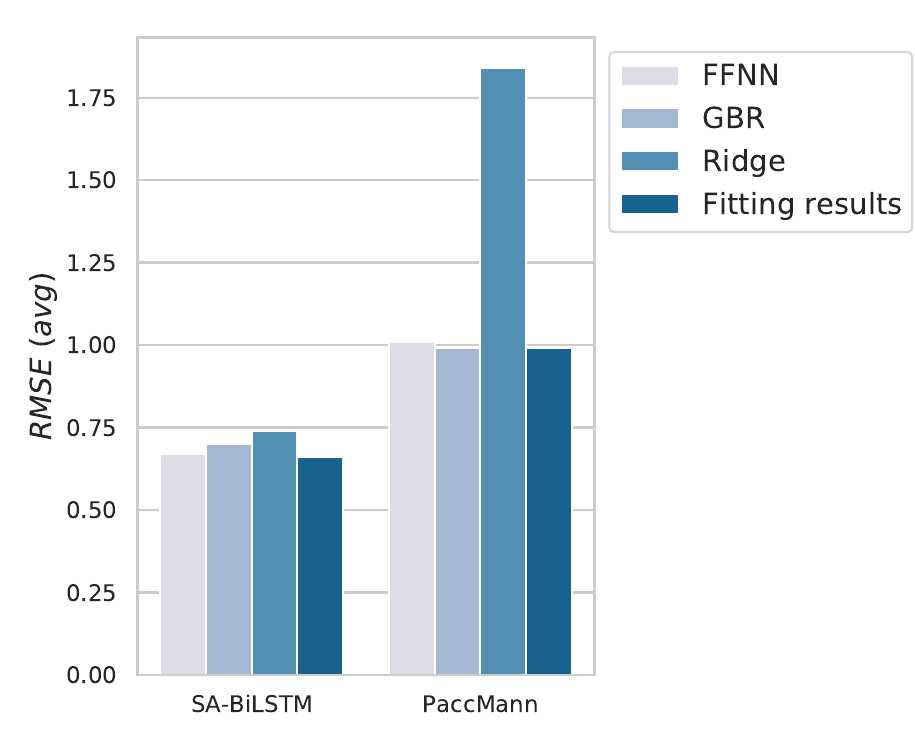} \\
(a) ESOL  & (b) FreeSolv & (c) Lipophilicity \\[3pt]
\end{tabular}
\caption{$RMSE$ results for the three regression datasets using supervised embeddings. \textit{SA-BiLSTM} obtained significantly better results than \textit{PaccMann}. The \textit{fitting} results of \textit{PaccMann} surpassed all other regression results obtained with the same embedding.}
\label{fig:q2_bars_regression}
\end{figure*}

When analyzing the unsupervised embeddings, for datasets \textit{ESOL} and \textit{FreeSolv} there were no significant differences between the results yielded by \textit{SMILESVec} and \textit{Seq2Seq} embeddings. In the case of the supervised embedding methods, as shown in Figure \ref{fig:q2_bars_classification} and \ref{fig:q2_bars_regression}, \textit{SA-BiLSTM} was significantly better than \textit{PaccMann} for all datasets. It is worth noticing that the \textit{fitting} results obtained by \textit{PaccMann} in all datasets were significantly better than those obtained by the \textit{PaccMann} embeddings using classifiers and regression methods. In the case of \textit{SA-BiLSTM}, its \textit{fitting} results were surpassed by at least one other classification technique in all datasets except for \textit{PCBA-686978} and \textit{Lipophilicity}. %However, this issue could be due to the number of nodes in the layers dedicated to property prediction in the \textit{SA-BiLSTM} model.

% FIGURA 8 VA AQUI
% \begin{figure*}[b]
%   % \centering{{\color{black!20}\rule{100pt}{30pt}}}
%     \centering \rulebox{Figure \ref{fig:q2_bars_classification} goes here}
% \caption{$F_1$ scores for the five classification datasets using supervised embeddings. \textit{SA-BiLSTM} obtained top results among the supervised embeddings, especially on FFNN. While \textit{PaccMann} results were significantly worse than those obtained by \textit{SA-BiLSTM}, its \textit{fitting} results---i.e., the classification results obtained from training the embedding model--- were usually better than those obtained on other classifiers.}
% \label{fig:q2_bars_classification}
% \end{figure*}

% CODIGO LATEX FIGURA 8 (DESCOMENTAR)
% FIN CODIGO LATEX FIGURA 8 (DESCOMENTAR)

% FIGURA 9 VA AQUI
% \begin{figure*}
%   % \centering{{\color{black!20}\rule{100pt}{30pt}}}
%     \centering \rulebox{Figure \ref{fig:q2_bars_regression} goes here}
% \caption{$RMSE$ results for the three regression datasets using supervised embeddings. \textit{SA-BiLSTM} obtained significantly better results than \textit{PaccMann}. The \textit{fitting} results of \textit{PaccMann} surpassed all other regression results obtained with the same embedding.}
% \label{fig:q2_bars_regression}
% \end{figure*}

% FIGURA 9

Our third research question (Q3) aimed to determine whether the canonical form of the SMILES formulas used during training or the embedding sizes had a significant impact on the classification performance of QSAR models. To answer this question, we analyzed the three unsupervised embedding techniques separately. These results can be seen in Figures \ref{fig:q1_scatter_classification} and \ref{fig:q1_scatter_regression}. 
We conducted separate two-way ANOVA tests and the corresponding pairwise Tukey tests to compare the results obtained by the following groups of molecular embeddings:
\begin{itemize}
    \item  \sloppy \textit{SMILESVec\_100, SMILESVec\_300, Deep\_SMILESVec\_100} and \textit{Deep\_SMILESVec\_300}: No significant differences were observed among the results in any of the five classification datasets, except for SVM in the case of the balanced datasets (\textit{SR-ARE}, \textit{SR-MMP} and \textit{PCBA-686978}), where all results were significantly different. We observed that the results of \textit{SMILESVec\_300} and \textit{Deep\_SMILESVec\_300} tended to be significantly better than those obtained by \textit{SMILESVEC\_100} and \textit{Deep\_SMILESVec\_100} when using FFNN for building the QSAR models on datasets \textit{SR-MMP} and \textit{SR-ATAD5}, as it can be seen in Figures \ref{fig:q1_scatter_classification} and \ref{fig:tukey}. In the case of regression tasks the differences were generally significant. Only for dataset \textit{Lipophilicity} we found no significant differences when varying the canonical form of the SMILES formulas in GBR. 
    \item \textit{Mol2Vec\_100} and \textit{Mol2Vec\_300}: The results obtained were significantly different when changing the size of the embedding from 100 to 300 throughout all datasets except for \textit{FreeSolv}. As it happened in the case of \textit{SMILESVec}, 300-dimension embeddings outperformed 100-dimension embeddings. This phenomenon was observed in all classifiers and regression models except for NB and Ridge, whose results were not significantly different.
    \item \textit{Seq2Seq\_100, Seq2Seq\_384, Deep\_Seq2Seq\_100} and \textit{Deep\_Seq2Seq\_384}: In the case of \textit{Seq2Seq}, while the changes in the canonicalization or the size of embeddings did not yield significantly different results in any of the five classification datasets, significant differences were observed in the results for regression datasets \textit{ESOL} and \textit{FreeSolv} when varying the canonical form of SMILES formulas.
\end{itemize}

In addition to our comparative analysis, we present the results obtained by the top-performing molecular representations on each dataset side by side to the results reported in the reference papers \cite{zheng2019identifying, jaeger2018mol2vec} and other state-of-the-art results for the datasets under study \cite{wu2018moleculenet, jiang2021could, wang2019smiles} in Tables \ref{tab:reference_results_classification} and \ref{tab:reference_results_regression}. In spite of these tables, which we present in an effort to connect with other results reported in the literature, it is worth mentioning that a direct comparison among our results and the reference ones should not be made, %because of the many different experimental conditions under which each paper reports its own results, and also 
considering that the vast majority of such papers report validation results on different sizes of fixed partitions. 

The referenced papers reported regression performance in terms of $RMSE$ for datasets \textit{ESOL, FreeSolv, Lipophilicity}, and classification performance in terms of $AUC$ for datasets \textit{SR-ARE, SR-MMP, SR-ATAD5, HIV}. In the case of dataset \textit{PCBA-686978}, the results were reported in terms of $Acc$ \cite{wang2019smiles}, even though such metric is not appropriate for imbalanced datasets \cite{sokolova2006beyond, chawla2002smote}. As shown in Table \ref{tab:reference_results_classification}, the results of our experiments on the classification datasets are either on a par or surpass the results obtained by the reference papers on the internal validation sets, which arguably shows that our results are sound and that the extensive model selection stage was crucial. In the case of the regression tasks, shown in Table \ref{tab:reference_results_regression}, we were able to match or surpass the results by one of the reference papers \cite{wu2018moleculenet}, but 
%we were not able to even reproduce the results reported in the other reference paper \cite{jiang2021could}.
not those reported in the other reference paper \cite{jiang2021could}. However, we reproduced their experiments on \textit{SVM} and \textit{XGBoost} using the source code provided by the authors, and the results appear to be on a par with our results.

\begin{figure*}
\scriptsize
\begin{tabular}{cc}
  \includegraphics[trim=0 0 10 0,clip,width=0.5\textwidth]{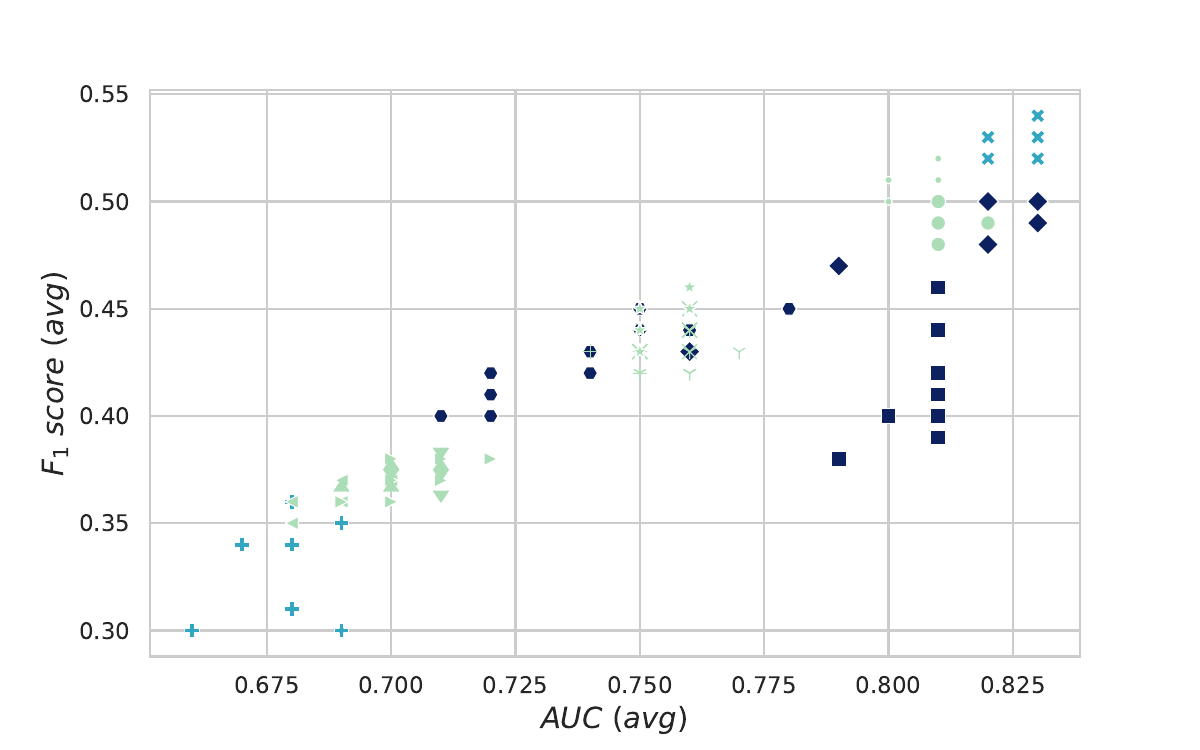} &   \includegraphics[trim=0 0 10 0,clip,width=0.5\textwidth]{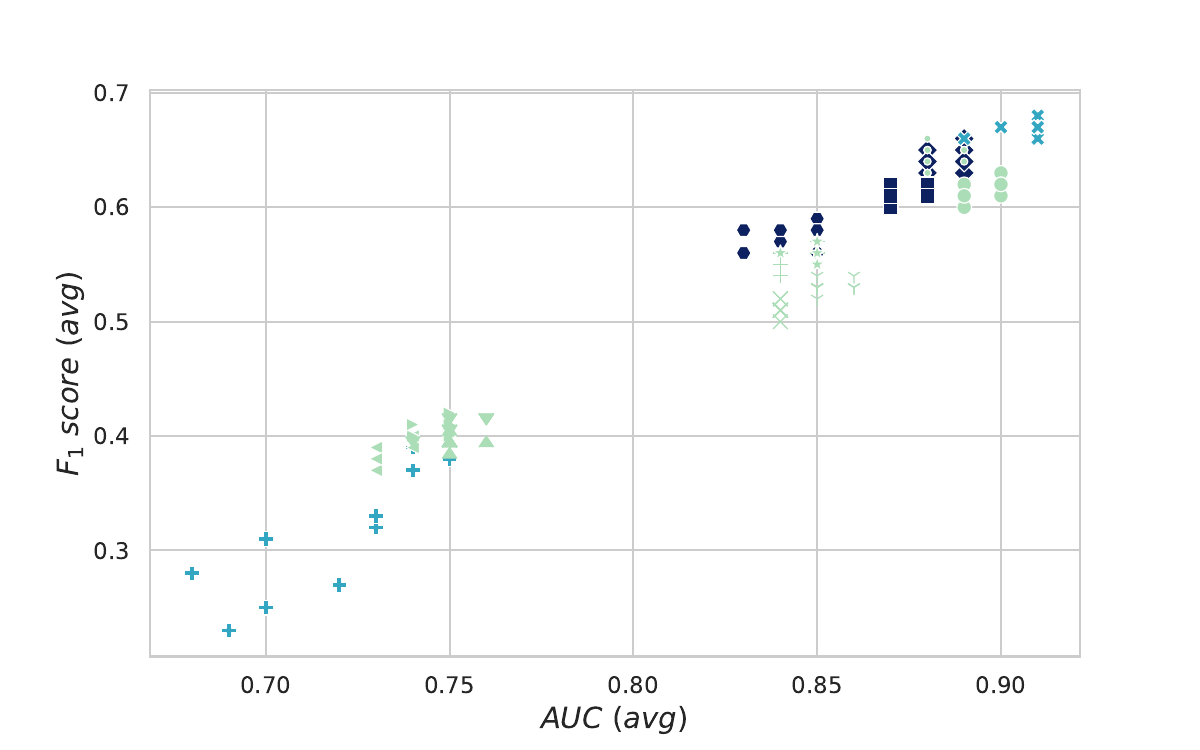} \\
(a) SR-ARE & (b) SR-MMP \\[3pt]
 \includegraphics[trim=0 0 10 0,clip,width=0.5\textwidth]{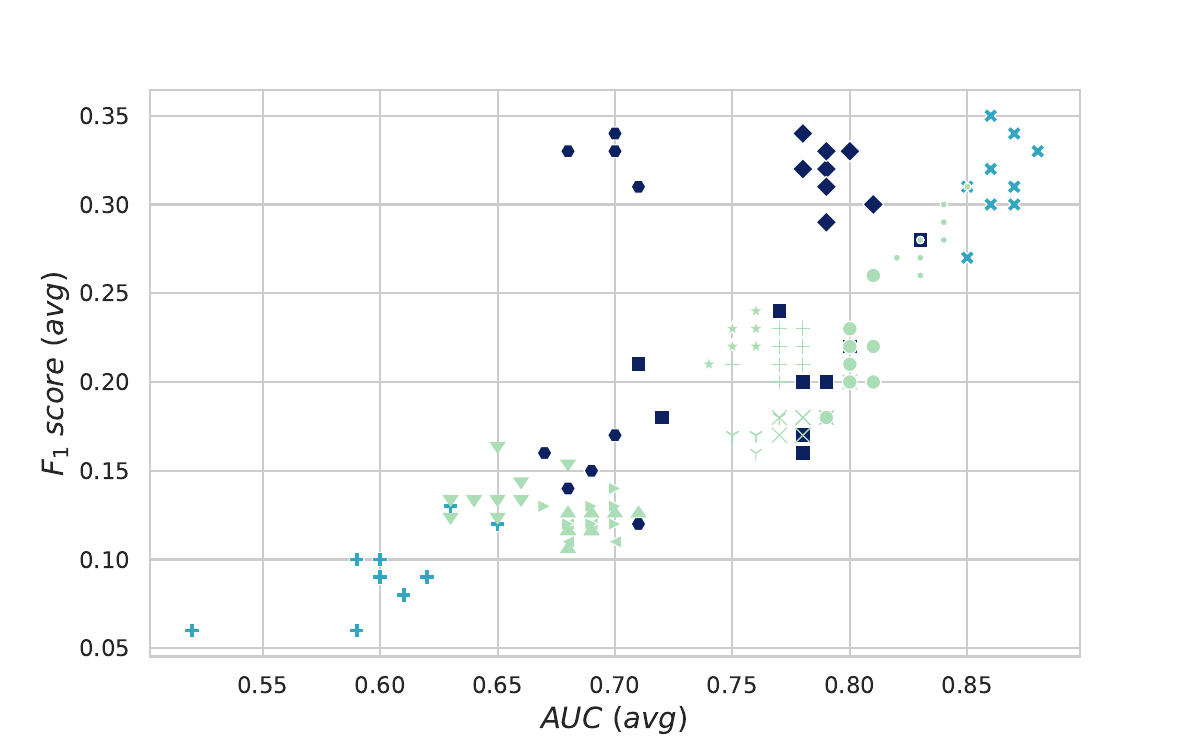} &   \includegraphics[trim=0 0 10 0,clip,width=0.5\textwidth]{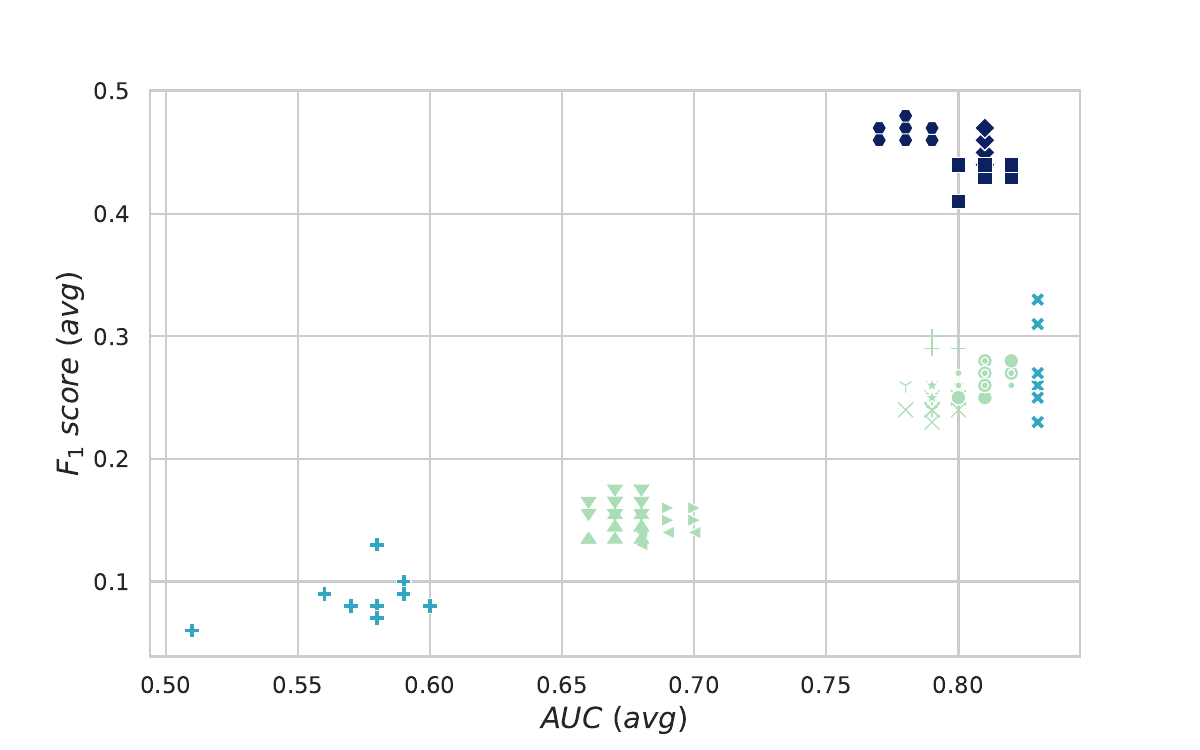} \\
(c) SR-ATAD5 & (d) HIV \\[3pt]
\includegraphics[trim=0 0 10 0,clip,width=0.5\textwidth]{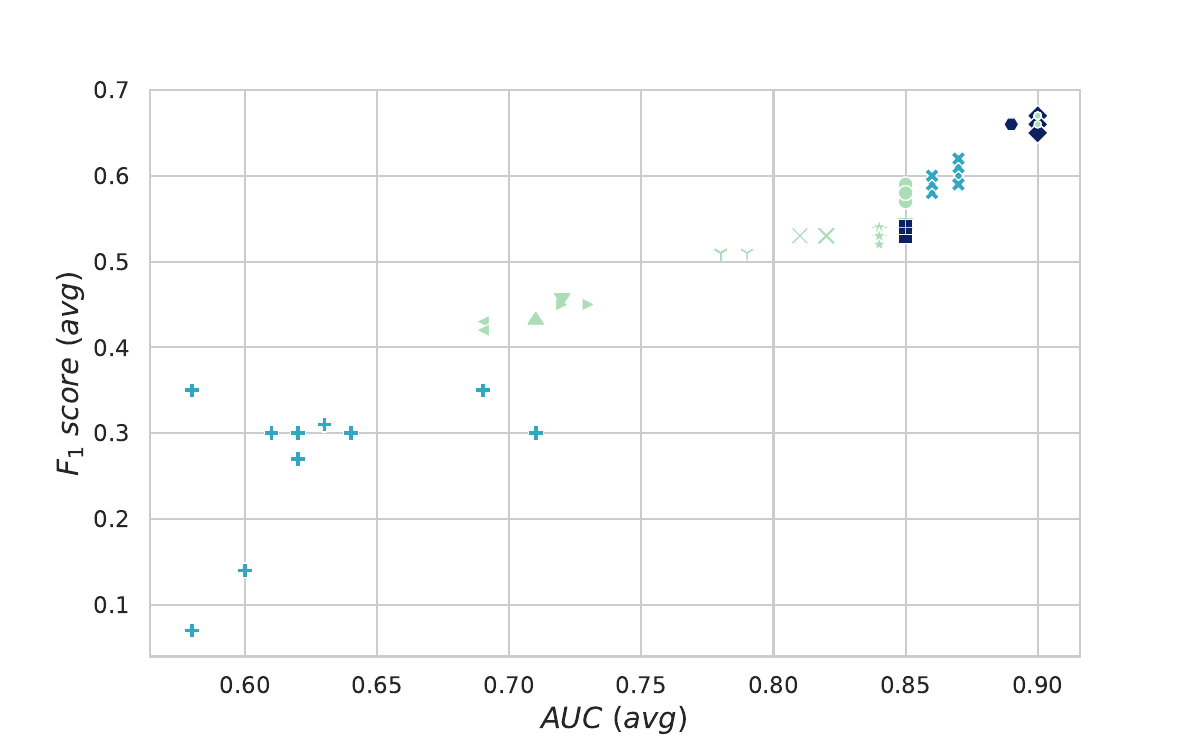} &
\includegraphics[trim=350 12 0 0,clip,width=0.22\textwidth]{q1/pcba_swarm_legend.pdf}\\
(e) PCBA-686978
\end{tabular}
\caption{Results in terms of $F_1$ score and $AUC$ for ten runs of FFNN classifiers, each using a different random initialization. \textit{SA-BiLSTM} obtained the best results in most classification datasets, often matched by \textit{molecular descriptors} or \textit{Mol2Vec} embeddings. Traditional representations yielded better $F_1$ score results than \textit{SA-BiLSTM} in the largest datasets (\textit{HIV} and \textit{PCBA-686978}).}
\label{fig:AUC_F1}
\end{figure*}

Finally, we present the classification results obtained by ten runs per molecular representation using FFNN classifiers, each using a different random initialization, in Figure \ref{fig:AUC_F1}. These results are expressed in terms of $F_1$ score and $AUC$, and they show the dispersion of the results of each molecular representation. Interestingly, while \textit{ECFP} attained the top results in terms of $F_1$ score for most classification datasets, \textit{SA-BiLSTM} matches or surpasses \textit{ECFP} when analyzing the $AUC$ results in all of these datasets except for \textit{PCBA-686978}. As it can be seen in Table \ref{tab:reference_results_classification} and in Figure \ref{fig:AUC_F1}, \textit{SA-BiLSTM} exhibits top $AUC$ results on FFNN, often on a par with \textit{molecular descriptors} and \textit{Mol2Vec} embeddings. There is low dispersion within different runs using the same molecular representations, with the exception of \textit{SR-ATAD5}, a fairly small and highly imbalanced dataset. When looking at the results in terms of $F_1$ score, traditional representations surpassed the majority of unsupervised embeddings in the case of the largest datasets, \textit{HIV} and \textit{PCBA-686978}. For \textit{PCBA-686978}, \textit{Mol2Vec\_300} obtained the top results using FFNN. When analyzing other metrics, such as $Precision$ and $Sp$ (available in the Supplementary Material), \textit{ECFP} attained better results than \textit{SA-BiLSTM} and \textit{Mol2Vec\_300}, which explains its superiority in terms of $F_1$ score.

% TABLA 4 VA AQUI
% \begin{table*}[b]
% \caption{Top results reported in the reference papers for the five classification datasets, in terms of $AUC$ and $Acc$, and top results obtained through our experimental workflow marked with an asterisk (*). The best results per dataset are highlighted in \textbf{bold}. \textit{na} denotes that the authors of the reference paper do not provide information about the confidence interval.}
% \label{tab:reference_results_classification}
%   \centering \rulebox{Table \ref{tab:reference_results_classification} goes here}
% \end{table*}

% CODIGO LATEX TABLA 4
\begin{table}[!t]
\resizebox{\textwidth}{!}{
\begin{tabular}{ccccccc}
\toprule
 & &  \multicolumn{4}{c}{AUC}                                   & Acc          \\ 
\cmidrule(r){3-6} \cmidrule(l){7-7}
Representation  & Classifier & SR-ARE       & SR-MMP       & SR-ATAD5     & HIV          & PCBA-686978  \\ 
\cmidrule(r){1-1} \cmidrule(r){2-2} \cmidrule(r){3-6} \cmidrule(){7-7}
            
SA-BiLSTM \cite{zheng2019identifying} & fitting & 0.81 $\pm$ \textit{na}        & 0.90 $\pm$ \textit{na}         & 0.86 $\pm$ \textit{na}         & 0.81 $\pm$ \textit{na}        & -            \\
Mol2Vec \cite{jaeger2018mol2vec} & RF & 0.83 $\pm$ 0.05 & 0.83 $\pm$ 0.05 & 0.83 $\pm$ 0.05 & -            & -            \\
Weave \cite{wu2018moleculenet}  & fitting   & \textbf{0.83 $\pm$ 0.01 }       & 0.83 $\pm$ 0.01        & 0.83 $\pm$ 0.01        & 0.74 $\pm$ 0.04         & -            \\
ECFP \cite{wu2018moleculenet}  & XGBoost   & 0.78 $\pm$ 0.02        & 0.78 $\pm$ 0.02        & 0.78 $\pm$ 0.02        & \textbf{0.84 $\pm$ 0.00}         & -            \\
SMILES-BERT \cite{wang2019smiles} & fine-tuning & -            & -            & -            & -            &\textbf{ 0.88 $\pm$ \textit{na}}         \\ 
\cmidrule(r){1-1} \cmidrule(lr){2-2} \cmidrule(lr){3-6} \cmidrule(l){7-7}
SA-BiLSTM (*) & FFNN   &\textbf{ 0.83 $\pm$ 0.02} &\textbf{ 0.91 $\pm$ 0.01 }&\textbf{ 0.87 $\pm$ 0.02} & 0.83 $\pm$ 0.01 & 0.80 $\pm$ 0.04 \\
Mol2Vec\_300 (*) & FFNN &  0.81 $\pm$ 0.01 & 0.88 $\pm$ 0.01 & 0.84 $\pm$ 0.01 & 0.81 $\pm$ 0.01 & 0.83 $\pm$ 0.00 \\
ECFP (*) & FFNN & 0.74 $\pm$ 0.02 & 0.84 $\pm$ 0.01 & 0.70 $\pm$ 0.04 & 0.78 $\pm$ 0.01 & 0.87 $\pm$ 0.00 \\ \bottomrule
\end{tabular}}
\caption{Top results reported in the reference papers for the five classification datasets, in terms of $AUC$ and $Acc$, and top results obtained through our experimental workflow marked with an asterisk (*). The best results per dataset are highlighted in \textbf{bold}. \textit{na} denotes that the authors of the reference paper do not provide information about the confidence interval.}
\label{tab:reference_results_classification}
\end{table}
% FIN CODIGO LATEX TABLA 4

% TABLA 5 VA AQUI
% \begin{table*}[b]
% \caption{Top results reported in the reference papers for the three regression datasets, in terms of $RMSE$, and top results obtained through our experimental workflow marked with an asterisk (*). The best results per dataset are highlighted in \textbf{bold}. \textit{na} denotes that the authors of the reference paper do not provide information about the confidence interval.}
% \label{tab:reference_results_regression}
%   \centering \rulebox{Table \ref{tab:reference_results_regression} goes here}
% \end{table*}

% CODIGO LATEX TABLA 5
\begin{table*}[!t]
\small
\begin{center}
\caption{Top results reported in the reference papers for the three regression datasets, in terms of $RMSE$, and top results obtained through our experimental workflow marked with an asterisk (*). The best results per dataset are highlighted in \textbf{bold}. \textit{na} denotes that the authors of the reference paper do not provide information about the confidence interval.}
\label{tab:reference_results_regression}
\begin{tabular*}{\textwidth}{@{\extracolsep{\fill}}ccccc@{\extracolsep{\fill}}}
\toprule
 & &  \multicolumn{3}{c}{RMSE}\\ 
\cmidrule(l){3-5} 
Representation  & Regression Model & ESOL       & FreeSolv       & Lipophilicity \\ 
\cmidrule(r){1-1} \cmidrule(lr){2-2} \cmidrule(l){3-5} 
Attentive FP \cite{jiang2021could} & fitting & \textbf{0.48 $\pm$ \textit{na}} & 0.52 $\pm$ \textit{na} & \textbf{0.52 $\pm$ \textit{na}} \\
MOE + FPs \cite{jiang2021could} & SVM & 0.62 $\pm$ \textit{na} &\textbf{ 0.42 $\pm$ \textit{na}} & 0.55 $\pm$ \textit{na} \\
MOE + FPs \cite{jiang2021could} & XGBoost & 0.51 $\pm$ \textit{na} & 0.69 $\pm$ \textit{na} & \textbf{0.52 $\pm$ \textit{na} }\\
MPNN \cite{wu2018moleculenet}  & fitting   & 0.55 $\pm$ 0.02 & 1.20 $\pm$ 0.02  & 0.76 $\pm$ 0.03 \\
Weave \cite{wu2018moleculenet}  & fitting   & 0.57 $\pm$ 0.04 & 1.19 $\pm$ 0.08 & 0.73 $\pm$ 0.01 \\
GC \cite{wu2018moleculenet}  & fitting   & 1.05 $\pm$ 0.15  & 1.35 $\pm$ 0.15  & 0.68 $\pm$ 0.04 \\
\cmidrule(r){1-1} \cmidrule(lr){2-2} \cmidrule(l){3-5}
SA-BiLSTM (*) & FFNN &  0.59 $\pm$ 0.03 & 0.83 $\pm$ 0.17 & 0.67 $\pm$ 0.01  \\
Mol2Vec\_300 (*) & FFNN & 0.66 $\pm$ 0.01 & 1.18 $\pm$ 0.09 & 0.68 $\pm$ 0.02 \\
Molecular descriptors (*) & FFNN & 0.62 $\pm$ 0.03 & 0.87 $\pm$ 0.08  & 0.69 $\pm$ 0.03  \\ 
Molecular descriptors (*) & GBR & 0.59 $\pm$ 0.04 & 1.05 $\pm$ 0.12 & 0.65 $\pm$ 0.02  \\
\bottomrule
\end{tabular*}
\end{center}
\end{table*}
% CODIGO LATEX TABLA 5

% FIGURA 10 VA AQUI
% \begin{figure*}
%   % \centering{{\color{black!20}\rule{100pt}{30pt}}}
%     \centering \rulebox{Figure \ref{fig:AUC_F1} goes here}
% \caption{Results in terms of $F_1$ score and $AUC$ for ten runs of FFNN classifiers, each using a different random initialization. \textit{SA-BiLSTM} obtained the best results in most classification datasets, often matched by \textit{molecular descriptors} or \textit{Mol2Vec} embeddings. Traditional representations yielded better $F_1$ score results than \textit{SA-BiLSTM} in the largest datasets (\textit{HIV} and \textit{PCBA-686978}).}
% \label{fig:AUC_F1}
% \end{figure*}

% CODIGO LATEX FIGURA 10
% FIN CODIGO LATEX FIGURA 10

Based on our observations we can conclude that, in general, learned molecular embeddings did not broadly surpass the results obtained by traditional molecular representations. Also, most of the unsupervised embeddings methods did not match the results obtained by traditional molecular representations. Among the unsupervised embedding techniques, \textit{Mol2Vec} yielded the best results, often performing significantly better than the results obtained using \textit{SMILESVec} or \textit{Seq2Seq}. This could be related to the preprocessing step on the SMILES formulas, which is based on the algorithm for computing \textit{ECFP} fingerprints, in contrast to the simple tokenization steps applied to SMILES formulas in the case of the other two techniques. \textit{Seq2Seq} obtained the lowest performance in terms of $F_1$ score and $RMSE$ of all representations, especially in the case of imbalanced classification datasets \textit{SR-ATAD5} and \textit{HIV}.

In the case of the supervised embeddings, on the one hand, \textit{SA-BiLSTM} yielded the best results among the learned representations, which were often on a par with those obtained using traditional molecular representations. On the other hand, \textit{PaccMann} did not yield good results in classification tasks. This could be explained by the high dimensionality of \textit{PaccMann} embeddings---ranging from $6976$ to $24,200$ dimensions, as shown in Table \ref{table:summary-supervised}---, which is often a cause of traditional machine learning models having difficulties in learning patterns from the data and, consequently, of a low predictive performance \cite{Goodarzi2012, alsenan2020autoencoder}. It is also worth noticing that the original \textit{PaccMann} embedding model proposed by 
%\citet{oskooei2018paccmann} 
Oskooei et al.~\cite{oskooei2018paccmann} was developed as a multimodal encoder for anticancer compound sensitivity prediction. Another possible reason for the significant difference between the results of the two supervised embedding techniques might be that \textit{SA-BiLSTM} consists of a multi-head self-attention model, whereas \textit{PaccMann} is a single-head self-attention model.

Our experiments were designed to attain the best possible performance for each molecular representation. We experimented with different classification and regression techniques, dataset sizes and class imbalance scenarios. We also tested different parameterizations and variations of the input data for each embedding method. The conclusions derived in this paper are presented after observing consistent and sound results throughout our analysis, which were also similar to other results reported in the literature \cite{zheng2019identifying,wu2018moleculenet,jiang2021could,jaeger2018mol2vec,wang2019smiles}. 
This pattern of learned molecular representations only matching the performance of traditional representations, while unexpected, can also been observed in other studies that experimented with learned molecular embeddings for QSAR modeling \cite{jiang2021could, jaeger2018mol2vec, gomez2018automatic, goh2017chemception, yang2019analyzing, erratumyang2019analyzing, chithrananda2020chemberta}. However, such results are not conclusive, as they either are not supported by any significance test, or there was not a systematic comparison where hyperparameters were tuned fairly.%, which might have potentially caused the severe overfitting observed in the training versus validation results \cite{wu2018moleculenet, jiang2021could} . %It also reminisces a common phenomenon in machine learning where simple modeling techniques are on a par with complex architectures \cite{musgrave2020metric, schick2020s}. We believe that our experiments motivate a rethinking of molecular embedding techniques and the potential role of learned representations or meta-learning algorithms for QSAR modeling.
We argue that our results prove the importance of conducting a thorough and careful experimental comparison of molecular embedding techniques and the potential role of learned representations in QSAR modeling.

Despite our findings, learned molecular embeddings might be suitable for many other tasks than QSAR modeling in the spectrum of drug design and virtual screening, which opens a wide range of experimental possibilities and future work \cite{chuang2020learning, elton2019deep, schneider2010virtual}. While traditional representations are computed following a standard algorithm and looking at a single molecule at a time, learned embeddings can be computed from large sets of compounds, potentially yielding richer representations that could be suitable for molecular similarity analysis \cite{chuang2020learning, huang2016communication}. Besides, techniques like self-attention might yield good embeddings for tasks like molecular substructure search, molecular docking \cite{chuang2020learning, elton2019deep, zheng2019identifying}, or for linking molecular substructures to bioactivity profiles \cite{zheng2019identifying}. Another possible direction would be to combine traditional representations with molecular embeddings, thus leveraging their strengths. 

\section{Conclusions}\label{sec:conclusion}

Many novel deep learning-based algorithms for learning molecular representations have been proposed in recent years. However, along with the proliferation of novel algorithms, it is crucial to conduct methodical and careful comparisons between different existing algorithms in order to shed light onto the key aspects of the embedding process. In this paper, we carried out an extensive comparison, which consisted of over $25,000$ trained models, using five different techniques for learning molecular representations. We tested their suitability for QSAR modeling in five classification and three regression tasks, considering aspects such as embedding size, SMILES canonicalization and deep learning algorithms, and compared them against traditional molecular representations. 

As a result of our experimental analysis, we found no evidence of molecular embeddings surpassing molecular descriptors or fingerprints, which might indicate that traditional representations are as good as molecular embeddings are at QSAR prediction tasks. We also found that unsupervised molecular embeddings exhibited lower performance than supervised techniques in general. Despite these results that contradict the current trend in molecular representations, drug discovery has numerous flourishing areas of research where there is room for learned representations, such as \textit{de novo} drug design, molecular docking or virtual screening \cite{chuang2020learning, elton2019deep}. Contrary to deterring the community from using and developing new algorithms for learning molecular representations, we hope that the results observed in this paper can serve as an incentive to design new ways of using molecular embeddings to leverage their potential for different tasks in drug discovery.


\section{Appendices}

\subsection{Appendix A: hyperparameters tested during grid search of supervised embedding methods}\label{apx:grid_search_supervised}

\begin{itemize}
    \item \textit{PaccMann}: in the self-attention layer, we varied the number of hidden units (i.e.~$16$, $50$ and $100$) and the attention depth value (i.e.~$20$, $50$, $100$ and $256$). In the feed-forward neural network used for property prediction, we varied the number of hidden units (i.e.~$[512, 256, 64, 16], [100, 50, 20, 5], [150, 50, 10]$ and $[100, 20, 5]$) and the activation functions (i.e.~\textit{ReLU} and \textit{sigmoid}) \cite{sharma2017activation} of the dense layers. We trained with and without a sinusoidal positional encoding function applied to the inputs \cite{vaswani2017attention}, and with and without a weighed loss function based on the imbalance in each class \cite{domingos1999metacost}. We varied the regularization hyperparameters such as dropout coefficients per layer (i.e.~$0.5$, $0.25$ and $0.15$) and \textit{lambda} values (i.e.~$0.0001$, $0.005$ and $0.001$) for L2-regularization \cite{moody1995simple}. The learning rate was fixed to $0.001$, and we used Adam optimizer \cite{kingma2014adam}. We used a minibatch size of $512$ for the small dataset (i.e.~\textit{SR-ARE, SR-MMP, SR-ATAD5}) and a minibatch size of $2048$ for the large datasets (i.e.~\textit{HIV, PCBA-686978}). As a result of this grid search process, we tested a total of $175$ different combinations of hyperparameters for each labeled dataset. 
    \item \textit{SA-BiLSTM}: since each run of this method takes a significant amount of time to train, we decided to focus on the hyperparameters tested by the authors in the original paper. We varied the number of attention heads (i.e.~$5$, $10$, $15$ and $18$) and the attention depth value (i.e.~$10$, $20$, $50$ and $100$) in the self-attention layer, as well as the number of hidden units in the Bi-LSTM RNN (i.e.~$64$ and $128$). Following the values indicated in the reference paper, we used a minibatch size of $64$ for all datasets and used a gradient clipping coefficient of $0.3$. The dropout coefficient per layer was set to $0.2$ and the $\lambda$ value for L2-regularization \cite{moody1995simple} to $0.01$. The models were trained for a maximum of 1000 epochs, as per the reference paper. We trained with and without applying a weighed loss function which compensated the class imbalance. As a result of this grid search stage, we tested a total of $64$ different combinations of hyperparameters for each labeled dataset.
\end{itemize}

\subsection{Appendix B: hyperparameters tested during grid search of classifiers and regression models} \label{apx:grid_search_classification}

\sloppy For each classification method, we conducted a hyperparameter grid search using a stratified data partition. In the case of the SVMs, we varied the regularization coefficient $c\in \{0.01, 0.05, 0.1, 0.25, 0.5, 1, 5, 10, 20\}$, and for the RF classifier we varied the maximum depth of the tree $md \in \{2, 3, 5, 8, 10, 20\}$. 
For the FFNNs, we varied the number of nodes in each layer (i.e.~$[150, 50, 10]$,  $[100, 50, 10]$  and $[100, 20, 5]$), the L2-regularization hyperparameter $\lambda \in \{0.0001, 0.005, 0.001\}$, the minibatch size $b \in \{64, 128, 256, 512\}$, the activation function used in the hidden layers (i.e.~ReLU and $tanh$) and the early-stopping $patience$ coefficient $p\in \{70, 100, 200, 500\}$. We also tested different learning rates $\alpha \in \{0.00001, 0.0001, 0.001\}$. %Further details about the parameterization of these classifiers are provided in the Supplementary Information.

\sloppy In the case of the regression models, for Ridge regression we varied the regularization coefficient $\alpha \in \{0.001, 0.05, 0.1, 0.5, 1, 10\}$, the tolerance $tol \in \{0.00001, 0.0001, 0.001\}$, and set the maximum number of iterations for convergence $max_iter= 10000$. For Gradient Boosting Regressor (GBR), we varied the number of estimators $n_estimators \in \{50, 100, 200, 500\}$, the maximum depth of the tree $max_depth \in \{2, 3, 5, 8 10, 20\}$ and the minimum number of samples required to split an internal node $min_samples_split \in \{2, 3\}$. Finally, for the FFNNs we varied the number of nodes in each layer (i.e.~$[200, 70, 15]$, $[150, 50, 10]$,  $[100, 50, 10]$ and $[100, 20, 5]$), the L2-regularization hyperparameter $\lambda \in \{0.0001, 0.005, 0.001, 0.01, 0.1\}$, the activation function used in the hidden layers (i.e.~ReLU and $tanh$), the early-stopping $patience$ coefficient $p\in \{100, 200, 500\}$ and the learning rate value $\alpha \in \{0.00001, 0.0001, 0.001\}$. We set the minibatch size $b = 200$, and also tested two loss functions: $RMSE$ and $MSE$.

% \end{appendices}

\section{Data availability statement}\label{sec:data_availability_statement}

In order to ensure the reproducibility of our experimental workflow, all resources, materials and source code used in this paper are either properly cited or we have made them publicly available.
The data, trained models and source code underlying this article are available in a public repository
%\footnote{\url{http://bit.ly/ME_BiB_2021}}.
\footnote{\url{https://csunseduar-my.sharepoint.com/:f:/g/personal/virginia_sabando_cs_uns_edu_ar/EjUkG4X2A31EgJ0Aj0EjveYBMcooO8mKIpQoHquoQtdUhw}}.

\section{Competing interests}\label{competing_interests}
The authors declare no competing interests.

% \section{Author contributions statement}
% Must include all authors, identified by initials, for example:
% S.R. and D.A. conceived the experiment(s),  S.R. conducted the experiment(s), S.R. and D.A. analysed the results.  S.R. and D.A. wrote and reviewed the manuscript.

\section{Funding}\label{sec:funding}
This work was supported by National Scientific and Technical Research Council (CONICET) [grant PIP 112-2017-0100829] (Argentina); by National Agency of Scientific and Technological Promotion (ANPCyT) [grant PICT-2019-03350] (Argentina); by Universidad Nacional del Sur [grant PGI 24/N042] (Argentina); by a Natural Sciences and Engineering Research Council (NSERC) Discovery grant (Canada); and by a Google Latin America Research Award 2020-2021.
This research was also partly supported by DeepSense\footnote{\url{https://www.deepsense.ca/}.}, ACENET\footnote{\url{https://www.ace-net.ca/}.}, Calcul Québec\footnote{\url{https://www.calculquebec.ca/}.} and Compute Canada\footnote{\url{https://www.computecanada.ca/}.}. 

\section{Acknowledgments}\label{sec:acknowledgments}
The authors thank Chris Whidden, Jason Newport and Lu Yang for their technical support with the DeepSense cluster.

\bibliography{bibliography}

\providecommand{\latin}[1]{#1}
\makeatletter
\providecommand{\doi}
  {\begingroup\let\do\@makeother\dospecials
  \catcode`\{=1 \catcode`\}=2 \doi@aux}
\providecommand{\doi@aux}[1]{\endgroup\texttt{#1}}
\makeatother
\providecommand*\mcitethebibliography{\thebibliography}
\csname @ifundefined\endcsname{endmcitethebibliography}
  {\let\endmcitethebibliography\endthebibliography}{}
\begin{mcitethebibliography}{72}
\providecommand*\natexlab[1]{#1}
\providecommand*\mciteSetBstSublistMode[1]{}
\providecommand*\mciteSetBstMaxWidthForm[2]{}
\providecommand*\mciteBstWouldAddEndPuncttrue
  {\def\EndOfBibitem{\unskip.}}
\providecommand*\mciteBstWouldAddEndPunctfalse
  {\let\EndOfBibitem\relax}
\providecommand*\mciteSetBstMidEndSepPunct[3]{}
\providecommand*\mciteSetBstSublistLabelBeginEnd[3]{}
\providecommand*\EndOfBibitem{}
\mciteSetBstSublistMode{f}
\mciteSetBstMaxWidthForm{subitem}{(\alph{mcitesubitemcount})}
\mciteSetBstSublistLabelBeginEnd
  {\mcitemaxwidthsubitemform\space}
  {\relax}
  {\relax}

\bibitem[Wu \latin{et~al.}(2020)Wu, Zhu, Kang, Leung, Lei, Shen, Jiang, Wang,
  Cao, and Hou]{wu2020we}
Wu,~Z.; Zhu,~M.; Kang,~Y.; Leung,~E. L.-H.; Lei,~T.; Shen,~C.; Jiang,~D.;
  Wang,~Z.; Cao,~D.; Hou,~T. Do we need different machine learning algorithms
  for QSAR modeling? A comprehensive assessment of 16 machine learning
  algorithms on 14 QSAR data sets. \emph{Briefings in Bioinformatics}
  \textbf{2020}, \relax
\mciteBstWouldAddEndPunctfalse
\mciteSetBstMidEndSepPunct{\mcitedefaultmidpunct}
{}{\mcitedefaultseppunct}\relax
\EndOfBibitem
\bibitem[Wu \latin{et~al.}(2021)Wu, Jiang, Hsieh, Chen, Liao, Cao, and
  Hou]{wu2021hyperbolic}
Wu,~Z.; Jiang,~D.; Hsieh,~C.-Y.; Chen,~G.; Liao,~B.; Cao,~D.; Hou,~T.
  Hyperbolic relational graph convolution networks plus: a simple but highly
  efficient QSAR-modeling method. \emph{Briefings in Bioinformatics}
  \textbf{2021}, \relax
\mciteBstWouldAddEndPunctfalse
\mciteSetBstMidEndSepPunct{\mcitedefaultmidpunct}
{}{\mcitedefaultseppunct}\relax
\EndOfBibitem
\bibitem[Cherkasov \latin{et~al.}(2014)Cherkasov, Muratov, Fourches, Varnek,
  Baskin, Cronin, Dearden, Gramatica, Martin, Todeschini, \latin{et~al.}
  others]{cherkasov2014qsar}
Cherkasov,~A.; Muratov,~E.~N.; Fourches,~D.; Varnek,~A.; Baskin,~I.~I.;
  Cronin,~M.; Dearden,~J.; Gramatica,~P.; Martin,~Y.~C.; Todeschini,~R.,
  \latin{et~al.}  QSAR modeling: where have you been? Where are you going to?
  \emph{Journal of Medicinal Chemistry} \textbf{2014}, \emph{57},
  4977--5010\relax
\mciteBstWouldAddEndPuncttrue
\mciteSetBstMidEndSepPunct{\mcitedefaultmidpunct}
{\mcitedefaultendpunct}{\mcitedefaultseppunct}\relax
\EndOfBibitem
\bibitem[Todeschini and Consonni(2009)Todeschini, and
  Consonni]{todeschini2009molecular}
Todeschini,~R.; Consonni,~V. \emph{Molecular descriptors for chemoinformatics:
  volume I: alphabetical listing/volume II: appendices, references}; John Wiley
  \& Sons, 2009; Vol.~41\relax
\mciteBstWouldAddEndPuncttrue
\mciteSetBstMidEndSepPunct{\mcitedefaultmidpunct}
{\mcitedefaultendpunct}{\mcitedefaultseppunct}\relax
\EndOfBibitem
\bibitem[Chuang \latin{et~al.}(2020)Chuang, Gunsalus, and
  Keiser]{chuang2020learning}
Chuang,~K.~V.; Gunsalus,~L.~M.; Keiser,~M.~J. Learning Molecular
  Representations for Medicinal Chemistry: Miniperspective. \emph{Journal of
  Medicinal Chemistry} \textbf{2020}, \emph{63}, 8705--8722\relax
\mciteBstWouldAddEndPuncttrue
\mciteSetBstMidEndSepPunct{\mcitedefaultmidpunct}
{\mcitedefaultendpunct}{\mcitedefaultseppunct}\relax
\EndOfBibitem
\bibitem[Elton \latin{et~al.}(2019)Elton, Boukouvalas, Fuge, and
  Chung]{elton2019deep}
Elton,~D.~C.; Boukouvalas,~Z.; Fuge,~M.~D.; Chung,~P.~W. Deep learning for
  molecular design—a review of the state of the art. \emph{Molecular Systems
  Design \& Engineering} \textbf{2019}, \emph{4}, 828--849\relax
\mciteBstWouldAddEndPuncttrue
\mciteSetBstMidEndSepPunct{\mcitedefaultmidpunct}
{\mcitedefaultendpunct}{\mcitedefaultseppunct}\relax
\EndOfBibitem
\bibitem[Chen \latin{et~al.}(2018)Chen, Engkvist, Wang, Olivecrona, and
  Blaschke]{chen2018rise}
Chen,~H.; Engkvist,~O.; Wang,~Y.; Olivecrona,~M.; Blaschke,~T. The rise of deep
  learning in drug discovery. \emph{Drug Discovery Today} \textbf{2018},
  \emph{23}, 1241--1250\relax
\mciteBstWouldAddEndPuncttrue
\mciteSetBstMidEndSepPunct{\mcitedefaultmidpunct}
{\mcitedefaultendpunct}{\mcitedefaultseppunct}\relax
\EndOfBibitem
\bibitem[Bouhedjar \latin{et~al.}(2020)Bouhedjar, Boukelia,
  Khorief~Nacereddine, Boucheham, Belaidi, and Djerourou]{bouhedjar2020natural}
Bouhedjar,~K.; Boukelia,~A.; Khorief~Nacereddine,~A.; Boucheham,~A.;
  Belaidi,~A.; Djerourou,~A. A natural language processing approach based on
  embedding deep learning from heterogeneous compounds for quantitative
  structure--activity relationship modeling. \emph{Chemical Biology \& Drug
  Design} \textbf{2020}, \emph{96}, 961--972\relax
\mciteBstWouldAddEndPuncttrue
\mciteSetBstMidEndSepPunct{\mcitedefaultmidpunct}
{\mcitedefaultendpunct}{\mcitedefaultseppunct}\relax
\EndOfBibitem
\bibitem[David \latin{et~al.}(2020)David, Thakkar, Mercado, and
  Engkvist]{david2020molecular}
David,~L.; Thakkar,~A.; Mercado,~R.; Engkvist,~O. Molecular representations in
  AI-driven drug discovery: a review and practical guide. \emph{Journal of
  Cheminformatics} \textbf{2020}, \emph{12}, 1--22\relax
\mciteBstWouldAddEndPuncttrue
\mciteSetBstMidEndSepPunct{\mcitedefaultmidpunct}
{\mcitedefaultendpunct}{\mcitedefaultseppunct}\relax
\EndOfBibitem
\bibitem[Weininger(1988)]{weininger1988smiles}
Weininger,~D. SMILES, a chemical language and information system. 1.
  Introduction to methodology and encoding rules. \emph{Journal of Chemical
  Information and Computer Sciences} \textbf{1988}, \emph{28}, 31--36\relax
\mciteBstWouldAddEndPuncttrue
\mciteSetBstMidEndSepPunct{\mcitedefaultmidpunct}
{\mcitedefaultendpunct}{\mcitedefaultseppunct}\relax
\EndOfBibitem
\bibitem[Wu \latin{et~al.}(2020)Wu, Pan, Chen, Long, Zhang, and
  Philip]{wu2020comprehensive}
Wu,~Z.; Pan,~S.; Chen,~F.; Long,~G.; Zhang,~C.; Philip,~S.~Y. A comprehensive
  survey on graph neural networks. \emph{IEEE Transactions on Neural Networks
  and Learning Systems} \textbf{2020}, \emph{32}, 950--957\relax
\mciteBstWouldAddEndPuncttrue
\mciteSetBstMidEndSepPunct{\mcitedefaultmidpunct}
{\mcitedefaultendpunct}{\mcitedefaultseppunct}\relax
\EndOfBibitem
\bibitem[{Vaswani} \latin{et~al.}(2017){Vaswani}, {Shazeer}, {Parmar},
  {Uszkoreit}, {Jones}, {Gomez}, {Kaiser}, and
  {Polosukhin}]{vaswani2017attention}
{Vaswani},~A.; {Shazeer},~N.; {Parmar},~N.; {Uszkoreit},~J.; {Jones},~L.;
  {Gomez},~A.~N.; {Kaiser},~L.; {Polosukhin},~I. Attention Is All You Need.
  \emph{arXiv e-prints} \textbf{2017}, arXiv:1706.03762\relax
\mciteBstWouldAddEndPuncttrue
\mciteSetBstMidEndSepPunct{\mcitedefaultmidpunct}
{\mcitedefaultendpunct}{\mcitedefaultseppunct}\relax
\EndOfBibitem
\bibitem[{Oskooei} \latin{et~al.}(2018){Oskooei}, {Born}, {Manica},
  {Subramanian}, {S{\'a}ez-Rodr{\'\i}guez}, and {Rodr{\'\i}guez
  Mart{\'\i}nez}]{oskooei2018paccmann}
{Oskooei},~A.; {Born},~J.; {Manica},~M.; {Subramanian},~V.;
  {S{\'a}ez-Rodr{\'\i}guez},~J.; {Rodr{\'\i}guez Mart{\'\i}nez},~M. {PaccMann:
  Prediction of anticancer compound sensitivity with multi-modal
  attention-based neural networks}. \emph{arXiv e-prints} \textbf{2018},
  arXiv:1811.06802\relax
\mciteBstWouldAddEndPuncttrue
\mciteSetBstMidEndSepPunct{\mcitedefaultmidpunct}
{\mcitedefaultendpunct}{\mcitedefaultseppunct}\relax
\EndOfBibitem
\bibitem[Zheng \latin{et~al.}(2019)Zheng, Yan, Yang, and
  Xu]{zheng2019identifying}
Zheng,~S.; Yan,~X.; Yang,~Y.; Xu,~J. Identifying structure--property
  relationships through SMILES syntax analysis with self-attention mechanism.
  \emph{Journal of Chemical Information and Modeling} \textbf{2019}, \emph{59},
  914--923\relax
\mciteBstWouldAddEndPuncttrue
\mciteSetBstMidEndSepPunct{\mcitedefaultmidpunct}
{\mcitedefaultendpunct}{\mcitedefaultseppunct}\relax
\EndOfBibitem
\bibitem[Wu \latin{et~al.}(2018)Wu, Ramsundar, Feinberg, Gomes, Geniesse,
  Pappu, Leswing, and Pande]{wu2018moleculenet}
Wu,~Z.; Ramsundar,~B.; Feinberg,~E.~N.; Gomes,~J.; Geniesse,~C.; Pappu,~A.~S.;
  Leswing,~K.; Pande,~V. MoleculeNet: a benchmark for molecular machine
  learning. \emph{Chemical Science} \textbf{2018}, \emph{9}, 513--530\relax
\mciteBstWouldAddEndPuncttrue
\mciteSetBstMidEndSepPunct{\mcitedefaultmidpunct}
{\mcitedefaultendpunct}{\mcitedefaultseppunct}\relax
\EndOfBibitem
\bibitem[Jiang \latin{et~al.}(2021)Jiang, Wu, Hsieh, Chen, Liao, Wang, Shen,
  Cao, Wu, and Hou]{jiang2021could}
Jiang,~D.; Wu,~Z.; Hsieh,~C.-Y.; Chen,~G.; Liao,~B.; Wang,~Z.; Shen,~C.;
  Cao,~D.; Wu,~J.; Hou,~T. Could graph neural networks learn better molecular
  representation for drug discovery? A comparison study of descriptor-based and
  graph-based models. \emph{Journal of cheminformatics} \textbf{2021},
  \emph{13}, 1--23\relax
\mciteBstWouldAddEndPuncttrue
\mciteSetBstMidEndSepPunct{\mcitedefaultmidpunct}
{\mcitedefaultendpunct}{\mcitedefaultseppunct}\relax
\EndOfBibitem
\bibitem[Jaeger \latin{et~al.}(2018)Jaeger, Fulle, and Turk]{jaeger2018mol2vec}
Jaeger,~S.; Fulle,~S.; Turk,~S. Mol2vec: unsupervised machine learning approach
  with chemical intuition. \emph{Journal of Chemical Information and Modeling}
  \textbf{2018}, \emph{58}, 27--35\relax
\mciteBstWouldAddEndPuncttrue
\mciteSetBstMidEndSepPunct{\mcitedefaultmidpunct}
{\mcitedefaultendpunct}{\mcitedefaultseppunct}\relax
\EndOfBibitem
\bibitem[G{\'o}mez-Bombarelli \latin{et~al.}(2018)G{\'o}mez-Bombarelli, Wei,
  Duvenaud, Hern{\'a}ndez-Lobato, S{\'a}nchez-Lengeling, Sheberla,
  Aguilera-Iparraguirre, Hirzel, Adams, and Aspuru-Guzik]{gomez2018automatic}
G{\'o}mez-Bombarelli,~R.; Wei,~J.~N.; Duvenaud,~D.;
  Hern{\'a}ndez-Lobato,~J.~M.; S{\'a}nchez-Lengeling,~B.; Sheberla,~D.;
  Aguilera-Iparraguirre,~J.; Hirzel,~T.~D.; Adams,~R.~P.; Aspuru-Guzik,~A.
  Automatic chemical design using a data-driven continuous representation of
  molecules. \emph{ACS Central Science} \textbf{2018}, \emph{4}, 268--276\relax
\mciteBstWouldAddEndPuncttrue
\mciteSetBstMidEndSepPunct{\mcitedefaultmidpunct}
{\mcitedefaultendpunct}{\mcitedefaultseppunct}\relax
\EndOfBibitem
\bibitem[{Goh} \latin{et~al.}(2017){Goh}, {Siegel}, {Vishnu}, {Hodas}, and
  {Baker}]{goh2017chemception}
{Goh},~G.~B.; {Siegel},~C.; {Vishnu},~A.; {Hodas},~N.~O.; {Baker},~N.
  {Chemception: A Deep Neural Network with Minimal Chemistry Knowledge Matches
  the Performance of Expert-developed QSAR/QSPR Models}. \emph{arXiv e-prints}
  \textbf{2017}, arXiv:1706.06689\relax
\mciteBstWouldAddEndPuncttrue
\mciteSetBstMidEndSepPunct{\mcitedefaultmidpunct}
{\mcitedefaultendpunct}{\mcitedefaultseppunct}\relax
\EndOfBibitem
\bibitem[Yang \latin{et~al.}(2019)Yang, Swanson, Jin, Coley, Eiden, Gao,
  Guzman-Perez, Hopper, Kelley, Mathea, \latin{et~al.}
  others]{yang2019analyzing}
Yang,~K.; Swanson,~K.; Jin,~W.; Coley,~C.; Eiden,~P.; Gao,~H.;
  Guzman-Perez,~A.; Hopper,~T.; Kelley,~B.; Mathea,~M., \latin{et~al.}
  Analyzing learned molecular representations for property prediction.
  \emph{Journal of Chemical Information and Modeling} \textbf{2019}, \emph{59},
  3370--3388\relax
\mciteBstWouldAddEndPuncttrue
\mciteSetBstMidEndSepPunct{\mcitedefaultmidpunct}
{\mcitedefaultendpunct}{\mcitedefaultseppunct}\relax
\EndOfBibitem
\bibitem[Yang \latin{et~al.}(2019)Yang, Swanson, Jin, Coley, Eiden, Gao,
  Guzman-Perez, Hopper, Kelley, Mathea, Palmer, Settels, Jaakkola, Jensen, and
  Barzilay]{erratumyang2019analyzing}
Yang,~K.; Swanson,~K.; Jin,~W.; Coley,~C.; Eiden,~P.; Gao,~H.;
  Guzman-Perez,~A.; Hopper,~T.; Kelley,~B.; Mathea,~M.; Palmer,~A.;
  Settels,~V.; Jaakkola,~T.; Jensen,~K.; Barzilay,~R. Correction to Analyzing
  Learned Molecular Representations for Property Prediction. \emph{Journal of
  Chemical Information and Modeling} \textbf{2019}, \emph{59}, 5304--5305,
  PMID: 31814400\relax
\mciteBstWouldAddEndPuncttrue
\mciteSetBstMidEndSepPunct{\mcitedefaultmidpunct}
{\mcitedefaultendpunct}{\mcitedefaultseppunct}\relax
\EndOfBibitem
\bibitem[{Chithrananda} \latin{et~al.}(2020){Chithrananda}, {Grand}, and
  {Ramsundar}]{chithrananda2020chemberta}
{Chithrananda},~S.; {Grand},~G.; {Ramsundar},~B. {ChemBERTa: Large-Scale
  Self-Supervised Pretraining for Molecular Property Prediction}. \emph{arXiv
  e-prints} \textbf{2020}, arXiv:2010.09885\relax
\mciteBstWouldAddEndPuncttrue
\mciteSetBstMidEndSepPunct{\mcitedefaultmidpunct}
{\mcitedefaultendpunct}{\mcitedefaultseppunct}\relax
\EndOfBibitem
\bibitem[Sterling and Irwin(2015)Sterling, and Irwin]{sterling2015zinc}
Sterling,~T.; Irwin,~J.~J. {ZINC} 15--ligand discovery for everyone.
  \emph{Journal of Chemical Information and Modeling} \textbf{2015}, \emph{55},
  2324--2337\relax
\mciteBstWouldAddEndPuncttrue
\mciteSetBstMidEndSepPunct{\mcitedefaultmidpunct}
{\mcitedefaultendpunct}{\mcitedefaultseppunct}\relax
\EndOfBibitem
\bibitem[{Sabando} \latin{et~al.}(2021){Sabando}, {Ulbrich}, {Selzer},
  {Byška}, {Mičan}, {Ponzoni}, {Soto}, {Ganuza}, and
  {Kozlíková}]{sabando2021chemva}
{Sabando},~M.~V.; {Ulbrich},~P.; {Selzer},~M.; {Byška},~J.; {Mičan},~J.;
  {Ponzoni},~I.; {Soto},~A.~J.; {Ganuza},~M.~L.; {Kozlíková},~B. ChemVA:
  Interactive Visual Analysis of Chemical Compound Similarity in Virtual
  Screening. \emph{IEEE Transactions on Visualization and Computer Graphics}
  \textbf{2021}, \emph{27}, 891--901\relax
\mciteBstWouldAddEndPuncttrue
\mciteSetBstMidEndSepPunct{\mcitedefaultmidpunct}
{\mcitedefaultendpunct}{\mcitedefaultseppunct}\relax
\EndOfBibitem
\bibitem[Cereto-Massagu{\'e} \latin{et~al.}(2015)Cereto-Massagu{\'e}, Ojeda,
  Valls, Mulero, Garcia-Vallv{\'e}, and Pujadas]{cereto2015molecular}
Cereto-Massagu{\'e},~A.; Ojeda,~M.~J.; Valls,~C.; Mulero,~M.;
  Garcia-Vallv{\'e},~S.; Pujadas,~G. Molecular fingerprint similarity search in
  virtual screening. \emph{Methods} \textbf{2015}, \emph{71}, 58--63\relax
\mciteBstWouldAddEndPuncttrue
\mciteSetBstMidEndSepPunct{\mcitedefaultmidpunct}
{\mcitedefaultendpunct}{\mcitedefaultseppunct}\relax
\EndOfBibitem
\bibitem[Grisoni \latin{et~al.}(2018)Grisoni, Consonni, and
  Todeschini]{grisoni2018impact}
Grisoni,~F.; Consonni,~V.; Todeschini,~R. \emph{Computational Chemogenomics};
  Springer New York: New York, NY, 2018; pp 171--209\relax
\mciteBstWouldAddEndPuncttrue
\mciteSetBstMidEndSepPunct{\mcitedefaultmidpunct}
{\mcitedefaultendpunct}{\mcitedefaultseppunct}\relax
\EndOfBibitem
\bibitem[Schneider(2010)]{schneider2010virtual}
Schneider,~G. Virtual screening: an endless staircase? \emph{Nature Reviews
  Drug Discovery} \textbf{2010}, \emph{9}, 273--276\relax
\mciteBstWouldAddEndPuncttrue
\mciteSetBstMidEndSepPunct{\mcitedefaultmidpunct}
{\mcitedefaultendpunct}{\mcitedefaultseppunct}\relax
\EndOfBibitem
\bibitem[Rogers and Hahn(2010)Rogers, and Hahn]{rogers2010extended}
Rogers,~D.; Hahn,~M. Extended-connectivity fingerprints. \emph{Journal of
  Chemical Information and Modeling} \textbf{2010}, \emph{50}, 742--754\relax
\mciteBstWouldAddEndPuncttrue
\mciteSetBstMidEndSepPunct{\mcitedefaultmidpunct}
{\mcitedefaultendpunct}{\mcitedefaultseppunct}\relax
\EndOfBibitem
\bibitem[Durant \latin{et~al.}(2002)Durant, Leland, Henry, and
  Nourse]{durant2002reoptimization}
Durant,~J.~L.; Leland,~B.~A.; Henry,~D.~R.; Nourse,~J.~G. Reoptimization of MDL
  keys for use in drug discovery. \emph{Journal of Chemical Information and
  Computer Sciences} \textbf{2002}, \emph{42}, 1273--1280\relax
\mciteBstWouldAddEndPuncttrue
\mciteSetBstMidEndSepPunct{\mcitedefaultmidpunct}
{\mcitedefaultendpunct}{\mcitedefaultseppunct}\relax
\EndOfBibitem
\bibitem[Seth and Roy(2020)Seth, and Roy]{seth2020qsar}
Seth,~A.; Roy,~K. QSAR modeling of algal low level toxicity values of different
  phenol and aniline derivatives using 2D descriptors. \emph{Aquatic
  Toxicology} \textbf{2020}, \emph{228}, 105627\relax
\mciteBstWouldAddEndPuncttrue
\mciteSetBstMidEndSepPunct{\mcitedefaultmidpunct}
{\mcitedefaultendpunct}{\mcitedefaultseppunct}\relax
\EndOfBibitem
\bibitem[Yang \latin{et~al.}(2020)Yang, Wang, Chang, Pan, Wei, Li, and
  Wang]{yang2020qsar}
Yang,~L.; Wang,~Y.; Chang,~J.; Pan,~Y.; Wei,~R.; Li,~J.; Wang,~H. QSAR modeling
  the toxicity of pesticides against Americamysis bahia. \emph{Chemosphere}
  \textbf{2020}, \emph{258}, 127217\relax
\mciteBstWouldAddEndPuncttrue
\mciteSetBstMidEndSepPunct{\mcitedefaultmidpunct}
{\mcitedefaultendpunct}{\mcitedefaultseppunct}\relax
\EndOfBibitem
\bibitem[Gao \latin{et~al.}(2020)Gao, Nguyen, Sresht, Mathiowetz, Tu, and
  Wei]{gao20202d}
Gao,~K.; Nguyen,~D.~D.; Sresht,~V.; Mathiowetz,~A.~M.; Tu,~M.; Wei,~G.-W. Are
  2D fingerprints still valuable for drug discovery? \emph{Physical Chemistry
  Chemical Physics} \textbf{2020}, \emph{22}, 8373--8390\relax
\mciteBstWouldAddEndPuncttrue
\mciteSetBstMidEndSepPunct{\mcitedefaultmidpunct}
{\mcitedefaultendpunct}{\mcitedefaultseppunct}\relax
\EndOfBibitem
\bibitem[Sabando \latin{et~al.}(2019)Sabando, Ponzoni, and
  Soto]{sabando2019neural}
Sabando,~M.~V.; Ponzoni,~I.; Soto,~A.~J. Neural-based approaches to overcome
  feature selection and applicability domain in drug-related property
  prediction. \emph{Applied Soft Computing} \textbf{2019}, \emph{85},
  105777\relax
\mciteBstWouldAddEndPuncttrue
\mciteSetBstMidEndSepPunct{\mcitedefaultmidpunct}
{\mcitedefaultendpunct}{\mcitedefaultseppunct}\relax
\EndOfBibitem
\bibitem[{Liu} \latin{et~al.}(2018){Liu}, {Furkan Demirel}, and
  {Liang}]{liu2018n}
{Liu},~S.; {Furkan Demirel},~M.; {Liang},~Y. {N-Gram Graph: Simple Unsupervised
  Representation for Graphs, with Applications to Molecules}. \emph{arXiv
  e-prints} \textbf{2018}, arXiv:1806.09206\relax
\mciteBstWouldAddEndPuncttrue
\mciteSetBstMidEndSepPunct{\mcitedefaultmidpunct}
{\mcitedefaultendpunct}{\mcitedefaultseppunct}\relax
\EndOfBibitem
\bibitem[Swann \latin{et~al.}(2018)Swann, Sun, Cleland, and
  Barnard]{swann2018representing}
Swann,~E.; Sun,~B.; Cleland,~D.; Barnard,~A. Representing molecular and
  materials data for unsupervised machine learning. \emph{Molecular Simulation}
  \textbf{2018}, \emph{44}, 905--920\relax
\mciteBstWouldAddEndPuncttrue
\mciteSetBstMidEndSepPunct{\mcitedefaultmidpunct}
{\mcitedefaultendpunct}{\mcitedefaultseppunct}\relax
\EndOfBibitem
\bibitem[{\"O}zt{\"u}rk \latin{et~al.}(2018){\"O}zt{\"u}rk, Ozkirimli, and
  {\"O}zg{\"u}r]{ozturk2018novel}
{\"O}zt{\"u}rk,~H.; Ozkirimli,~E.; {\"O}zg{\"u}r,~A. A novel methodology on
  distributed representations of proteins using their interacting ligands.
  \emph{Bioinformatics} \textbf{2018}, \emph{34}, i295--i303\relax
\mciteBstWouldAddEndPuncttrue
\mciteSetBstMidEndSepPunct{\mcitedefaultmidpunct}
{\mcitedefaultendpunct}{\mcitedefaultseppunct}\relax
\EndOfBibitem
\bibitem[Xu \latin{et~al.}(2017)Xu, Wang, Zhu, and Huang]{xu2017seq2seq}
Xu,~Z.; Wang,~S.; Zhu,~F.; Huang,~J. Seq2seq fingerprint: An unsupervised deep
  molecular embedding for drug discovery. Proceedings of the 8th ACM
  International Conference on Bioinformatics, Computational Biology, and Health
  Informatics. 2017; pp 285--294\relax
\mciteBstWouldAddEndPuncttrue
\mciteSetBstMidEndSepPunct{\mcitedefaultmidpunct}
{\mcitedefaultendpunct}{\mcitedefaultseppunct}\relax
\EndOfBibitem
\bibitem[Kuzminykh \latin{et~al.}(2018)Kuzminykh, Polykovskiy, Kadurin,
  Zhebrak, Baskov, Nikolenko, Shayakhmetov, and Zhavoronkov]{kuzminykh20183d}
Kuzminykh,~D.; Polykovskiy,~D.; Kadurin,~A.; Zhebrak,~A.; Baskov,~I.;
  Nikolenko,~S.; Shayakhmetov,~R.; Zhavoronkov,~A. 3d molecular representations
  based on the wave transform for convolutional neural networks.
  \emph{Molecular Pharmaceutics} \textbf{2018}, \emph{15}, 4378--4385\relax
\mciteBstWouldAddEndPuncttrue
\mciteSetBstMidEndSepPunct{\mcitedefaultmidpunct}
{\mcitedefaultendpunct}{\mcitedefaultseppunct}\relax
\EndOfBibitem
\bibitem[Shi \latin{et~al.}(2019)Shi, Yang, Huang, Chen, Kuang, Heng, and
  Mei]{shi2019molecular}
Shi,~T.; Yang,~Y.; Huang,~S.; Chen,~L.; Kuang,~Z.; Heng,~Y.; Mei,~H. Molecular
  image-based convolutional neural network for the prediction of ADMET
  properties. \emph{Chemometrics and Intelligent Laboratory Systems}
  \textbf{2019}, \emph{194}, 103853\relax
\mciteBstWouldAddEndPuncttrue
\mciteSetBstMidEndSepPunct{\mcitedefaultmidpunct}
{\mcitedefaultendpunct}{\mcitedefaultseppunct}\relax
\EndOfBibitem
\bibitem[{{\"O}z{\c{c}}elik} \latin{et~al.}(2018){{\"O}z{\c{c}}elik},
  {{\"O}zt{\"u}rk}, {{\"O}zg{\"u}r}, and {Ozkirimli}]{ozturk2018chemical}
{{\"O}z{\c{c}}elik},~R.; {{\"O}zt{\"u}rk},~H.; {{\"O}zg{\"u}r},~A.;
  {Ozkirimli},~E. {ChemBoost: A chemical language based approach for
  protein-ligand binding affinity prediction}. \emph{arXiv e-prints}
  \textbf{2018}, arXiv:1811.00761\relax
\mciteBstWouldAddEndPuncttrue
\mciteSetBstMidEndSepPunct{\mcitedefaultmidpunct}
{\mcitedefaultendpunct}{\mcitedefaultseppunct}\relax
\EndOfBibitem
\bibitem[{Mikolov} \latin{et~al.}(2013){Mikolov}, {Chen}, {Corrado}, and
  {Dean}]{mikolov2013efficient}
{Mikolov},~T.; {Chen},~K.; {Corrado},~G.; {Dean},~J. {Efficient Estimation of
  Word Representations in Vector Space}. \emph{arXiv e-prints} \textbf{2013},
  arXiv:1301.3781\relax
\mciteBstWouldAddEndPuncttrue
\mciteSetBstMidEndSepPunct{\mcitedefaultmidpunct}
{\mcitedefaultendpunct}{\mcitedefaultseppunct}\relax
\EndOfBibitem
\bibitem[Segler \latin{et~al.}(2018)Segler, Kogej, Tyrchan, and
  Waller]{segler2018generating}
Segler,~M.~H.; Kogej,~T.; Tyrchan,~C.; Waller,~M.~P. Generating focused
  molecule libraries for drug discovery with recurrent neural networks.
  \emph{ACS Central Science} \textbf{2018}, \emph{4}, 120--131\relax
\mciteBstWouldAddEndPuncttrue
\mciteSetBstMidEndSepPunct{\mcitedefaultmidpunct}
{\mcitedefaultendpunct}{\mcitedefaultseppunct}\relax
\EndOfBibitem
\bibitem[Popova \latin{et~al.}(2018)Popova, Isayev, and
  Tropsha]{popova2018deep}
Popova,~M.; Isayev,~O.; Tropsha,~A. Deep reinforcement learning for de novo
  drug design. \emph{Science Advances} \textbf{2018}, \emph{4}, eaap7885\relax
\mciteBstWouldAddEndPuncttrue
\mciteSetBstMidEndSepPunct{\mcitedefaultmidpunct}
{\mcitedefaultendpunct}{\mcitedefaultseppunct}\relax
\EndOfBibitem
\bibitem[{Cho} \latin{et~al.}(2014){Cho}, {van Merrienboer}, {Gulcehre},
  {Bahdanau}, {Bougares}, {Schwenk}, and {Bengio}]{Cho2014}
{Cho},~K.; {van Merrienboer},~B.; {Gulcehre},~C.; {Bahdanau},~D.;
  {Bougares},~F.; {Schwenk},~H.; {Bengio},~Y. {Learning Phrase Representations
  using RNN Encoder-Decoder for Statistical Machine Translation}. \emph{arXiv
  e-prints} \textbf{2014}, arXiv:1406.1078\relax
\mciteBstWouldAddEndPuncttrue
\mciteSetBstMidEndSepPunct{\mcitedefaultmidpunct}
{\mcitedefaultendpunct}{\mcitedefaultseppunct}\relax
\EndOfBibitem
\bibitem[Joshi(2020)]{joshi2020transformers}
Joshi,~C. Transformers are Graph Neural Networks.
  \url{https://thegradient.pub/transformers-are-gaph-neural-networks/ },
  2020\relax
\mciteBstWouldAddEndPuncttrue
\mciteSetBstMidEndSepPunct{\mcitedefaultmidpunct}
{\mcitedefaultendpunct}{\mcitedefaultseppunct}\relax
\EndOfBibitem
\bibitem[Hochreiter and Schmidhuber(1997)Hochreiter, and
  Schmidhuber]{hochreiter1997long}
Hochreiter,~S.; Schmidhuber,~J. Long short-term memory. \emph{Neural
  computation} \textbf{1997}, \emph{9}, 1735--1780\relax
\mciteBstWouldAddEndPuncttrue
\mciteSetBstMidEndSepPunct{\mcitedefaultmidpunct}
{\mcitedefaultendpunct}{\mcitedefaultseppunct}\relax
\EndOfBibitem
\bibitem[Wang \latin{et~al.}(2019)Wang, Guo, Wang, Sun, and
  Huang]{wang2019smiles}
Wang,~S.; Guo,~Y.; Wang,~Y.; Sun,~H.; Huang,~J. SMILES-BERT: large scale
  unsupervised pre-training for molecular property prediction. Proceedings of
  the 10th ACM International Conference on Bioinformatics, Computational
  Biology and Health Informatics. 2019; pp 429--436\relax
\mciteBstWouldAddEndPuncttrue
\mciteSetBstMidEndSepPunct{\mcitedefaultmidpunct}
{\mcitedefaultendpunct}{\mcitedefaultseppunct}\relax
\EndOfBibitem
\bibitem[{Devlin} \latin{et~al.}(2018){Devlin}, {Chang}, {Lee}, and
  {Toutanova}]{devlin2018bert}
{Devlin},~J.; {Chang},~M.-W.; {Lee},~K.; {Toutanova},~K. {BERT: Pre-training of
  Deep Bidirectional Transformers for Language Understanding}. \emph{arXiv
  e-prints} \textbf{2018}, arXiv:1810.04805\relax
\mciteBstWouldAddEndPuncttrue
\mciteSetBstMidEndSepPunct{\mcitedefaultmidpunct}
{\mcitedefaultendpunct}{\mcitedefaultseppunct}\relax
\EndOfBibitem
\bibitem[Lipinski(2004)]{lipinski2004lead}
Lipinski,~C.~A. Lead- and drug-like compounds: the rule-of-five revolution.
  \emph{Drug Discovery Today: Technologies} \textbf{2004}, \emph{1},
  337--341\relax
\mciteBstWouldAddEndPuncttrue
\mciteSetBstMidEndSepPunct{\mcitedefaultmidpunct}
{\mcitedefaultendpunct}{\mcitedefaultseppunct}\relax
\EndOfBibitem
\bibitem[Landrum(2016)]{landrum2016rdkit}
Landrum,~G. RDKit: open-source cheminformatics http://www. rdkit. org.
  2016\relax
\mciteBstWouldAddEndPuncttrue
\mciteSetBstMidEndSepPunct{\mcitedefaultmidpunct}
{\mcitedefaultendpunct}{\mcitedefaultseppunct}\relax
\EndOfBibitem
\bibitem[Delaney(2004)]{delaney2004esol}
Delaney,~J.~S. ESOL: estimating aqueous solubility directly from molecular
  structure. \emph{Journal of chemical information and computer sciences}
  \textbf{2004}, \emph{44}, 1000--1005\relax
\mciteBstWouldAddEndPuncttrue
\mciteSetBstMidEndSepPunct{\mcitedefaultmidpunct}
{\mcitedefaultendpunct}{\mcitedefaultseppunct}\relax
\EndOfBibitem
\bibitem[Mobley and Guthrie(2014)Mobley, and Guthrie]{mobley2014freesolv}
Mobley,~D.~L.; Guthrie,~J.~P. FreeSolv: a database of experimental and
  calculated hydration free energies, with input files. \emph{Journal of
  computer-aided molecular design} \textbf{2014}, \emph{28}, 711--720\relax
\mciteBstWouldAddEndPuncttrue
\mciteSetBstMidEndSepPunct{\mcitedefaultmidpunct}
{\mcitedefaultendpunct}{\mcitedefaultseppunct}\relax
\EndOfBibitem
\bibitem[Bento \latin{et~al.}(2014)Bento, Gaulton, Hersey, Bellis, Chambers,
  Davies, Kr{\"u}ger, Light, Mak, McGlinchey, \latin{et~al.}
  others]{bento2014chembl}
Bento,~A.~P.; Gaulton,~A.; Hersey,~A.; Bellis,~L.~J.; Chambers,~J.; Davies,~M.;
  Kr{\"u}ger,~F.~A.; Light,~Y.; Mak,~L.; McGlinchey,~S., \latin{et~al.}  The
  ChEMBL bioactivity database: an update. \emph{Nucleic acids research}
  \textbf{2014}, \emph{42}, D1083--D1090\relax
\mciteBstWouldAddEndPuncttrue
\mciteSetBstMidEndSepPunct{\mcitedefaultmidpunct}
{\mcitedefaultendpunct}{\mcitedefaultseppunct}\relax
\EndOfBibitem
\bibitem[Dalke(2018)]{dalke2018deepsmiles}
Dalke,~A. DeepSMILES: An Adaptation of SMILES for Use in Machine-Learning of
  Chemical Structures. 2018\relax
\mciteBstWouldAddEndPuncttrue
\mciteSetBstMidEndSepPunct{\mcitedefaultmidpunct}
{\mcitedefaultendpunct}{\mcitedefaultseppunct}\relax
\EndOfBibitem
\bibitem[Schwaller \latin{et~al.}(2018)Schwaller, Gaudin, Lanyi, Bekas, and
  Laino]{schwaller2018found}
Schwaller,~P.; Gaudin,~T.; Lanyi,~D.; Bekas,~C.; Laino,~T. “Found in
  Translation”: predicting outcomes of complex organic chemistry reactions
  using neural sequence-to-sequence models. \emph{Chemical Science}
  \textbf{2018}, \emph{9}, 6091--6098\relax
\mciteBstWouldAddEndPuncttrue
\mciteSetBstMidEndSepPunct{\mcitedefaultmidpunct}
{\mcitedefaultendpunct}{\mcitedefaultseppunct}\relax
\EndOfBibitem
\bibitem[Moriwaki \latin{et~al.}(2018)Moriwaki, Tian, Kawashita, and
  Takagi]{moriwaki2018mordred}
Moriwaki,~H.; Tian,~Y.-S.; Kawashita,~N.; Takagi,~T. Mordred: a molecular
  descriptor calculator. \emph{Journal of Cheminformatics} \textbf{2018},
  \emph{10}, 1--14\relax
\mciteBstWouldAddEndPuncttrue
\mciteSetBstMidEndSepPunct{\mcitedefaultmidpunct}
{\mcitedefaultendpunct}{\mcitedefaultseppunct}\relax
\EndOfBibitem
\bibitem[Sch{\"o}lkopf \latin{et~al.}(2004)Sch{\"o}lkopf, Tsuda, and
  Vert]{scholkopf2004kernel}
Sch{\"o}lkopf,~B.; Tsuda,~K.; Vert,~J.-P. \emph{Kernel methods in computational
  biology}; MIT press, 2004\relax
\mciteBstWouldAddEndPuncttrue
\mciteSetBstMidEndSepPunct{\mcitedefaultmidpunct}
{\mcitedefaultendpunct}{\mcitedefaultseppunct}\relax
\EndOfBibitem
\bibitem[Pedregosa \latin{et~al.}(2011)Pedregosa, Varoquaux, Gramfort, Michel,
  Thirion, Grisel, Blondel, Prettenhofer, Weiss, Dubourg, Vanderplas, Passos,
  Cournapeau, Brucher, Perrot, and Duchesnay]{scikit-learn}
Pedregosa,~F. \latin{et~al.}  Scikit-learn: Machine Learning in {P}ython.
  \emph{Journal of Machine Learning Research} \textbf{2011}, \emph{12},
  2825--2830\relax
\mciteBstWouldAddEndPuncttrue
\mciteSetBstMidEndSepPunct{\mcitedefaultmidpunct}
{\mcitedefaultendpunct}{\mcitedefaultseppunct}\relax
\EndOfBibitem
\bibitem[Chollet(2015)]{chollet2015keras}
Chollet,~F. Keras. 2015; {https://github.com/fchollet/keras}, online
  July~2020\relax
\mciteBstWouldAddEndPuncttrue
\mciteSetBstMidEndSepPunct{\mcitedefaultmidpunct}
{\mcitedefaultendpunct}{\mcitedefaultseppunct}\relax
\EndOfBibitem
\bibitem[Abadi \latin{et~al.}(2015)Abadi, Agarwal, Barham, Brevdo, Chen, Citro,
  Corrado, Davis, Dean, Devin, Ghemawat, Goodfellow, Harp, Irving, Isard, Jia,
  Jozefowicz, Kaiser, Kudlur, Levenberg, Man\'{e}, Monga, Moore, Murray, Olah,
  Schuster, Shlens, Steiner, Sutskever, Talwar, Tucker, Vanhoucke, Vasudevan,
  Vi\'{e}gas, Vinyals, Warden, Wattenberg, Wicke, Yu, and
  Zheng]{tensorflow2015-whitepaper}
Abadi,~M. \latin{et~al.}  {TensorFlow}: Large-Scale Machine Learning on
  Heterogeneous Systems. 2015; \url{http://tensorflow.org/}, Software available
  from tensorflow.org\relax
\mciteBstWouldAddEndPuncttrue
\mciteSetBstMidEndSepPunct{\mcitedefaultmidpunct}
{\mcitedefaultendpunct}{\mcitedefaultseppunct}\relax
\EndOfBibitem
\bibitem[Baumann and Baumann(2014)Baumann, and Baumann]{baumann2014reliable}
Baumann,~D.; Baumann,~K. Reliable estimation of prediction errors for QSAR
  models under model uncertainty using double cross-validation. \emph{Journal
  of cheminformatics} \textbf{2014}, \emph{6}, 1--19\relax
\mciteBstWouldAddEndPuncttrue
\mciteSetBstMidEndSepPunct{\mcitedefaultmidpunct}
{\mcitedefaultendpunct}{\mcitedefaultseppunct}\relax
\EndOfBibitem
\bibitem[Sokolova \latin{et~al.}(2006)Sokolova, Japkowicz, and
  Szpakowicz]{sokolova2006beyond}
Sokolova,~M.; Japkowicz,~N.; Szpakowicz,~S. Beyond accuracy, F-score and ROC: a
  family of discriminant measures for performance evaluation. Australasian
  Joint Conference on Artificial Intelligence. 2006; pp 1015--1021\relax
\mciteBstWouldAddEndPuncttrue
\mciteSetBstMidEndSepPunct{\mcitedefaultmidpunct}
{\mcitedefaultendpunct}{\mcitedefaultseppunct}\relax
\EndOfBibitem
\bibitem[Chawla \latin{et~al.}(2002)Chawla, Bowyer, Hall, and
  Kegelmeyer]{chawla2002smote}
Chawla,~N.~V.; Bowyer,~K.~W.; Hall,~L.~O.; Kegelmeyer,~W.~P. SMOTE: synthetic
  minority over-sampling technique. \emph{Journal of Artificial Intelligence
  Research} \textbf{2002}, \emph{16}, 321--357\relax
\mciteBstWouldAddEndPuncttrue
\mciteSetBstMidEndSepPunct{\mcitedefaultmidpunct}
{\mcitedefaultendpunct}{\mcitedefaultseppunct}\relax
\EndOfBibitem
\bibitem[Tukey \latin{et~al.}(1977)Tukey, \latin{et~al.}
  others]{tukey1977exploratory}
Tukey,~J.~W., \latin{et~al.}  \emph{Exploratory data analysis}; Reading, Mass.,
  1977; Vol.~2\relax
\mciteBstWouldAddEndPuncttrue
\mciteSetBstMidEndSepPunct{\mcitedefaultmidpunct}
{\mcitedefaultendpunct}{\mcitedefaultseppunct}\relax
\EndOfBibitem
\bibitem[Goodarzi \latin{et~al.}(2012)Goodarzi, Dejaegher, and {Vander
  Heyden}]{Goodarzi2012}
Goodarzi,~M.; Dejaegher,~B.; {Vander Heyden},~Y. {Feature selection methods in
  QSAR studies.} \emph{Journal of AOAC International} \textbf{2012}, \emph{95},
  636--51\relax
\mciteBstWouldAddEndPuncttrue
\mciteSetBstMidEndSepPunct{\mcitedefaultmidpunct}
{\mcitedefaultendpunct}{\mcitedefaultseppunct}\relax
\EndOfBibitem
\bibitem[Alsenan \latin{et~al.}(2020)Alsenan, Al-Turaiki, and
  Hafez]{alsenan2020autoencoder}
Alsenan,~S.; Al-Turaiki,~I.; Hafez,~A. Autoencoder-based Dimensionality
  Reduction for QSAR Modeling. 2020 3rd International Conference on Computer
  Applications \& Information Security (ICCAIS). 2020; pp 1--4\relax
\mciteBstWouldAddEndPuncttrue
\mciteSetBstMidEndSepPunct{\mcitedefaultmidpunct}
{\mcitedefaultendpunct}{\mcitedefaultseppunct}\relax
\EndOfBibitem
\bibitem[Huang and Von~Lilienfeld(2016)Huang, and
  Von~Lilienfeld]{huang2016communication}
Huang,~B.; Von~Lilienfeld,~O.~A. Communication: Understanding molecular
  representations in machine learning: The role of uniqueness and target
  similarity. 2016\relax
\mciteBstWouldAddEndPuncttrue
\mciteSetBstMidEndSepPunct{\mcitedefaultmidpunct}
{\mcitedefaultendpunct}{\mcitedefaultseppunct}\relax
\EndOfBibitem
\bibitem[Sharma(2017)]{sharma2017activation}
Sharma,~S. Activation functions in neural networks. 2017\relax
\mciteBstWouldAddEndPuncttrue
\mciteSetBstMidEndSepPunct{\mcitedefaultmidpunct}
{\mcitedefaultendpunct}{\mcitedefaultseppunct}\relax
\EndOfBibitem
\bibitem[Domingos(1999)]{domingos1999metacost}
Domingos,~P. Metacost: A general method for making classifiers cost-sensitive.
  Proceedings of the Fifth ACM SIGKDD International Conference on Knowledge
  Discovery and Data Mining. 1999; pp 155--164\relax
\mciteBstWouldAddEndPuncttrue
\mciteSetBstMidEndSepPunct{\mcitedefaultmidpunct}
{\mcitedefaultendpunct}{\mcitedefaultseppunct}\relax
\EndOfBibitem
\bibitem[Moody \latin{et~al.}(1995)Moody, Hanson, Krogh, and
  Hertz]{moody1995simple}
Moody,~J.; Hanson,~S.; Krogh,~A.; Hertz,~J.~A. A simple weight decay can
  improve generalization. \emph{Advances in Neural Information Processing
  Systems} \textbf{1995}, \emph{4}, 950--957\relax
\mciteBstWouldAddEndPuncttrue
\mciteSetBstMidEndSepPunct{\mcitedefaultmidpunct}
{\mcitedefaultendpunct}{\mcitedefaultseppunct}\relax
\EndOfBibitem
\bibitem[{Kingma} and {Ba}(2014){Kingma}, and {Ba}]{kingma2014adam}
{Kingma},~D.~P.; {Ba},~J. {Adam: A Method for Stochastic Optimization}.
  \emph{arXiv e-prints} \textbf{2014}, arXiv:1412.6980\relax
\mciteBstWouldAddEndPuncttrue
\mciteSetBstMidEndSepPunct{\mcitedefaultmidpunct}
{\mcitedefaultendpunct}{\mcitedefaultseppunct}\relax
\EndOfBibitem
\end{mcitethebibliography}

\end{document}